\documentclass[aps,pre,amsmath,amsfonts,amssymb,nofootinbib]{revtex4}

\usepackage{graphicx}
\usepackage{epsfig}
\usepackage{epstopdf}

\def\s{\sigma}
\def\us{\underline{\sigma}}
\def\t{\tau}
\def\ut{\underline{\tau}}

\def\eqd{\overset{\rm d}{=}}
\def\neqd{\overset{\rm d}{\neq}}
\def\E{\mathbb{E}}
\def\I{\mathbb{I}}

\def\N{{\cal N}}

\def\da{\partial a}
\def\di{\partial i}

\def\cN{{\cal N}}
\def\hW{{\mathfrak h}}
\def\thW{\widetilde{{\mathfrak h}}}
\def\uW{{\mathfrak u}}
\def\vW{{\mathfrak v}}

\def\da{\partial a}
\def\dami{\partial a \setminus i}
\def\di{\partial i}
\def\dima{\partial i \setminus a}
\def\dpima{\partial_+ i(a)}
\def\dmima{\partial_- i(a)}
\def\dpi{\partial_+ i}
\def\dmi{\partial_- i}
\def\dsi{\partial_\s i}

\def\homega{\widehat{\omega}}
\def\teta{\widetilde{\eta}}
\def\th{\widetilde{h}}

\def\WPeta{\mathfrak{n}}
\def\WPnu{\mathfrak{v}}
\def\WPteta{\widetilde{\mathfrak{n}}}
\def\WPf{\mathfrak{f}}
\def\WPg{\mathfrak{g}}

\def\dd{{\rm d}}
\def\ve{\varepsilon}

\begin{document}
\date{\today}

\title{On the cavity method for decimated random constraint satisfaction 
problems and the analysis of belief propagation guided decimation algorithms} 

\author{Federico Ricci-Tersenghi$^{\,1}$ and Guilhem Semerjian$^{\,2}$}
\affiliation{$^{1\,}$
Dipartimento di Fisica, Universit\`a di Roma La Sapienza,
P. A. Moro 2, 00185 Roma, Italy,
}
\affiliation{$^{2\,}$
LPTENS, Unit\'e Mixte de Recherche (UMR 8549) du CNRS et de l'ENS
associ\'ee \`a l'universit\'e Pierre et Marie Curie, 24 Rue Lhomond, 75231 
Paris Cedex 05, France.
}


\begin{abstract}

We introduce a version of the cavity method for diluted mean-field
spin models that allows the computation of thermodynamic quantities
similar to the Franz-Parisi quenched potential in sparse random graph
models. This method is developed in the particular case of partially
decimated random constraint satisfaction problems. This allows to
develop a theoretical understanding of a class of algorithms for
solving constraint satisfaction problems, in which elementary degrees
of freedom are sequentially assigned according to the results of a
message passing procedure (belief-propagation). We confront this
theoretical analysis to the results of extensive numerical simulations.

\end{abstract}

\maketitle

\section{Introduction}

In constraint satisfaction problems (CSP) a set of variables is
required to simultaneously satisfy a series of constraints. One can
equivalently define an energy function as the number of unsatisfied
constraints of a given assignment of the variables, and rephrase the
CSP as the quest for a zero energy groundstate configuration. This
analogy with low temperature physics triggered an intensive research
effort within the statistical mechanics community. More precisely, one
line of approach for these problems consists in looking for typical
properties of randomly generated large instances. This translates into
the presence of quenched disorder in the corresponding physical
model. The constraints to be satisfied are generally contradicting
each other, and the definition of the random instances does not
involve an underlying finite dimensional space; in consequence these
problems fall into the category of mean-field spin glasses, for which
a set of analytical tools have been developed during the last
decades~\cite{Beyond,Talagrand,IPC}.

The most famous example of random CSP is the random $k$-sat ensemble. 
Statistical mechanics studies have led to two kind of
results for this problem. On the one hand, qualitative and quantitative 
predictions
have been made about the various phase transitions encountered for the
typical behaviour of large instances when the control parameter
governing the amount of frustration is varied. The satisfiability
transition marks the sudden disappearance of the solutions (zero-energy
groundstates). There exist rigorous results on the properties of this
transition~\cite{Friedgut} and bounds~\cite{lb,ub} on its possible
location. Statistical mechanics has complemented these results with an
heuristic way to compute this threshold hence yielding quantitative
conjectures on its value~\cite{MeZe,MePaZe,MeMeZe}. Another important
contribution has been the suggestion of other phase transitions in the
satisfiable regime, that concerns the geometrical organization of the
set of solutions inside the configuration
space~\cite{BiMoWe,MeZe,MePaZe,pnas}.  For large enough frustration,
but below the satisfiability transition, the solutions can be grouped
in clusters of nearby solutions, each cluster being separated from the
others.

On the other hand attention has also been paid to algorithmic issues,
that is to procedures aiming at solving CSP, by finding their
solutions or proving that no solution exist. These algorithms can be
roughly divided in two broad categories: local search and sequential
assignment procedures.  In the first one, which has also been studied
with statistical mechanics methods~\cite{wsat1,wsat2,wsat3}, a random
walk is performed in the configuration space, with transition rules
tuned to bias the walk towards the solutions. This kind of algorithm
is called incomplete: it cannot prove the absence of solution if it
fails to find one. The second category proceeds differently: at some
step of the algorithm only one part of the variables has a definite
value, the others being still free. Each step thus corresponds to the
choice of one free variable and of the value it will be assigned to,
the CSP on the remaining variables being consequently simplified. The
heuristics guiding these choices can be more or less elaborate. In the
simplest cases one only takes into account simple properties of the
free variables, such as the number of their occurences in the
remaining CSP. A rigorous analysis of these simple (``myopic'')
approaches is possible and is at the basis of most of the lower bounds
on the satisfiability transition~\cite{lb}.  Such algorithms can be
made complete if backtracking is allowed, i.e. choices which have led
to a contradiction can be corrected in a systematic way. This complete
version of the procedure is called the DPLL algorithm~\cite{DPLL}.

One outcome of the statistical mechanics studies of random CSP has
been the proposal of an incomplete sequential assignment algorithm
called Survey Propagation inspired decimation~\cite{MeZe,SID}, which
proved to be very efficient on satisfiable random instances close to
the satisfiability transition. This algorithm relies on the clustering
picture of the solution space in the satisfiable regime. Unfortunately
its theoretical analysis is much more difficult than for myopic
ones. Indeed the heuristics of choice of the variables to be assigned
is based on the result of a message passing iterative computation
which depends on the whole remaining CSP in an intricate way. More
generally the analysis of message-passing decimation procedures is
difficult and there are few results on this issue, with the notable
exception of~\cite{MosselPlanted,unif_sat,unif_sat2} for the so-called
Warning Propagation algorithm on overconstrained satisfiability
formulas.

In this paper we provide an analytical description of an algorithm
similar to Survey Propagation, yet simpler. It has been studied
numerically in~\cite{Pretti,Aurell}. A part of our results were published
in~\cite{allerton}, the method developed there was also applied to
another family of CSP in~\cite{ZdMe}. 
In the sequential assignment procedure under
investigation the choice of the value of the assigned variable is made
at each step according to the Belief Propagation message passing
algorithm (instead of Survey Propagation). It aims at mimicking the
following ideal procedure. After a certain number of variables has
been assigned, one can define the uniform probability measure over the
solutions of the CSP which are still compatible with the previous
choices. If one were able to compute the marginal probabilities of
this (conditional) probability measure and use them to draw the value
of the newly assigned variable at each step, one would construct an
uniform sampler of the solutions of the original CSP, and this would
in particular lead to an algorithm for finding one solution of the
CSP. The computation of these marginal probabilities is a
computationally intractable task; Belief Propagation is a fast
heuristic algorithm, widely used for inference
problems~\cite{factor,Yedidia}, which is often able to compute good
approximations of these marginal probabilities. Analyzing the
behaviour of the Belief inspired decimation procedure thus amounts to
control the error which accumulates at each step by using the BP
approximate estimates of the marginal probabilities instead of the
exact ones. A theoretical understanding and quantitative description of
the deviations between exact and BP-computed marginal probabilities for
graphical models is a formidable open problem that we shall not attack
directly in this paper. We will instead perform
a theoretical analysis of the putative algorithm based
on an hypothetic exact marginal computation. This analysis will be
obtained by a generalization of the cavity method which is able to
deal with the partially decimated CSP encountered along the execution
of the algorithm, and to compute the extended phase diagram of these
problems.  This approach is technically similar to the computation of
Franz-Parisi quenched potentials~\cite{potential}. The relevance of
this theoretical analysis for the understanding of the approximate BP
implementation will then be argued on the basis of a comparison with
extensive numerical simulations.

The paper is organized as follows. In Sec.~\ref{sec_thought} we give a
more precise definition of the ideal decimation procedure sketched
above and explain how an approximate realization of this idea can be
performed in practice. Sec.~\ref{sec_cav} is devoted to the cavity
method for decimated formulas that provides an analytical description
of the ideal decimation procedure. In the next two sections we apply
this formalism to two specific CSP, and compare its predictions to the
results of numerical simulations of the BP guided decimation
algorithm. We begin in Sec.~\ref{sec_xor} with the xor-satisfiability
problem, a well-studied simple example for which many results can be
checked with alternative techniques.  We then turn to the case of
$k$-satisfiability random instances in Sec.~\ref{sec_sat}. We draw our
conclusions in Sec.~\ref{sec_conclu}. More technical details are
deferred to an Appendix.

\section{A thought experiment}
\label{sec_thought}

\subsection{Definition of random CSPs and a brief review of their properties}

A Constraint Satisfaction Problem (CSP) is defined on a set of $N$ variables
$\s_1,\dots,\s_N$, taking values in a finite alphabet. We shall
denote $\us=(\s_1,\dots,\s_N)$ the global configuration of the variables,
and for a subset $S$ of the indices $\{1,\dots,N\}$ we call 
$\us_S=\{ \s_i , i \in S\}$ the partial configuration of the variables in
$S$. The solutions of the CSP are the configurations which simultaneously
satisfy $M$ constraints (also called clauses in the following), 
each of them being specified by a function
$\psi_a(\us_{\da})$ of a subset $\da$ of the variables. The function 
$\psi_a$ takes value 1 (resp. 0) whenever the constraint is satisfied 
(resp. unsatisfied). A CSP admits a natural representation in terms
of a factor graph~\cite{factor}, i.e. a bipartite graph where one type 
of vertex 
(variable node) is associated to each variable $i=1,\dots,N$, another type
(function node) to each constraint $a=1,\dots,M$. An edge links the $i$'th
variable node with the $a$'th function node whenever the constraint $\psi_a$
depends on $\s_i$, i.e. in the notation introduced above whenever 
$i \in \da$. We shall similarly denote $\di$ the set of function nodes which
depend on the $i$'th variable, and define the distance between two variable
nodes $i$ and $j$ as the minimal number of constraint nodes encountered on 
a path of the factor graph joining $i$ and $j$.

In the following we will concentrate on two examples of CSP, 
both on binary variables that we shall represent by Ising spins, 
$\s_i = \pm 1$:
\begin{itemize}

\item[$\bullet$] $k$-xorsat. Each constraint $a$ depends on $k$ distinct
variables $\da= \{ i_1^a,\dots,i_k^a \}$, and requires the product of the 
corresponding spins to take a given value $J^a \in \{-1,+1\}$:
\begin{equation}
\psi_a (\us_{\da}) = \I \left( \prod_{i \in \da} \s_i = J^a \right) \ ,
\label{eq_psi_xorsat}
\end{equation}
where here and in the following $\I(\cdot)$ is the indicator function of
an event. This condition is easily seen to be equivalent to a constraint on 
the value of the eXclusive OR of $k$ boolean variables, hence the name of the
problem.

\item[$\bullet$] $k$-sat. The constraint $a$ depends again on 
$\da= \{ i_1^a,\dots,i_k^a \}$, but imposes the configuration of these
$k$ variables to avoid one out of the $2^k$ possible ones,
\begin{equation}
\psi_a (\us_{\da}) = 1 - \I \left( \s_i = J_i^a \ \forall i \in \da \right) \ ,
\label{eq_psi_sat}
\end{equation}
where $(J_{i_1^a}^a,\dots,J_{i_k^a}^a) \in \{-1,+1 \}^k$ are fixed constants 
defining the constraint. Equivalently, one requires the logical OR of
$k$ literals (a boolean variable or its negation) to evaluate to TRUE.
\end{itemize}

From a computational complexity point of view these two problems are very 
different. The decision version of a CSP consists in determining whether
it admits at least one solution, i.e. one configuration satisfying all
constraints simultaneously. The $k$-xorsat decision problem belongs to the
easy, polynomial P complexity class~\cite{NP} for any value of $k$. 
One can indeed use Gaussian elimination to check if the associated system of 
linear equations modulo 2 is solvable. $k$-sat is on the contrary NP-complete 
for all $k \ge 3$: no algorithm able to decide the satisfiability of every
$k$-sat formula in a time bounded by a polynomial of the formula size is
known.

Despite this deep difference in the worst-case point of view, these two
families of problems share common features in their ``average complexity'' 
behavior. By this we mean the random ensembles of instances that have been
extensively studied in the computer science and statistical physics literature
and that are defined as follows. A random $k$-xorsat formula is generated
by drawing in an independent, identical way $M$ constraints; the $k$-uplet
of indices $\da$ is drawn uniformly among the $\binom{N}{k}$ possible ones,
and the coupling constant $J^a$ is taken to be $\pm 1$ with equal probability
one half. The generation of a random $k$-sat formula is similar, with for
all constraints $a$ the $k$ constants $\{J_i^a\}_{i \in \da}$ being taken
independently equal to $\pm 1$ with equal probability. These random ensembles
exhibit a rich phenomenology in the thermodynamic limit 
$N,M \to \infty$ with $\alpha = M /N$ fixed. In particular a satisfiability
phase transition occurs at a value $\alpha_{\rm s}$ (which depends on the
value of $k$ and on the problem, sat or xorsat, under consideration): random
formulas with $\alpha<\alpha_{\rm s}$ are, with high probability, satisfiable,
whereas for $\alpha > \alpha_{\rm s}$ they are unsatisfiable. Here and in the
following ``with high probability'' (w.h.p.) means with a probability going
to one in the above stated thermodynamic limit. To be
more precise, for xorsat this statement has been proven and the values of
$\alpha_{\rm s}$ have been computed~\cite{xor1,xor2}. For sat random formulas
this satisfiability transition is, strictly speaking, only a conjecture. The
existence of a tight threshold $\alpha_{\rm s}(N)$ has been proven 
in~\cite{Friedgut}, but not the convergence of $\alpha_{\rm s}(N)$, that could
in principle oscillate between the bounds on its possible 
location~\cite{lb,ub}. It is however most probable that the values of 
$\alpha_{\rm s}$ computed within the statistical mechanics 
framework~\cite{MeZe,MePaZe,MeMeZe} are exact.

Besides the satisfiability transition other interesting phenomena occurs
in the satisfiable phase $\alpha<\alpha_{\rm s}$. In this regime the
formula are w.h.p. satisfiable and in fact admits an exponential number
of solutions; however there are structural phase transitions at which
the properties of the set of solutions change qualitatively. To describe
the set of solutions it is convenient to introduce the uniform probability
measure on this set,
\begin{equation}
\mu(\us) = \frac{1}{Z} \prod_{a=1}^M \psi_a(\us_{\da}) \ ,
\label{eq_def_mu}
\end{equation}
where the normalizing factor $Z$ is the number of solutions of the CSP.
Note that this probability measure is itself a random object, as it depends
on the instance of the CSP under study, and that it is defined only for
the satisfiable instances, which is the case w.h.p. in the regime
$\alpha < \alpha_{\rm s}$ we are considering here.

In the xorsat ensemble of random formulas
there is a single structural phase transition in the satisfiable
regime~\cite{xor1,xor2}, known as the clustering transition with a threshold
denoted $\alpha_{\rm d}$. For lower values of the connectivity $\alpha$,
the set of solutions is well-connected. On the contrary for 
$\alpha_{\rm d} < \alpha < \alpha_{\rm s}$ the exponential number of solutions
gets splitted in an exponential number $\exp[N\Sigma]$ of clusters, 
separated from each other in the configuration space. The rate of growth 
$\Sigma$ of the number of clusters is usually called the complexity,
it decreases when $\alpha$ grows and vanishes at the satisfiability threshold. 
In the clustered phase each cluster contains the same exponential number of 
solutions (with a rate of growth called internal entropy). The structure of 
xorsat is sufficiently simple for a clear-cut definition of the clusters to be 
possible. In fact the clustering transition corresponds to a percolation 
transition in the associated factor graph, where an extensive 2-core 
discontinuously appears.

The structure of the satisfiable phase of the random satisfiability
ensemble is richer~\cite{pnas,sat_long}.
At the clustering transition~\cite{BiMoWe,MeZe} the exponentially numerous
clusters have, contrarily to xorsat, a large diversity of internal entropies.
This leads, for $k \ge 4$, to another transition, the condensation one
at $\alpha_{\rm c}$. When $\alpha_{\rm d} < \alpha < \alpha_{\rm c}$
the measure (\ref{eq_def_mu}) is splitted in an exponential number of
clusters, while for $\alpha_{\rm c} < \alpha < \alpha_{\rm s}$ almost all
solutions are contained in a sub-exponential number of clusters.
These various phase transitions can be characterized in terms of the
strength of the correlations between variables. The clustering 
$\alpha_{\rm d}$ is related to the appearance of long-range point-to-set
correlations, or in other words to the possibility of reconstruction
of the value of a variable given the values of all the variables at a large
distance from it~\cite{MeMo}. At the condensation transition non trivial
correlations are already revealed by the correlation functions between a finite
number of variables~\cite{pnas}.

\subsection{Oracle guided algorithm and ensemble of decimated CSPs}
\label{sec_oracle}

The study presented in this paper is based on the analysis of an ideal
procedure to find the solutions of a CSP, that we discuss here more 
precisely. Consider a satisfiable
CSP instance, and the uniform measure $\mu(\cdot)$ over its solutions
defined in~(\ref{eq_def_mu}).
Let us also introduce a subset $D$ of the variables, and a partial 
configuration of these variables $\ut_D$ compatible with at least one
solution of the instance. We can thus define a conditional version of $\mu$,
$\mu(\cdot | \ut_D)$, which is the uniform measure over the solutions of
the formula compatible with $\ut_D$:
\begin{equation}
\mu(\us|\ut_D) = \begin{cases}
\frac{1}{Z(\ut_D)} \prod_{a=1}^M \psi_a(\us_{\da}) & \text{if} \
\us_D = \ut_D \\ 0 & \text{otherwise} 
\end{cases} \ .
\label{eq_def_mu_cond}
\end{equation}
The normalization $Z(\ut_D)$ counts the number of solutions compatible with the 
partial assignment of the variables in $D$.

A possible procedure for sampling from $\mu(\cdot)$ goes as follows.
Choose
arbitrarily a permutation of $\{1,\dots,N\}$, denoted $i(1),\dots,i(N)$,
and call $D_t = \{i(1),\dots,i(t)\}$ for $t=1,\dots,N$, $D_0=\emptyset$.
Construct now sequentially a configuration $\ut$, assigning at time $t$
the value $\t_{i(t)}$; to this aim draw $\s_{i(t)}$ according to the
marginal of the conditioned measure $\mu(\cdot | \ut_{D_{t-1}})$, and set
$\t_{i(t)}=\s_{i(t)}$. It is easy to see that after $t$ steps of the
algorithm the partial configuration $\ut_{D_t}$ is distributed according 
to the marginal law of $\mu(\cdot)$. In particular the final configuration
$\ut$ obtained when the $N$ variables are assigned is a uniformly chosen
solution of the CSP.

This simple algorithm would thus provide a uniform sampler of the solution
set of any CSP; it is however only meant as a thought experiment. Indeed,
computing exactly the probabilities $\mu(\s_{i(t)} | \ut_{D_{t-1}})$ is in general
a $\#$P-complete problem, with no polynomial algorithm known until now, and we
shall thus content ourselves with faster yet approximative means for
computing these marginal probabilities. Before introducing them let us
discuss further the idealized procedure.

The analytical description of the dynamics followed by this ideal process
seems very difficult: at each time step the probability of the evolution
$\ut_{D_{t-1}} \to \ut_{D_t}$ depends in a non-trivial way on all the choices
made in the previous steps. However the description of the process at a given
point of its evolution is very simple. As noted above $\ut_{D_t}$ is 
distributed according to the marginal of $\mu(\cdot)$. One can state this
in a slightly different way: $\ut_{D_t}$ can be obtained by drawing uniformly
a solution $\ut$ from $\mu(\cdot)$, retaining the configuration of the 
variables in $D_t$, and discarding the rest of the configuration. 
We shall further
assume that the permutation $i(1),\dots,i(N)$ is drawn uniformly at random,
such that $D_t$ is a random set of $t$ variables among $N$. In the 
thermodynamic limit we shall define $\theta=t/N$, the fraction of assigned 
variables, and consider for simplicity that $D_{\theta N}$ is built
by retaining independently each variable with probability $\theta$ (we only
make an error of order $1/\sqrt{N}$ on the fraction of variables thus 
included in $D$).

These considerations lead us to the definition of an ensemble of CSP instances
parametrized by $\alpha$ and $\theta$, generalizing the original one which
corresponds to $\theta=0$.  Explicitly this
ensemble of formulas corresponds to the following generation process: 
\begin{enumerate}
\item draw a satisfiable CSP with parameter $\alpha$ 
\item draw a uniform solution $\ut$ of this CSP 
\item choose a set $D$ by retaining each variable independently
with probability $\theta$ 
\item consider the residual formula on the variables outside $D$ 
obtained by imposing the allowed configurations to coincide with $\ut$ on $D$.
\end{enumerate}
Let us emphasize that, apart from simple cases like the
xorsat model, these ensembles do not coincide in general with randomly 
uniform formulas
conditioned on their degree distributions. The fact that the generation of the
configuration $\ut$ depends on the initial CSP induces non-trivial
correlations in the structure of the final formula.

We shall see in the following how to adapt
the statistical mechanics techniques to compute the typical properties of
such generalized formulas, and in particular to determine the phase 
transition thresholds in the $(\alpha,\theta)$ plane. One characterization
of these random ensembles is the quenched average residual entropy,
\begin{equation}
\omega(\theta) = \lim_{N \to \infty} \frac{1}{N} \E_F \E_{\ut}  \E_D
[\ln Z(\ut_D) ] \ , \qquad 
Z(\ut_D) = \sum_{\us} \prod_{a=1}^M \psi_a(\us_{\da}) \, \I(\us_D = \ut_D ) \ ,
\label{eq_omega}
\end{equation}
where the three expectation values correspond to the three steps of the 
definition above. This quantity is similar, yet distinct, from the 
Franz-Parisi quenched potential~\cite{potential}. 
The definition of the latter also involves a ``thermalized'' 
reference configuration $\ut$, but is given by the free-energy of the
measure on the configurations at a given Hamming distance from $\ut$. In
other words the two real replicas $\us$ and $\ut$ are coupled uniformly 
across the variables in Franz-Parisi quenched potential,
whereas in the definition of $\omega$ they are coupled infinitely strongly
on $D$ where they are forced to coincide, and not at all outside $D$. The
computations presented in the rest of the paper can however be easily adapted
to obtain the usual quenched potential.

We shall characterize the reduced measure $\mu(\cdot | \ut_D)$ more precisely
by computing other quantities besides $\omega(\theta)$. The existence of 
clusters in this measure will be tested by the computation of the long-range 
point-to-set correlations and the complexity of the typical clusters.

\subsection{Bethe-Peierls approximation for decimated CSPs}

We recall in this section the Bethe-Peierls approximation for statistical
models defined on factor graphs and show how to adapt it to partially 
decimated CSPs. 
Let us first consider a probability measure with a weight funcion which can
be factorized as in Eq.~(\ref{eq_def_mu}), with $\psi_a$ some a priori
arbitrary positive functions. The Bethe approximation for the computation
of the partition function $Z$ consists in extremizing the following
expression,
\begin{equation}
\ln Z = - \sum_{i, a \in \di} \ln \left(
\sum_{\s_i} \nu_{a \to i} (\s_i) \eta_{i \to a}(\s_i) \right)
+ \sum_a \ln \left( \sum_{\us_{\da}} \psi_a(\us_{\da}) 
\prod_{i \in \da} \eta_{i \to a}(\s_i) \right)
+ \sum_i \ln \left( \sum_{\s_i} \prod_{a \in \di} \nu_{a \to i}(\s_i) \right) 
\label{eq_lnZ}
\end{equation}
over the unknown $\{\nu_{a \to i},\eta_{i \to a}\}$. These are probability 
measures on the alphabet of $\s_i$, defined on the directed edges of the 
factor 
graph, which we shall call messages for reasons that will become clear below. 
The extremization of the Bethe approximation for $\ln Z$ leads to
a set of equations between the messages,
\begin{equation}
\nu_{a \to i}(\s_i) = f(\{\eta_{j \to a}\}_{j \in \dami} )
\ , \qquad
\eta_{i \to a}(\s_i) = g(\{\nu_{b \to i} \}_{b \in \dima} ) \ ,
\label{eq_BP1}
\end{equation}
where the (edge-dependent) functions $f$ and $g$ are defined by
\begin{equation}
\nu_{a \to i}(\s_i) = \frac{1}{z_{a \to i}} 
\sum_{\us_{\dami}} \psi_a(\us_{\da}) 
\prod_{j \in \dami} \eta_{j \to a}(\s_j) \ , \qquad
\eta_{i \to a}(\s_i) = \frac{1}{z_{i \to a}} 
\prod_{b \in \dima} \nu_{b \to i}(\s_i) \ ,
\label{eq_BP2}
\end{equation}
with $z_{a \to i}$ and $z_{i \to a}$ ensuring the normalization of 
$\nu_{a \to i}$ and $\eta_{i \to a}$. 
When the factor graph is a tree the log partition
function is exactly given by~(\ref{eq_lnZ}) evaluated on the unique solution
of the stationarity equations~(\ref{eq_BP2}), see for instance~\cite{factor}.
The messages $\nu_{a \to i}$ (resp. $\eta_{i \to a}$) are then the marginal
probabilities for $\s_i$
of a modified measure corresponding to a factor graph where all 
factor nodes around $i$ except $a$ have been removed (resp. only $a$ has been
removed). From the knowledge of the messages solution of~(\ref{eq_BP2}) one
can compute the marginal probability of the variables in the full factor
graph law~(\ref{eq_def_mu}), for instance the marginal probability of 
variable $i$ reads
\begin{equation}
\frac{1}{z_i} \prod_{a \in \di} \nu_{a \to i}(\s_i) \ , 
\label{eq_marginal}
\end{equation}
with again $z_i$ fixed by normalization. In general factor graphs do contain
loops, in that case (\ref{eq_lnZ},\ref{eq_BP2},\ref{eq_marginal}) are only
approximations, at the basis of the so-called Belief Propagation algorithm
discussed in more details below.

The Bethe approximation can be easily adapted to the case where 
the configuration is forced to the value $\ut_D$ on a subset of the sites
$i \in D$, that is to the conditional measure~(\ref{eq_def_mu_cond}). 
The estimation of the conditioned log partition function follows 
from~(\ref{eq_lnZ}):
\begin{equation}
\ln Z(\ut_D) = - \sum_{i \notin D, a \in \di} \ln \left( 
\sum_{\s_i} \nu_{a \to i}^{\ut_D} (\s_i) \eta_{i \to a}^{\ut_D}(\s_i) \right)
+ \sum_a \ln \left( \sum_{\us_{\da}} \psi_a(\us_{\da}) 
\prod_{i \in \da} \eta_{i \to a}^{\ut_D}(\s_i) \right)
+ \sum_{i \notin D} \ln \left(
\sum_{\s_i} \prod_{a \in \di} \nu_{a \to i}^{\ut_D}(\s_i) \right) \ ,
\label{eq_lnZ_ut_D}
\end{equation}
where the messages $\{\eta_{i \to a}^{\ut_D} , \nu_{a \to i}^{\ut_D} \}$
depend on the imposed partial configuration $\ut_D$. They indeed obey the same 
equations~(\ref{eq_BP2}), complemented with the boundary conditions
$\eta_{i \to a}^{\ut_D}(\s_i) =\delta_{\s_i,\t_i}$ when $i \in D$. 

\subsection{Practical approximate implementation of the thought experiment}
\label{sec_BPguided}
The ideal sampling algorithm described in Sec.~\ref{sec_oracle} cannot
be practically implemented, because the computation of the marginals
of the probability law $\mu(\us|\ut_D)$ has generically a cost exponential
in the number of variables. One can however mimic this procedure, using
a faster yet approximate estimation of the marginals of $\mu(\us|\ut_D)$
by means of the Belief Propagation algorithm. This modification
of the ideal sampler, which will be called BP guided decimation in the
following, thus corresponds to (for a given CSP instance):

\begin{enumerate}
\item choose a random order of the variables, $i(1),\dots,i(N)$, call
$D_0=\emptyset$, $D_t=\{i(1),\dots,i(t)\}$

\item for $t=1,\dots,N$:
\begin{enumerate}
\item find a fixed point of the BP equations (\ref{eq_BP1}) with the boundary
conditions $\eta_{i \to a}(\s_i)=\delta_{\s_i,\t_i}$ when $i \in D_{t-1}$
\item draw $\s_{i(t)}$ according to the BP estimation of 
$\mu(\s_i|\ut_{D_{t-1}})$ given in (\ref{eq_marginal})
\item set $\t_{i(t)}=\s_{i(t)}$
\end{enumerate}
\end{enumerate}

The Belief Propagation part of the algorithm corresponds to step 2.(a). It
amounts to search for a stationary point of the Bethe approximation for
the log partition function, in an iterative manner. The unknowns of the
Bethe expression, $\eta_{i \to a}$ (resp. $\nu_{a \to i}$), are considered as
messages passed from a variable to a neighboring clause (resp. from a clause to
a variable). In a random sequential order a message, say $\eta_{i \to a}$, 
updates itself by recomputing its value from the current messages sent
by its neighbors $\{\nu_{b \to i} \}_{b \in \dima}$, according to the equation
in~(\ref{eq_BP2}). If the factor graph of the formula were a tree, these 
iterations would converge in a finite number of updates to the unique fixed 
point solution of~(\ref{eq_BP2}). On generic factor graphs there is no 
guarantee of convergence of these iterations, in practical implementations
one has thus to precise the definition of the algorithm, giving criterions
to stop the iterations of the BP updates, we shall come back to this point
in Sec.~\ref{sec_sat_results}.

The definition of the probability measure conditioned on the choice of
the reference configuration $\ut_D$~(\ref{eq_def_mu_cond}) and the subsequent 
derivation of the BP equations only make sense if the formula admits at 
least one solution compatible with $\ut_D$. In the analysis of the ideal
algorithm this is automatically the case as soon as the initial formula is
satisfiable. However this can fail in the course of the BP guided decimation 
algorithm because the marginals used to generate the configuration $\ut$ are
only approximate. The BP equations are no longer well defined when there are
no solutions of the formula compatible with $\ut_D$. This shows up for instance
in the computation of the message sent by a variable $i$ to a clause $a$; 
whenever the product $\prod_{b \in \dima} \nu_{b \to i}(\s_i)$ vanishes for
all possible values of $\s_i$ the message $\eta_{i \to a}$ can no longer be 
normalized, a contradiction has occured between the strong requests imposed
by the clauses in $\dima$. The BP guided decimation algorithm has then to 
stop and fails to construct a solution of the formula.

This mechanism which unveils the contradictions in the choice of $\ut_D$, and
the fact that no solution is compatible with it, is actually equivalent to the
Unit Clause Propagation (UCP) algorithm well known in computer 
science. For concreteness let us recall its functioning in the case of 
satisfiability formulas. UCP takes in input a list of variables and a list of
clauses. If all clauses have length greater or equal to 2 it stops. 
Otherwise it chooses one of the unit clauses (i.e. of length 1).
The variable in this unit clause must be fixed to the value satisfying the
clause for the constructed configuration of variables to be a solution of the 
formula. The logical implications of this assignment are then
drawn. All the clauses where the fixed variable appeared with the same
sign as in the unit clause can be removed from the formula, as they
are automatically satisfied. All clauses where it appeared with the
opposite sign are effectively reduced in length, as the fixed variable will 
never satisfy them. This process
is iterated as long as unit clauses are present in the formula. If clauses
of length 0 are produced during the propagation of logical implications, then
the input formula was not satisfiable: the logical implications required at 
least one of the variables to take simultaneously its two possible values.
If no contradictions occur, the set of variables that appeared in unit clauses
during the process are termed logically implied, their value being uniquely
determined by the input formula.

It turns out that an equivalent formulation of the UCP rule for drawing
logical implications can be given in terms of a message passing procedure
known as Warning Propagation (WP)~\cite{SID}. The messages 
$\{\WPeta_{i \to a} , \WPnu_{a \to i} \}$ 
that WP sends 
along edges of the factor graph are projections of those of BP, where the only 
informations retained is whether a value of the variable $\s_i$ is authorized
($\eta_{i\to a}(\s_i)>0$) or not ($\eta_{i\to a}(\s_i)=0$):
\begin{equation}
\WPeta_{i \to a}(\s_i)=\I (\eta_{i\to a}(\s_i)>0  ) \ , \qquad
\WPnu_{a \to i}(\s_i)=\I (\nu_{a\to i}(\s_i)>0  ) \ .
\label{eq_projection}
\end{equation}
The projection of the BP equations~(\ref{eq_BP1}) leads to recurrence
equations on the WP messages,
\begin{equation}
\WPnu_{a \to i} = \WPf(\{\WPeta_{j \to a}\}_{j \in \dami} )
\ , \qquad
\WPeta_{i \to a} = 
\WPg(\{\WPnu_{b \to i} \}_{b \in \dima} ) \ .
\label{eq_WP}
\end{equation}
Monotonicity arguments can be used to show 
that these recurrence equations, initialized with the permissive value
of the messages $\WPeta_{i \to a}(\s_i)=1$, converge to a unique fixed
point independent on the order of updates of the messages. Moreover
this fixed point contains the same information as revealed by UCP: a
contradiction occurs in UCP if and only if there is a variable $i$ such that
\begin{equation}
\prod_{a \in \di} \WPnu_{a \to i}(\s_i) = 0 \ \ \forall \ \s_i \ .
\end{equation}
If no contradiction occurs the set of variables logically implied by UCP
corresponds to the variables $i$ such that there is only one authorized 
value $\s_i$ for it,
\begin{equation}
\exists\ ! \ \s_i \ : \ \prod_{a \in \di} \WPnu_{a \to i}(\s_i) = 1 \ .
\end{equation}
This correspondance was explicitly shown in the case of satisfiability 
formulas in~\cite{allerton}.
In our practical implementation of the BP guided decimation algorithm we 
check, after each assignment of a variable $\t_i(t)$, whether a contradiction 
in the partial configuration $\ut_{D_t}$ can be detected by UCP/WP. According
to the pseudocode above we do not immediately assign the variables which
are logically implied by $\ut_{D_t}$; please note however that this does not
modify at all the subsequent steps of the algorithm, as the BP equations
effectively take into account the effect of these logical implications.
We also emphasize the fact that in general there are variables which can only
take one value under the law $\mu(\cdot|\ut_D)$, yet that are not unveiled as
logically implied by UCP/WP; if this were the case UCP would always be
succesful on any satisfiable formula (and incidentally one would have P=NP),
which is of course well known to be wrong.

As a final remark on the BP guided decimation algorithm, let us emphasize
another difference with the theoretical analysis. In practice we apply
the algorithm to uniformly generated formula of CSP, with 
$\alpha<\alpha_{\rm s}$, which are typically satisfiable in the thermodynamic
limit, but we cannot systematically exclude unsatisfiable instances as in the
analysis of the ideal algorithm.

\section{Analysis of the thought experiment with the cavity method}
\label{sec_cav}

\subsection{Reminder on the usual cavity method}
\label{sec_cav_usual}

The goal of the cavity method is to compute the typical properties in the
thermodynamic limit of graphical models defined on random factor graphs, and
in particular the average entropy of the associated random CSP. In the
simplest situation, known as the replica symmetric (RS) case, the hypothesis
of the method is that the Bethe-Peierls approximation is asymptotically exact
for the large random factor graphs of the ensemble considered. For
concreteness we explain it in a setting encompassing the $k$-(xor)sat
formulas, that is where each of the $M=\alpha N$ constraints imply $k$
randomly chosen variables.
The prediction for the average log partition function then
follows by averaging~(\ref{eq_lnZ}),
\begin{eqnarray}
\lim_{N \to \infty} \frac{1}{N} \E[\ln Z] = 
&-&\alpha k \, \E\left[\ln\left(\sum_\s \nu(\s) \eta(\s)  \right)\right]  
\nonumber \\
&+& \alpha \, \E\left[\ln\left(\sum_{\s_1,\dots,\s_k} \psi(\s_1,\dots,\s_k) \,
\eta_1(\s_1) \dots \eta_k(\s_k)  \right)\right]
+ \E\left[\ln\left(\sum_\s \nu_1(\s) \dots \nu_l(\s) \right) \right] \ .
\label{eq_lnZ_av}
\end{eqnarray}
Let us detail the justification and the meaning of the right hand side of this
relation. We have first used the translational invariance in the definition of
the random ensemble: each term in the sums of Eq.~(\ref{eq_lnZ}) contributes
on average in the same way. Moreover we have introduced the random variable
$\eta$ (resp. $\nu$), whose distribution can be constructed as follows: 
drawing a random factor graph, finding the solution of the BP 
equations~(\ref{eq_BP1}), picking a random edge $i-a$ of the factor graph,
and setting $\eta=\eta_{i \to a}$ (resp. $\nu=\nu_{a \to i}$).
The averages in the right hand side of (\ref{eq_lnZ_av})
are thus over independent copies of $\eta$ and $\nu$,
over the random constraint function $\psi$, and over a Poisson random variable
$l$ of mean $\alpha k$. This last quantity is the degree of an uniformly
chosen variable inside such a factor graph.

The equations fixing the distributions of the random variables $\eta$ and $\nu$
can be obtained either looking for the stationary points of~(\ref{eq_lnZ_av})
or interpreting the BP equations~(\ref{eq_BP1}) 
in the random graph perspective.
Both reasoning leads to the distributional equations
\begin{equation}
\nu \eqd f(\eta_1,\dots,\eta_{k-1}) \ , \qquad 
\eta \eqd g(\nu_1,\dots,\nu_l) \ .
\label{eq_msg_RS}
\end{equation}
These equations have to be interpreted as equalities between distributions
of random variables. The $\eta_i$ (resp. $\nu_i$) are independent copies 
of the random variable $\eta$ (resp. $\nu$), $l$ is a Poissonian random
variable of parameter $\alpha k$, and $f$ and $g$ have been defined
in~(\ref{eq_BP1},\ref{eq_BP2}), $f$ being itself
random because of the choice of the constraint function $\psi$. 

The RS cavity method is based on the assumption of asymptotic correctness of
the Bethe-Peierls approximation, which can be rephrased as the existence of a
single pure state (or cluster) in the probability law $\mu$. A complementary
statement of the RS method concerns the local description of the law
$\mu$. Let us consider an arbitrary variable node $i_0$, and its depth $L$
neighborhood, that is the set of variables at graph distance smaller or equal
than $L$ from $i_0$. In such random graph ensembles this neighborhood is, with
high probability, a Galton-Watson random tree with Poissonian offspring of
mean $\alpha k$ for the variable nodes (the constraint nodes being of course
always of degree $k$). In the RS case the marginal of the law $\mu$ for the
variables in this neighborhood converges in distribution to the law of a
finite tree of depth $L$, the only effect of the rest of the graph being
summarized in a single message $\eta$ acting on the boundary (the depth $L$
variables), drawn independently with the fixed point distribution solution of
Eq.~(\ref{eq_msg_RS}). In other words the depth $L$ variables would be
considered independent if the constraints inside the depth $L$ neighborhood
around $i_0$ were removed.

In the context of random CSPs the RS assumption is only valid for small values
of the density of constraints $\alpha$. When this parameter grows the solution
space splits into a large number of pure states (clusters), which induces
non-trivial 
correlations between the variables of the graphical model. The cavity method
at the level of the first step of replica symmetry breaking (1RSB) is able to
describe this situation~\cite{MePa}. Let us sketch some of its important
features without entering into technical details. The assumption of the 1RSB
method is that the Bethe-Peierls approximation can still be used, but only for
the probability measures restricted to a single pure state; to each such pure
state is associated a solution of the BP equations~(\ref{eq_BP1}). In order to
handle the proliferation of pure states one introduces, on each directed edge
of a given factor graph, a probability distribution, with respect to the
choice of the pure states, of the corresponding BP messages. The 1RSB
equations linking these distributions on adjacent edges depend on the Parisi
parameter $m$, which controls the relative importance of the pure states
in the sampling of the corresponding BP messages, according to their internal
sizes. A slightly more explicit interpretation of the 1RSB equations has been
given in~\cite{pnas,sat_long}. There it was shown that the distribution over
pure states can be viewed as a distribution over 'far-away' boundary
conditions, with a specific probability measure over these boundary conditions
depending on $m$. A special role is played by the value $m=1$. In this
case the 'far-away' boundary conditions are themselves drawn from the original
Gibbs measure. The clustering transition, that is the appearance of a
non-trivial solution of the 1RSB equations at $m=1$, can thus be related to
the existence of long-range correlations of a particular type (so called
point-to-set) in the Gibbs measure, as first discussed in~\cite{MeMo}.
These correlations measure the influence on a variable $i_0$ of fixing a subset
$B$ of variables, for instance $B(i_0,\ell)$ the set of variables at distance
exactly $\ell$ from $i_0$, to a reference value drawn from the equilibrium
probability measure. Using the notation of conditional probability defined in
Eq.~(\ref{eq_def_mu_cond}) the typical long range point-to-set correlation
reads 
\begin{equation}
C_\infty = 
\lim_{\ell \to \infty} \lim_{N \to \infty} 
\E \sum_{\s_{i_0}} |\mu( \s_{i_0} | \ut_{B(i_0,\ell)} ) - \mu(\s_{i_0})| \ ,
\label{eq_correl}
\end{equation}
where the expectation is over the equilibrium configuration $\ut$ and the
factor graph model. In an unclustered regime the influence of the distant
boundary vanishes, $C_\infty=0$, while in the presence of clustering
$C_\infty>0$. In the latter case one can moreover compute the complexity of
relevant clusters, that is the logarithm of the degeneracy of the clusters
which bears the vast majority of the weights of the probability measures (at
the leading exponential precision) by computing the difference in entropy of
the conditional and original measures, $\mu( \cdot | \ut_{B(i_0,\ell)} )$ and
$\mu(\cdot)$. The so-called condensation transition is signalled by a
vanishing of this complexity. We shall come back in the next subsection on the
technical details of the 1RSB $m=1$ computations, generalizing it to the case
of partially decimated formulas.

\subsection{Cavity method for decimated CSP}
\label{sec_cav_dec}

We turn now to the extension of the cavity method to partially decimated
factor graphs. Our goal is to compute the residual entropy (\ref{eq_omega})
by averaging the Bethe expression (\ref{eq_lnZ_ut_D}) with respect to
the distribution of the conditional messages 
$\{\eta_{i \to a}^{\ut_D} , \nu_{a \to i}^{\ut_D} \}$. The randomness of these
objects has several origins: (i) the choice of the factor graph (ii) the
generation of the reference configuration $\ut$ from the uniform measure
$\mu$ (iii) the selection of the decimated variables in $D$, each 
independently with probability $\theta$. The difficulty of this computation
arises from the correlation between (i) and (ii), the measure $\mu$ being 
itself defined in terms of the factor graph. This dependence can however be
handled within the context of the cavity method. Let us suppose indeed that the
local properties of the original measure $\mu(\cdot)$ are well described by
the assumptions of replica symmetry, that is for $\alpha<\alpha_{\rm
  c}$\footnote{the local properties are indeed of a RS type also in the
  clustered uncondensed regime $\alpha \in [\alpha_{\rm d},\alpha_{\rm
    c}]$~\cite{pnas}.}. We can thus perform the computation in the random tree
model that corresponds locally to the random graph one. In this case
the generation of the reference configuration $\ut$ of a tree factor graph
model can be done recursively, in 
a broadcasting way, thanks to the Markov property of such probability laws. 
The value of the root $\t_{i_0}$ is first drawn with its marginal probability 
computed from the incoming messages according to (\ref{eq_marginal}). Then
the configuration of the neighbors of $i_0$ can be drawn conditioned on the
value of $\t_{i_0}$, and the process can be iterated away from the root. 
Once the
reference configuration $\ut$ has been generated in such a way, the messages
$\{\eta_{i \to a}^{\ut_D} , \nu_{a \to i}^{\ut_D} \}$ can be computed, their
dependence on $\ut$ arising from the condition 
$\eta_{i \to a}^{\ut_D}(\s_i) =\delta_{\s_i,\t_i}$ for variables $i$ in the 
decimated set $D$. At this point, for a given tree, reference
configuration $\ut$ and set $D$, each directed edge of the factor graph bears
a pair of messages, for instance $(\nu_{a \to i},\nu_{a \to i}^{\ut_D})$ 
on the edge from constraint $a$ to variable $i$. We can now define the
random variable $(\nu,\nu^\t)_\ell$ which has the distribution of 
$(\nu_{a \to i},\nu_{a \to i}^{\ut_D})$ when one takes into account the 
randomness in the choice of the tree, of the set $D$ and of the reference
configuration $\ut$, the latter being conditioned on $\t_i=\t$. Moreover the
positive integer $\ell$ indexes the depth of the random tree construction.
We shall also introduce random variables having the same distribution as
$(\eta_{i \to a},\eta_{i \to a}^{\ut_D})$. For the sake of clarity in the
following we actually introduce two versions of these random variables,
$(\eta,\eta^\t)_\ell$ and $(\eta,\teta^{\, \t})_\ell$, the former being
additionally conditioned on $i \notin D$. The equations defining these random
variables by recurrence on $\ell$ can now easily be written,
\begin{equation}
(\eta,\teta^{\, \t})_\ell \eqd \begin{cases} 
(\eta,\eta^\t)_\ell & \text{with probability} \ 1-\theta \\
(\eta, \delta^\t) & \text{otherwise} 
\end{cases} \ ,
\label{eq_pair_teta}
\end{equation}
where we defined $\delta^\t(\s)=\delta_{\t,\s}$, and
\begin{equation}
(\eta,\eta^\t)_\ell \eqd (g(\nu_1,\dots,\nu_l),g(\nu_1^\t,\dots,\nu_l^\t) )
\ ,
\label{eq_pair_eta}
\end{equation}
where $l$ is a Poisson random variable of parameter $\alpha k$ and
$(\nu_1,\nu_1^\t),\dots,(\nu_l,\nu_l^\t)$ are independent copies of 
$(\nu,\nu^\t)_\ell$. Finally one has
\begin{equation}
(\nu,\nu^\t)_{\ell +1} \eqd (f(\eta_1,\dots,\eta_{k-1}) , 
f(\teta_1^{\, \t_1},\dots,\teta_{k-1}^{\, \t_{k-1}} )) \ ,
\label{eq_pair_nu}
\end{equation}
where the $(\eta_i,\teta_i^{\, \t_i})$ are independent copies of 
$(\eta,\teta^{\, \t_i})_\ell$, and the configuration 
$\t_1,\dots,\t_{k-1}$ is drawn according to
\begin{equation}
P[\t_1,\dots,\t_{k-1} | \t ] = \frac{1}{z} \psi(\t, \t_1,\dots,\t_{k-1}) 
\eta_1(\t_1) \dots \eta_{k-1}(\t_{k-1}) \ .
\label{eq_broadcast}
\end{equation}
Let us emphasize that the function $\psi$ used in the broadcasting generation
of $\t_1,\dots,\t_{k-1}$ is the same (random) constraint function as the one
used to compute $f$ in (\ref{eq_pair_nu}), and that the messages $\eta_i$ are
the same in (\ref{eq_pair_nu}) and (\ref{eq_broadcast}).
We shall discuss a numerical procedure for solving these equations on the
example of satisfiability formulas in Sec.~\ref{sec_sat}.
The prediction of the cavity method for the average residual 
entropy~(\ref{eq_omega}) can finally be expressed in terms of these random
variables,
\begin{eqnarray}
\omega = &-&\alpha k (1-\theta) \, \E \left[ \sum_\t 
\frac{\nu(\t) \eta(\t)}{\sum_{\t'}\nu(\t') \eta(\t')} \ln\left( 
\sum_\s \nu^\t(\s) \eta^\t(\s) \right) \right] \nonumber \\
&+& \alpha \, \E \left[ \sum_{\t_1,\dots,\t_k} \frac{\psi(\t_1,\dots,\t_k) 
\eta_1(\t_1) \dots \eta_k(\t_k)}{\sum_{\t'_1,\dots,\t'_k} 
\psi(\t'_1,\dots,\t'_k) 
\eta_1(\t'_1) \dots \eta_k(\t'_k) } \ln \left( 
\sum_{\s_1,\dots,\s_k} \psi(\s_1,\dots,\s_k) \teta_1^{\, \t_1}(\s_1) \dots 
\teta_k^{\, \t_k}(\s_k) \right) \right] \nonumber \\
&+& (1-\theta) \, \E \left[ \sum_\t 
\frac{\nu_1(\t) \dots \nu_l(\t)}{\sum_{\t'}\nu_1(\t') \dots \nu_l(\t')}
\ln\left( \sum_\s \nu_1^\t(\s) \dots \nu_l^\t(\s) \right)
\right] \ ,
\label{eq_omega_cav}
\end{eqnarray}
where as before $l$ is a Poisson random variable of parameter $\alpha k$ and
the various $(\eta_i,\eta_i^{\t_i})$ are independent copies of the
corresponding random variables, with the $\ell \to \infty$ limit kept 
understood.

We shall now discuss an important issue that we kept under silence, namely the
definition of the initial condition $(\eta,\eta^\t)_{\ell=0}$. Let us first
make the connection between the computation just presented and the $m=1$ 1RSB
description of non-decimated formulas, that is consider for a while the case
$\theta=0$. We shall call $I_0$ the initial condition 
$(\eta,\eta^\t)_{\ell=0} \eqd (\eta,\eta)$, with $\eta$ a random variable 
solution of the RS fixed point equation (\ref{eq_msg_RS}). 
It is easy to show that with such an initial condition 
$(\eta,\eta^\t)_\ell \eqd (\eta,\eta)$ for all values of $\ell$, and that 
(\ref{eq_omega_cav}) reduces to the RS prediction (\ref{eq_lnZ_av}) for the
average entropy. If on the contrary one considers the initial condition $I_1$
defined by 
$(\eta,\eta^\t)_{\ell=0} \eqd (\eta,\delta^\t)$ and iterates the recursion 
equations
(\ref{eq_pair_teta},\ref{eq_pair_eta},\ref{eq_pair_nu}), one reproduces the
$m=1$ computations in the form of~\cite{MeMo,sat_long}. This should not be a
surprise~: the interpretation of the $m=1$ 1RSB formalism presented above
was precisely the study of the effect of a far away boundary, which is
implemented here in this initial condition and in the limit $\ell \to \infty$.
In an unclustered phase the two initial conditions lead to the same random
variable $(\eta,\eta)$ in the large $\ell$ limit, the effect of the far away
boundary vanishes at long distance and the correlation function
(\ref{eq_correl}) is equal to 0. If on the contrary the two initial conditions
do not lead to the same limit when $\ell$ diverges the correlation function
remains positive, signaling the presence of clustering in the solution space.
In that case the value of (\ref{eq_omega_cav}) computed with the initial 
condition $I_1$ is the internal entropy of the
relevant clusters, the difference between (\ref{eq_lnZ_av}) and 
(\ref{eq_omega_cav}) is ascribed to the complexity, the degeneracy of the
relevant clusters. We now come back to the decimated case $\theta>0$. The two
initial conditions are still relevant and have the same interpretation with
the measure $\mu(\cdot)$ replaced by $\mu(\cdot|\ut_D)$. Suppose indeed that
the point-to-set correlation criterion were to be tested for the typical
properties of $\mu(\cdot|\ut_D)$. Then one would compute the generalization of
(\ref{eq_correl}),
\begin{equation}
C_\infty (\theta) = 
\lim_{\ell \to \infty} \lim_{N \to \infty} 
\E \sum_{\s_{i_0}} |\mu( \s_{i_0} | \ut_{D \cup B(i_0,\ell) } ) - 
\mu(\s_{i_0}|\ut_D)| \ ,
\label{eq_correl2}
\end{equation}
where the expectation would include an average over the set $D$ of $\theta N$
variables. This is precisely what is realized by the two initial conditions
followed by the recursion relations
(\ref{eq_pair_teta},\ref{eq_pair_eta},\ref{eq_pair_nu}) for this value of
$\theta$. Again we shall conclude on the absence of clustering in 
$\mu(\cdot|\ut_D)$ when the two initial conditions have the same $\ell \to
\infty$ limit, and obtain the typical complexity of the clusters from the
difference in the values of (\ref{eq_omega_cav}) for the two different initial
conditions.

\subsection{The cavity computation of the average number of logically implied
  variables}
\label{sec_cav_phi}

Another quantity which can be interesting to compute is the amount of logical
implications, in the UCP/WP sense explained in Sec.~\ref{sec_BPguided},
induced by the choice of the partial reference solution $\ut_D$.
Let us define as
\begin{equation}
\phi(\theta) = \theta + \lim_{N \to \infty} \frac{1}{N} \E_F \E_{\ut} \E_D
[\text{nb. directly implied variables} ] \ ,
\end{equation}
the average fraction of variables which have been explicitly assigned or which
can be logically deduced from these assignments. A motivation for the study of
this quantity, as explained in~\cite{allerton}, is that 
$\frac{\dd \phi}{\dd \theta}-1$ is the average number of the newly implied
variables as a single assignment step is performed. The size of this set of
variables, and in particular its possible divergence with the formula size
when $\phi(\theta)$ is discontinuous, should thus be a measure of the
sensitivity of the decimation procedure with respects to small errors made by
the BP version of the procedure with respects to the ideal one.

The computation of $\phi(\theta)$ can be performed, within the RS assumptions
on the local structure of the probability law $\mu(\cdot)$, in the locally
equivalent random tree model. For a generic CSP one can reproduce the
reasoning above, replacing the conditional messages $\eta_{i \to a}^{\ut_D}$
by their WP counterparts $\WPeta_{i \to a}^{\ut_D}$ according to the
projection defined in~(\ref{eq_projection}). This leads to the similar 
definition of sequences of random pairs of variables, 
$(\eta,\WPteta^\t)_\ell$, $(\eta,\WPeta^\t)_\ell$ and $(\eta,\WPnu^\t)_\ell$, 
which obeys recursion equations analogous to 
Eqs.~(\ref{eq_pair_teta},\ref{eq_pair_eta},\ref{eq_pair_nu}),
\begin{equation}
(\eta,\WPteta^\t)_\ell \eqd \begin{cases} 
(\eta,\WPeta^\t)_\ell & \text{with probability} \ 1-\theta \\
(\eta, \delta^\t) & \text{otherwise} 
\end{cases} \ , \qquad
(\eta,\WPeta^\t)_\ell \eqd 
(g(\nu_1,\dots,\nu_l),\WPg(\WPnu_1^\t,\dots,\WPnu_l^\t) ) \ ,
\end{equation}
\begin{equation}
(\nu,\WPnu^\t)_{\ell +1} \eqd (f(\eta_1,\dots,\eta_{k-1}) , 
\WPf(\WPteta_1^{\t_1},\dots,\WPteta_{k-1}^{\t_{k-1}} )) \ .
\end{equation}
The function $\phi(\theta)$ can then be computed as
\begin{equation}
\phi(\theta) = \E \left[ \sum_\t \eta(\tau) \I(\WPteta^\t = \delta^\t ) \right]
\ ,
\end{equation}
in the $\ell \to \infty$ limit.
This form of the computation is valid for any CSP; in some particular cases
one can however devise a more efficient (and numerically more precise) 
formulation, using the specific form of
the constraints rules to replace the WP message in the second entry of the
random pair by a probability of implication. We refer the reader
to~\cite{allerton} for further details on this computation for
random satisfiability formulas.

\section{Application to the XORSAT ensemble}
\label{sec_xor}

We apply in this section the general formalism developed above to the ensemble
of xor-satisfiability formulas. As we shall see great simplifications occur
for this simple model, and the study of both the decimated ensemble of random
formulas and of the BP guided decimation algorithm can in fact be performed
with simpler methods~\cite{xor1,xor2,xor3}. 
It is however an instructive toy model to begin with
before handling the more involved case of satisfiability formulas,
and has deep connections with information theory, in particular with the 
Low Density Parity Check codes. The analysis of the ``Maxwell decoder'' 
in~\cite{Maxwell} is actually very tightly related with the content of this 
section.

\subsection{BP equations}

The constraints of a xorsat CSP have been defined in Eq.~(\ref{eq_psi_xorsat}).
The set of solutions of such a CSP exhibits some simplifying symmetries, even
in the decimated case where some variables are fixed to a given value. It is
indeed easy to realize that a non-decimated variable is either fixed to +1 in
all the solutions, or fixed to -1, otherwise it takes the value +1 in exactly
half of the solutions and -1 in the other half. Consequently the BP messages
$\nu_{a \to i}$ and $\eta_{i \to a}$ can be restricted to a set of three
possible types, that we shall encode with three-valued numbers 
$u_{a \to i}$ and $h_{i \to a}$ according to the following correspondence:
\begin{equation}
\nu_{a \to i}(\s_i) = \begin{cases}
\delta_{\s_i,+1} & \Leftrightarrow u_{a \to i}=1 \\
\delta_{\s_i,-1} & \Leftrightarrow u_{a \to i}=-1 \\
\frac{1}{2} & \Leftrightarrow u_{a \to i}=0 
\end{cases} \ , \qquad
\eta_{i \to a}(\s_i) = \begin{cases}
\delta_{\s_i,+1} & \Leftrightarrow h_{i \to a}=1 \\
\delta_{\s_i,-1} & \Leftrightarrow h_{i \to a}=-1 \\
\frac{1}{2} & \Leftrightarrow h_{i \to a}=0 
\end{cases} \ .
\end{equation}
With these notations the BP equations (\ref{eq_BP2}) can be rewritten as
\begin{equation}
u_{a\to i} = J_a \prod_{j\in \dami} h_{j \to a} \ , \qquad 
h_{i \to a} = \begin{cases}
0 & \text{if} \ u_{b \to i} = 0 \ \ \ \forall b \in \dima \\
+1 &  \text{if} \ \exists b\in \dima \ \text{with} \ u_{b \to i}=1 \\
-1 & \text{if} \ \exists b\in \dima \ \text{with} \ u_{b \to i}=-1 
\end{cases} \ .
\label{eq_BP_xorsat}
\end{equation}
The second expression is well defined as long as no contradictions are 
detected, which means that the conditions in the last two lines are not
fulfilled simultaneously.
The boundary condition for a decimated variable $i\in D$ is 
$h_{i \to a}^{\ut_D}=\t_i$ for all neighboring interactions $a \in \di$.

Due to the symmetry of the model the BP equations (\ref{eq_BP_xorsat}) can
actually be regarded as WP equations that express the simplifications of 
the Unit Propagation rule. The first equation reflects the fact that a
constraint imposes the value of one of its variables if and only if the $k-1$
other variables are fixed (either by decimation or by propagation of logical
implications), the second means that a variable is fixed as soon as one of its
neighboring clauses imposes its value.

\subsection{The cavity method computations}

Following the general formalism introduced in Sec.~\ref{sec_cav} for the
treatment of decimated CSPs and the parametrization of the BP messages in
terms of $u$ and $h$, we have to find the distributions of the random
variables $(h,h^\t)_\ell$, $(u,u^\t)_\ell$ and $(h,\th^\t)_\ell$ 
for $\t=\pm 1$. These are the solutions of 
Eqs.~(\ref{eq_pair_teta},\ref{eq_pair_eta},\ref{eq_pair_nu}) specialized to
the functions $f$, $g$ and $\psi$ of the xorsat model. The relevant solution
of this equation takes a particularly simple form which allows for an analytic
solution. The results of~\cite{xor1,xor2} imply indeed that the set of
solutions of a non-decimated $k$-xorsat formulas is described by the trivial
solution of the RS equation, $u=h=0$. Moreover the value of the conditional
message $h^\t$ cannot be equal to $-\t$~: by definition $h^\t$ describes
the measure where the reference solution has been drawn conditional on its
value at the root being $\t$, whereas $h^\t=-\t$ would mean that all 
solutions compatible with the values of the reference on some subset of
variables $D$ have $-\t$ at the root, which is in contradiction with the 
hypothesis. In consequence the solution of 
(\ref{eq_pair_teta},\ref{eq_pair_eta},\ref{eq_pair_nu}) can be looked for under
the form:
\begin{equation}
(h,h^\t)_\ell \eqd \begin{cases} (0,0) & \text{with probability} \ 1-x_\ell \\
(0,\t) & \text{with probability} \ x_\ell
\end{cases} \ , 
(u,u^\t)_\ell \eqd \begin{cases} (0,0) & \text{w.p.} \ 1-y_\ell \\
(0,\t) & \text{w.p.} \ y_\ell
\end{cases} \ , 
(h,\th^\t)_\ell \eqd \begin{cases} (0,0) & \text{w.p.} \ 1-\phi_\ell \\
(0,\t) & \text{w.p.} \ \phi_\ell
\end{cases} \ .
\label{eq_xor_ansatz}
\end{equation}
Inserting these forms in 
(\ref{eq_pair_teta},\ref{eq_pair_eta},\ref{eq_pair_nu}) leads to the 
following equations~:
\begin{equation}
\phi_\ell = \theta + (1-\theta) x_\ell \ , \qquad 
x_\ell =1-\exp[-\alpha k y_\ell] \ , \qquad 
y_{\ell+1} = \phi_\ell^{k-1} \ , 
\label{eq_xor_x_y_phi}
\end{equation}
which can be closed under a single recursion equation on $\phi_\ell$,
\begin{equation}
\phi_{\ell+1} = 
\theta + (1-\theta) \left(1 - e^{-\alpha k \phi_\ell^{k-1}} \right) \ .
\label{eq_xor_phi}
\end{equation}
\begin{figure}
\includegraphics[width=8cm]{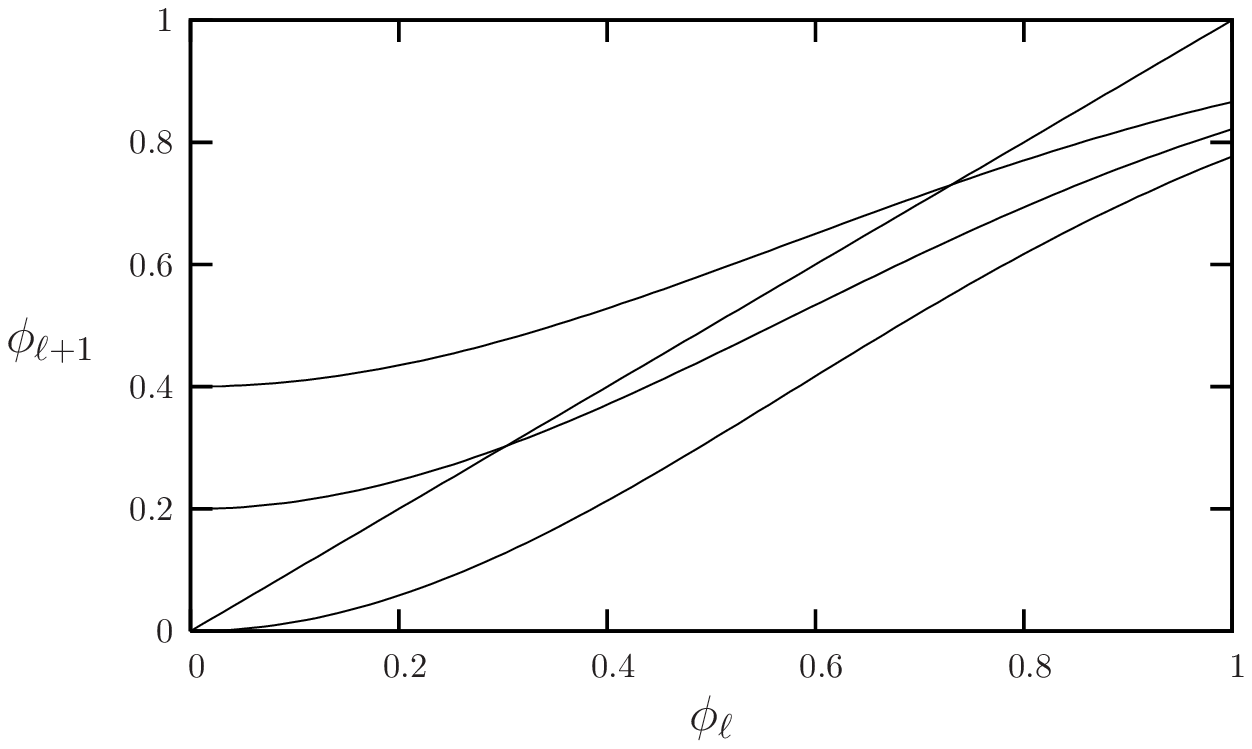}
\hspace{1cm}
\includegraphics[width=8cm]{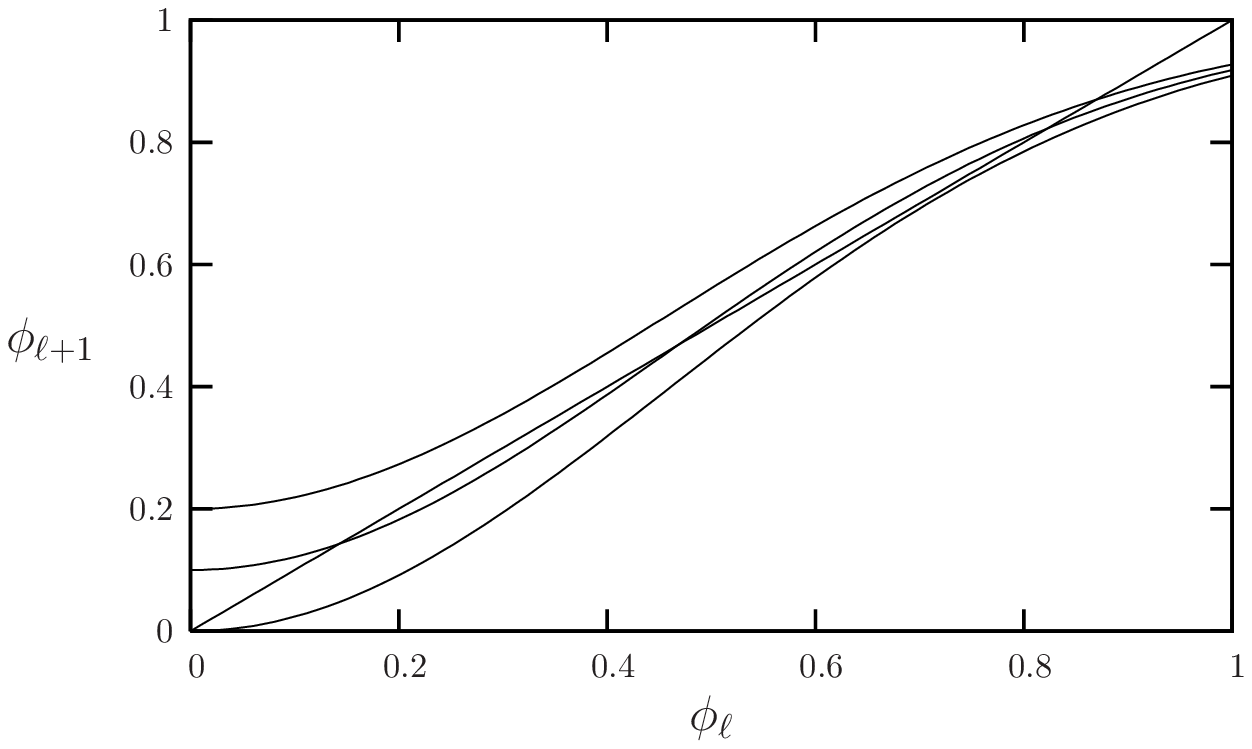}
\caption{Illustration of the recurrence equation (\ref{eq_xor_phi}) for
3-xorsat. Left panel: $\alpha=0.5 < \alpha_*(k=3)=2/3$, 
right panel: $\alpha=0.8 > \alpha_*$. 
On each of the plot the three curves are for
different values of $\theta$ (growing from bottom to top).}
\label{fig_xor_recurs}
\end{figure}
The fixed point equation $\phi_{\ell +1} = \phi_\ell$ 
has between one and three distinct
solutions on $[0,1]$, depending on the values of $\alpha$ 
and $\theta$ (examples of the
various situations are provided on Fig.~\ref{fig_xor_recurs}).
A quick analysis of the equation shows that for $\alpha < \alpha_*$, with
\begin{equation}
\alpha_* = \frac{1}{k} \left(\frac{k-1}{k-2} \right)^{k-2} \ ,
\label{eq_xor_alpha_*}
\end{equation}
the equation (\ref{eq_xor_phi}) admits a single solution for all values of 
$\theta$. If on the contrary $\alpha > \alpha_*$, there exists a range of
$\theta$, denoted $[\theta_-(\alpha),\theta_+(\alpha)]$, where 
Eq.~(\ref{eq_xor_phi}) admits three solutions in $[0,1]$. In that case we
shall call $\phi(\theta)$ (resp. $\psi(\theta)$) the smallest (resp. the
largest) of these three solutions. 
Some examples of these curves are shown in Fig.~\ref{fig_xor_phi},
and the lines $\theta_\pm(\alpha)$ are displayed in Fig.~\ref{fig_xor_lines}.

\begin{figure}
\includegraphics[width=8cm]{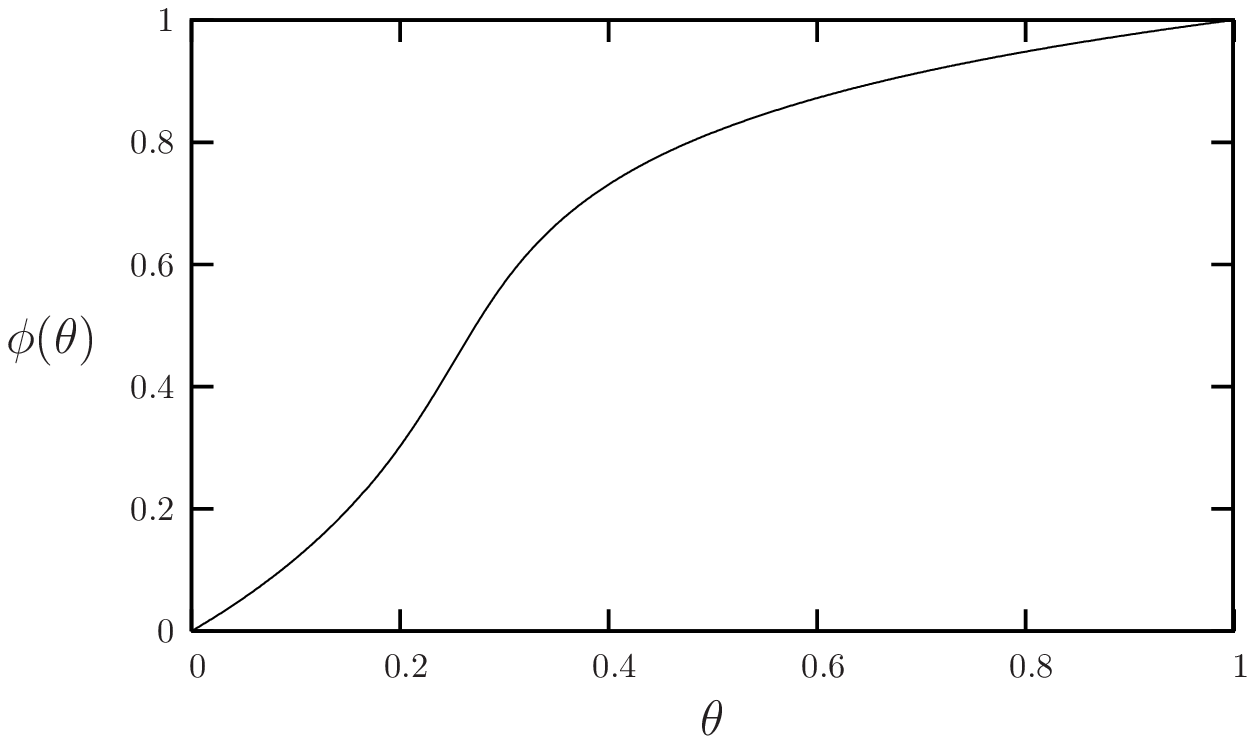}
\hspace{1cm}
\includegraphics[width=8cm]{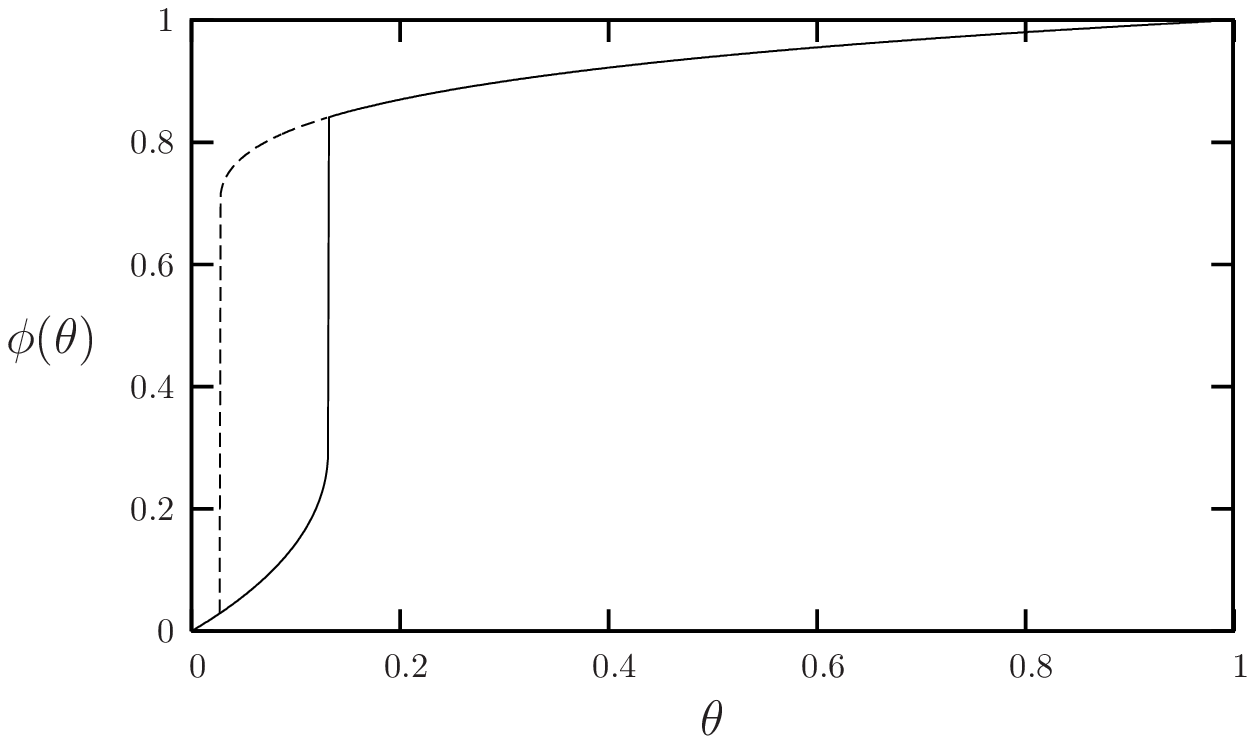}
\caption{Fixed point solution(s) of Eq.~(\ref{eq_xor_phi}) for $k=3$. 
Left panel: $\alpha=0.5 < \alpha_*(k=3)=2/3$, 
right panel: $\alpha=0.8 > \alpha_*$. 
The solid line is $\phi(\theta)$, the dashed line in the
right panel is $\psi(\theta)$, the largest fixed point solution of 
Eq.~(\ref{eq_xor_phi}).}
\label{fig_xor_phi}
\end{figure}

\begin{figure}
\includegraphics[width=8cm]{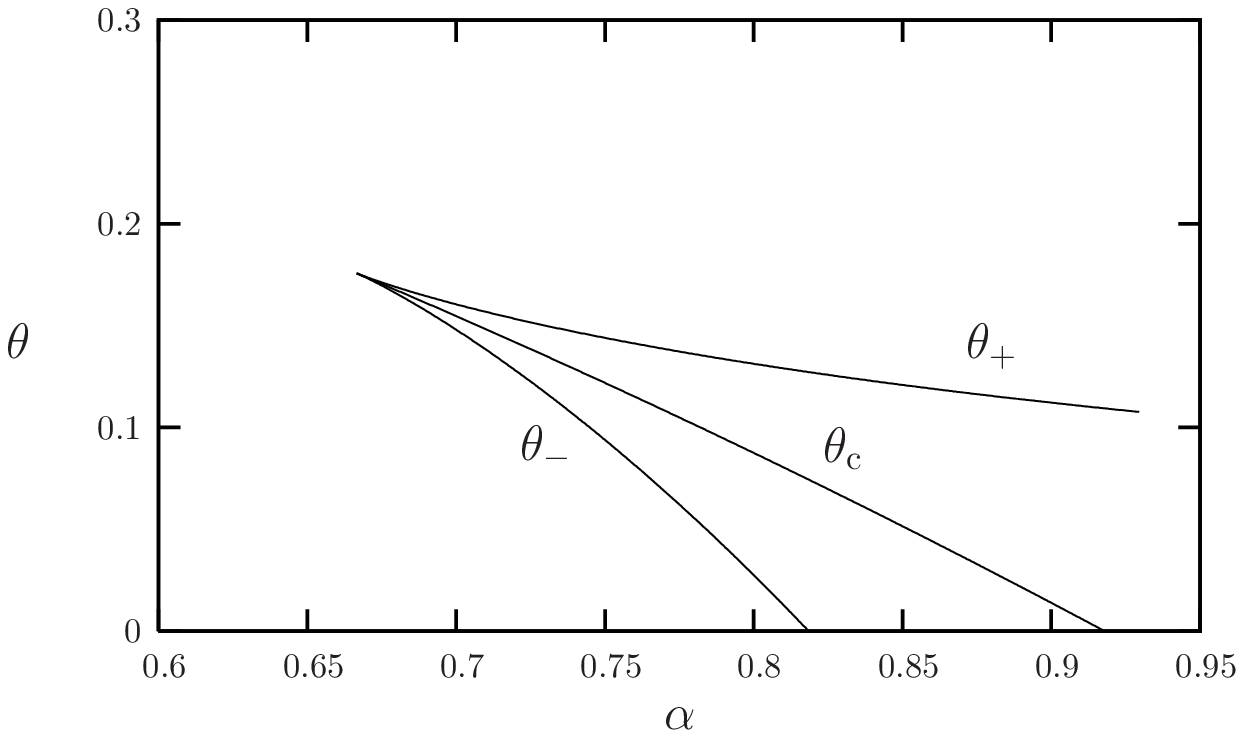}
\caption{Phase diagram in the $(\alpha,\theta)$ plane for the ensemble of
3-xorsat decimated random formulas, the three critical lines meet at 
$\alpha_*(k=3)=2/3$.}
\label{fig_xor_lines}
\end{figure}

\begin{figure}
\includegraphics[width=8cm]{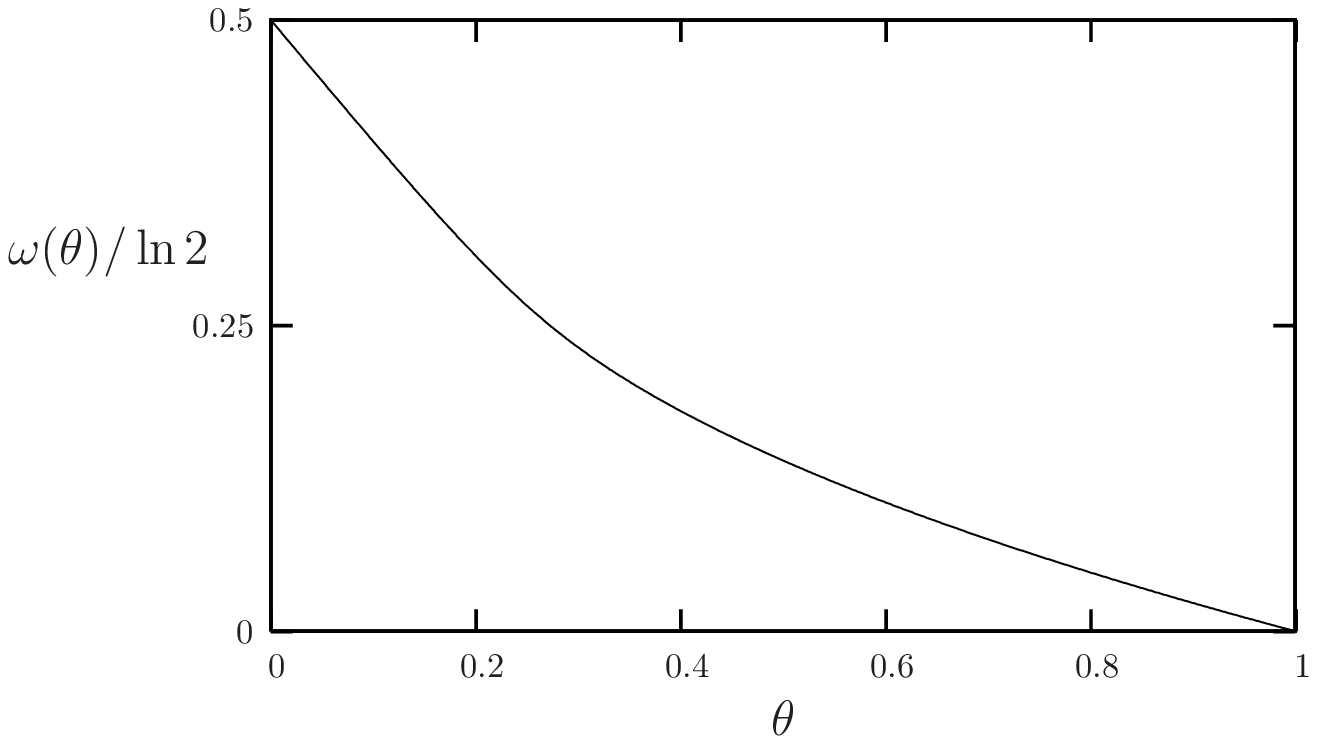}
\hspace{1cm}
\includegraphics[width=8cm]{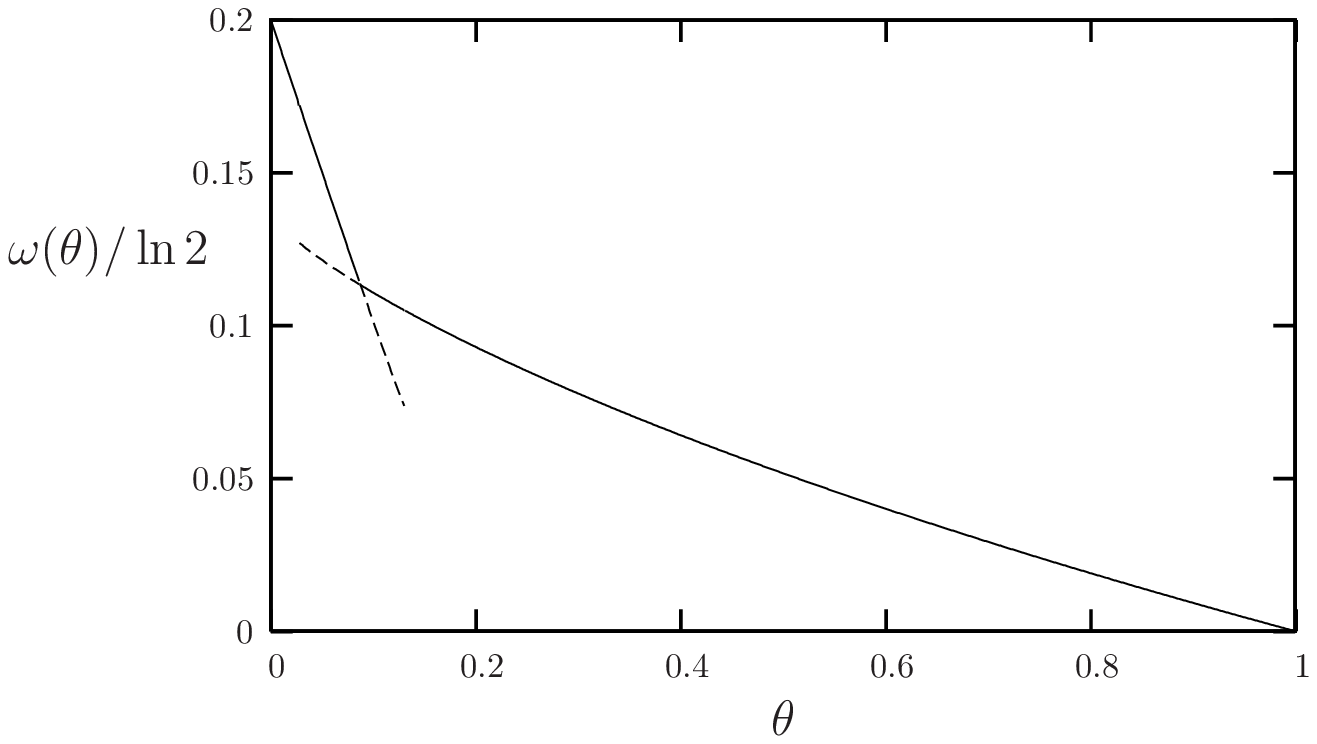}
\caption{Residual entropy $\omega(\theta)$ for 3-xorsat. Left panel: 
$\alpha=0.5$, right panel: $\alpha=0.8$. 
The solid line is the true entropy 
$\omega(\theta)=\max[\homega(\phi(\theta)),\homega(\psi(\theta))]$, the dashed
lines are the two irrelevant branches, 
$\min[\homega(\phi(\theta)),\homega(\psi(\theta))]$.}
\label{fig_xor_omega}
\end{figure}

The expression of $\omega$ given in Eq.~(\ref{eq_omega_cav}) can 
be computed using the ansatz (\ref{eq_xor_ansatz}),
\begin{equation}
\omega =  (\ln 2) [\alpha k (1-\theta)(1-xy) - \alpha(1-\phi^k) - 
(1-\theta) ((\alpha k)(1-y) - \exp[-\alpha k y])] \ ,
\end{equation}
where the limit $\ell \to \infty$ is kept understood.
Using the relations between $x$, $y$ and $\phi$ stated in 
(\ref{eq_xor_x_y_phi}), one can express this residual entropy in terms of 
$\phi$,
\begin{equation}
\homega(\phi) = (\ln 2) [1 - \phi - \alpha + \alpha k (1-\phi) \phi^{k-1} 
+ \alpha \phi^k] \ .
\label{eq_xor_homega}
\end{equation} 
When $\alpha < \alpha_*$ the fixed point solution of (\ref{eq_xor_phi}) is
unique, hence $\omega(\theta) = \homega(\phi(\theta))$ is a smoothly 
decreasing function of $\theta$, as plotted on the left panel of 
Fig.~\ref{fig_xor_omega}.
For larger values of $\alpha$, i.e. $\alpha > \alpha_*$, we have seen above
that there exists a range of parameter 
$\theta \in [\theta_-(\alpha),\theta_+(\alpha)]$ where two solutions
of (\ref{eq_xor_phi}), $\phi(\theta)$ and $\psi(\theta)$, coexists.
On the right panel of Fig.~\ref{fig_xor_omega} one can see that the two 
branches of the entropy, $\homega(\phi(\theta))$ and $\homega(\psi(\theta))$
cross each other at an intermediate value 
$\theta_{\rm c}(\alpha)\in [\theta_-(\alpha),\theta_+(\alpha)]$, which is also
plotted in function of $\alpha$ in the phase diagram Fig.~\ref{fig_xor_lines}.
It is natural (and we shall argue in the following that it is the correct 
choice) to consider that in the region of coexistence the relevant branch is
the one leading to the largest entropy, 
$\omega(\theta)=\max[\homega(\phi(\theta)),\homega(\psi(\theta))]$,
which thus exhibits a discontinuity in its slope when $\theta$ crosses 
$\theta_{\rm c}$.

A direct justification of this choice will be given in the next subsection,
here we argue in its favour on the basis of the cavity method. 
The two boundary conditions discussed in Sec.~\ref{sec_cav_dec} corresponds
to $\phi_{\ell=0}=0$ ($I_0$) and $\phi_{\ell=0}=1$ ($I_1$).
When the fixed point solution of (\ref{eq_xor_phi}) is unique both initial
conditions lead to the same limit $\phi(\theta)$ in the large $\ell$ limit,
which leads to the conclusion that there is no clustering in the solution
space of the decimated formula for these values of $\alpha$ and $\theta$.

In the region of coexistence these two initial
conditions yield, respectively, $\phi_\ell \to \phi(\theta)$ and 
$\phi_\ell \to \psi(\theta)$ as $\ell$ diverges. One is thus led to
assign the difference $\homega(\phi(\theta))-\homega(\psi(\theta))$ 
to the complexity
of the decimated formula, that is the contribution of
the entropy due to the presence of clusters in the measure 
$\mu(\cdot|D_{\theta N})$. This interpretation is valid only when the
complexity is positive, that is in the range
$[\theta_-,\theta_{\rm c}]$. A condensation transition occurs when the
threshold $\theta_{\rm c}$ is crossed. In the region 
$[\theta_{\rm c},\theta_+]$ only a subextensive number of clusters are
relevant, and the total entropy is equal to their internal entropy. The latter
being given by the thermodynamic computation with the initial condition $I_1$,
one concludes that $\omega(\theta)=\homega(\psi(\theta))$ for 
$\theta \in [\theta_{\rm c},\theta_+]$.

\subsection{A more direct computation and its interpretation}
\label{sec_xor_direct}
It is instructive to rederive the above results on xor-satisfiability 
decimated formulas by more direct means. Let us first show that for this
computation one can 
assume that the formula is unfrustrated (i.e. $J_a=1$ for all constraints) 
and that in the reference solution $\ut$ all variables are fixed to $\t_i=1$. 
Suppose indeed that the formula has
been generated with random $J_a=\pm 1$. As we have conditioned the ensemble
of formulas on satisfiable ones, there is at least one solution, call it
$\us^{(0)}$. By the gauge transformation 
$\s_i \leftrightarrow \s'_i = \s_i \s_i^{(0)}$,
the solutions $\us$ of the original problem are in bijection with the 
solutions $\us'$ of the unfrustrated model. Consider furthermore a reference 
solution $\ut$ of the unfrustrated model and a set $D$ of variables, such that
the decimated problem to solve reads
\begin{equation}
\prod_{i \in \da} \s'_i = 1 \ \ \ \forall a \ , \qquad 
\s'_i = \t_i \ \ \ \forall i \in D \ .
\end{equation}
Applying now the gauge transformation 
$\s'_i \leftrightarrow \s''_i = \s'_i \t_i$, noting that $\ut$ is a solution
of the unfrustrated model, one reduces the problem to
\begin{equation}
\prod_{i \in \da} \s''_i = 1 \ \ \ \forall a \ , \qquad 
\s''_i = 1 \ \ \ \forall i \in D \ .
\end{equation}
Having get rid of the signs in the constraints and in the reference solution,
the size and the structure of the set of solutions of the decimated problem
can be deduced from the underlying hypergraph of
constraints~\cite{xor1,xor2,xor3}.

The initial hypergraph is drawn uniformly with $\alpha N$ clauses of length 
$k$ among $N$ variables. A fraction $\theta$ of the variables are fixed to
+1, and can thus be eliminated from the constraints which are reduced in size.
Unit Clause Propagation can then be run to propagate these simplifications.
The details of this computation are deferred to Appendix~\ref{app_xor},
we only quote here the results. When UCP stops, there are $N(1-\phi(\theta))$ 
variables unassigned, with $\phi(\theta)$ the smallest fixed point solution of 
(\ref{eq_xor_phi}). The simplified formula contains constraints of all 
lengths $\kappa \in [2,k]$, more precisely there are
$\alpha N \binom{k}{\kappa} (1-\phi)^\kappa \phi^{k-\kappa}$ clauses of
length $\kappa$. The unassigned variables have a Poisson degree distribution 
with average $\alpha k (1-\phi^{k-1})$.

At this point the structure of the solutions of this reduced formula can be
studied with the leaf removal algorithm~\cite{xor1,xor2}. The
details are again deferred to Appendix~\ref{app_xor}. One finds that the 
presence of an extensive 2-core is equivalent to Eq.~(\ref{eq_xor_phi})
admitting more than one solution, i.e. if $\alpha > \alpha_*$ and 
in the interval $\theta \in [\theta_-,\theta_+]$. If this is the case,
the larger solution $\psi$ gives the fraction of the variables which are
either fixed at the end of UCP, or in the backbone of the UCP-reduced formula.
The difference between the number of variables and the number of clauses 
in the 2-core is $N(\homega(\phi) - \homega(\psi))/(\ln 2)$. In the
interval $\theta \in [\theta_-,\theta_{\rm c}]$ this quantity is positive,
hence it is interpreted as the entropy of the number of solutions of the
2-core, i.e. the complexity of the reduced formula. In 
$[\theta_{\rm c},\theta_+]$ the negative complexity is due to
rare events, typically the 2-core only contains a sub-exponential number of
solutions, hence the discontinuity in slope of $\omega(\theta)$ at this
condensation transition $\theta_{\rm c}$.
For $\alpha \ge \alpha_{\rm d}$ (the usual dynamical threshold) 
the original formula already has a 2-core, hence $\theta_-=0$ in this case.
Similarly $\theta_{\rm c}=0$ for $\alpha \ge \alpha_{\rm c}$, the 
satisfiability transition of the standard ensemble.


Let us remark that the density of clauses of length 2 in the UCP-reduced 
formula
reads $(1/2) \alpha k (k-1) (1-\phi) \phi^{k-2}$. When $\theta$ reaches 
$\theta_+$ from below this density reaches $1/2$ and thus the sub-formula
made of length 2 clauses percolates. Indeed $\theta_+$ is the point of
disappearance of the solution $\phi(\theta)$ from the Eq.~(\ref{eq_xor_phi}),
hence by the implicit function theorem the derivatives with respect to
$\phi$ of the two sides of Eq.~(\ref{eq_xor_phi}) are equal at that point.

\subsection{Numerical experiments on BP guided decimation}

We present in this section the results of numerical experiments
performed with the BP guided decimation algorithm. According to the 
definitions given in the general setting, these experiments consisted in
generating a random xorsat formula (with $J_a = \pm 1$ with probability one 
half) and assigning step by step the value of the variables. The variables 
were assigned in an uniformly random order. Each time a variable
is assigned the BP equations (\ref{eq_BP_xorsat}) are iterated until 
convergence is reached or a contradiction is detected (that is a variable $i$
receives at least two contradicting messages $u=+1$ and $u=-1$ from the
neighboring clauses). As long as no contradiction is found, the
value of the next variable to be assigned is drawn according to the BP 
estimation of its marginal probability. In this simple model this BP marginal
is either completely unbiased (when all incoming messages $u$ from the 
neighboring clauses vanish), in which case the value of the variable
is $\pm 1$ with equal probability, or completely biased, and the assignment
is nothing more than the validation of an implication of previous choices.
A run of this algorithm is successful if it assigns the value of the $N$ 
variables without encountering any contradiction, the configuration obtained
at the end of the process is then a solution of the formula.
\begin{figure}
\includegraphics[width=8cm]{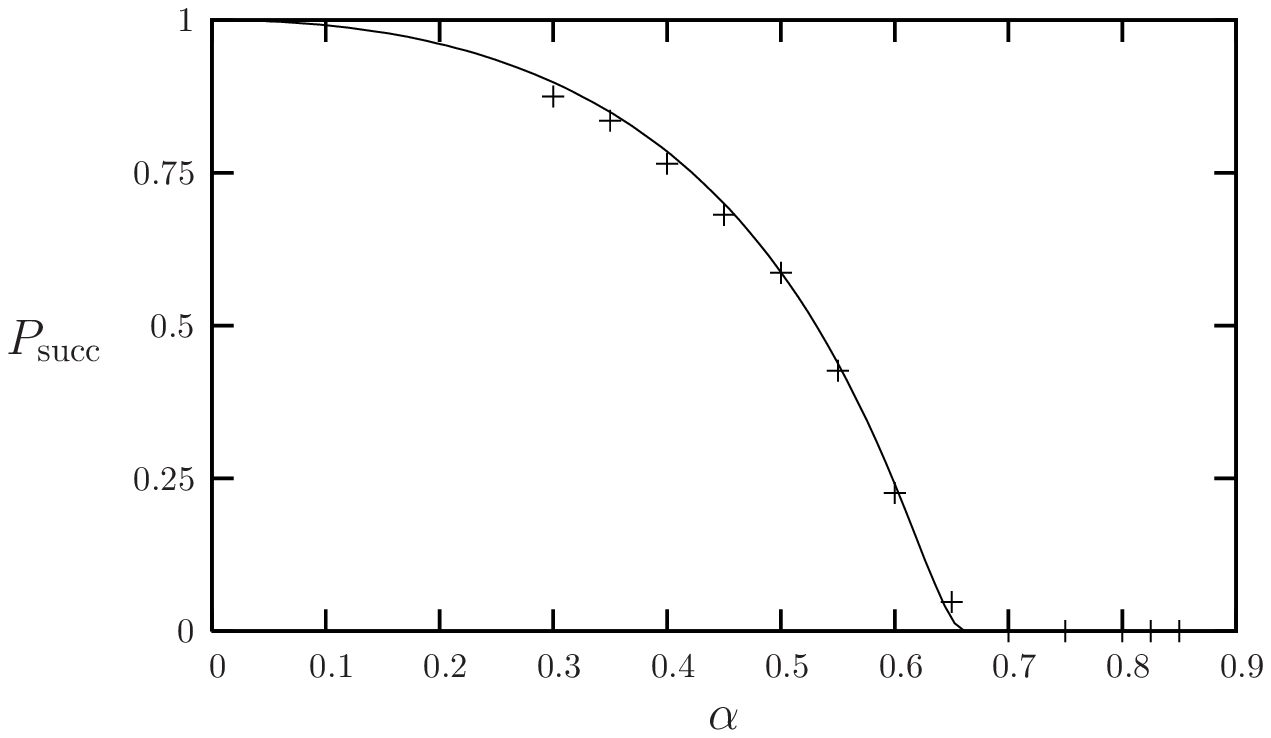}
\hspace{1cm}
\includegraphics[width=8cm]{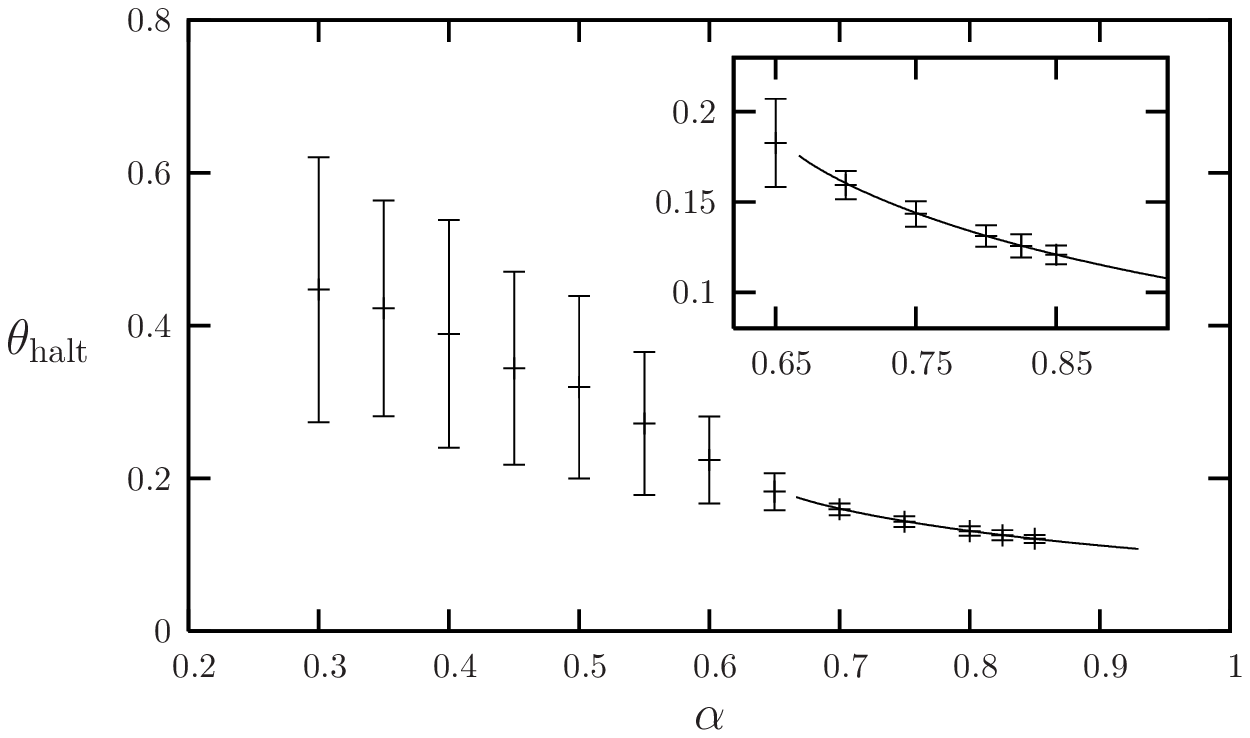}
\caption{Left panel: probability of success of BP guided decimation
on 3-xorsat random formulas,
the solid line is the analytical prediction~\cite{FrSu,DeMo} 
in the infinite size limit (cf. Eq.(\ref{eq_xor_prob})),
which vanishes for $\alpha > \alpha_*$, the symbols are the results
of numerical simulations on 800 formulas of size $N=2 \cdot 10^4$ variables
for each value of $\alpha$.
Right panel: the solid line is the curve $\theta_+(\alpha)$,
symbols indicate the mean and variance (over unsuccesful runs) 
of the fraction of variables assigned before a contradiction is detected,
from numerical simulations on 800 formulas of size $N=2 \cdot 10^4$ variables
for each value of $\alpha$.}
\label{fig_xor_prob_halt}
\end{figure}

In the left panel of Fig.~\ref{fig_xor_prob_halt} we present the probability
of successful runs, with respect to the choice of the formula and to the 
randomness in the course of the run (order of the variables and free choices 
for unbiased marginals). It goes to a finite value in the thermodynamic limit 
(which can be computed analytically, see below) for $\alpha<\alpha_*$, and
to 0 for $\alpha > \alpha_*$. A further piece of information is
given in the right part of Fig.~\ref{fig_xor_prob_halt} about the number
of steps performed by the algorithm (i.e. the number of variables assigned)
before it stops. Let us call this random variable $T_{\rm halt}$ and
the associated fraction $\theta_{\rm halt}=T_{\rm halt} /N$. 
We plotted the mean and variance (represented by error bars) of 
$\theta_{\rm halt}$, computed only on unsuccesful runs. For $\alpha<\alpha_*$
one finds that $\theta_{\rm halt}$ converges in the thermodynamic limit to 
a non trivial random variable. On the contrary in the regime where the
algorithm fails w.h.p. (i.e. for $\alpha > \alpha_*$) the variance of 
$\theta_{\rm halt}$ vanishes at large $N$ (this result was obtained by 
performing the simulations at various sizes, which is not shown on the plot).
In this case $\theta_{\rm halt}$ concentrates around its mean, which is found
to coincide with the function $\theta_+(\alpha)$ defined above (see in 
particular the inset of the right panel of Fig.~\ref{fig_xor_prob_halt}). 

These numerical observations can now be interpreted in the light of the
analytical computations performed above, which were mimicking the decimation
process using the perfect marginals instead of the BP estimation. The 
threshold $\alpha_*$ above which the BP guided decimation algorithm fails 
w.h.p. coincides with the point where the evolution in the $(\alpha,\theta)$
phase diagram has to cross the transition lines drawn on 
Fig.~\ref{fig_xor_lines}, and in particular to penetrate the region 
$[\theta_{\rm c}(\alpha),\theta_+(\alpha)]$ where the 2-core of the residual
formula only admits a sub-exponential number of solutions when the perfect
marginals are used for the decimation. In this case the algorithm is naturally
very sensitive to the small mistakes made by the BP algorithm, which destroy
the few solutions of the 2-core. The fact that the residual formula is no
longer satisfiable remains however unnoticed until a fraction $\theta_+(\alpha)$
of the variables have been assigned. At this point the fraction $\phi(\theta)$
of decimated and logically implied variables has a finite discontinuity
(see right panel of Fig.~\ref{fig_xor_phi}), which means that the assignment
of a few new variables triggers an avalanche of implications of extensive
size. The extensive subgraph of newly implied variables will contain 
implication cycles which, if some mistakes have been done in the previous
assignment steps, will lead to contradictions. More quantitatively, it was 
underlined above that $\theta_+(\alpha)$ marked the percolation of the 
subformula of length 2 clauses which supports the propagation of the logical
implications.

The simplifying symmetries of the xor-satisfiability formulas are such that
BP guided decimation is here almost equivalent to the Unit Clause Propagation
algorithm with random heuristic (and also to the random pivoting Gaussian
elimination algorithm of~\cite{global}). The only slight difference between
the two lies in the order in which the variables are treated, the logical
implications being propagated as soon as they are detected in UCP. In the BP
description of the algorithm the implication is effectively taken into account
by the propagation of the messages, even if the variable is not explicitely
declared as assigned. The behavior of UCP on xor-satisfiability formulas 
has been studied in~\cite{xor3}, the results we just found are in agreement
with the ones of this paper. In particular the phase diagram in
Fig.~\ref{fig_xor_lines} reproduces the left panel of Fig. 3 
in~\cite{xor3}, apart from the difference in the definition of the
vertical time axis explained above. The equivalence with UCP allows also
the computation of the probability of success in the thermodynamic limit
for $\alpha<\alpha_*$. A detailed derivation for satisfiability formulas
can be found in~\cite{FrSu,DeMo}, we state here the result without proof, 
\begin{equation}
P_{\rm succ} = \exp\left[
-\int_0^1 \frac{\dd t}{4(1-t)}\frac{f(t)^2}{1-f(t)} \right] \qquad
\text{with} \quad 
f(t) = \alpha k (k-1) t^{k-2} (1-t) \ .
\end{equation}
For $k=3$ this expression can be further simplified,
\begin{equation}
P_{\rm succ} = \exp\left[\frac{3 \alpha}{4} - \frac{1}{2} 
\frac{1}{\sqrt{\frac{\alpha_*}{\alpha} -1 }} 
\arctan \left( \frac{1}{\sqrt{\frac{\alpha_*}{\alpha} -1 }}\right)
\right] \ .
\label{eq_xor_prob}
\end{equation}
This function is plotted as a solid line in the left panel of 
Fig.~\ref{fig_xor_prob_halt} and agrees with the results of the numerical
experiments.

\section{Application to the SAT ensemble}
\label{sec_sat}

We turn in this section to the case of random satisfiability formulas. We
shall first apply the analytical cavity formalism to this particular model, 
then present the results of numerical experiments with the BP guided 
decimation algorithm and confront the two approaches.

\subsection{BP equations}

Let us begin by expliciting the BP equations~(\ref{eq_BP2}) for the 
satisfiability constraints defined in~(\ref{eq_psi_sat}). As the variables
$\s_i$ are binary the messages $\{ \nu_{a \to i}(\s_i),\eta_{i \to a}(\s_i) \}$
can be parametrized with a single real for each, that we shall denote
$\{u_{a \to i}, h_{i \to a} \}$, under the form:
\begin{equation}
\nu_{a \to i}(\s_i) = \frac{1 - J_i^a \s_i \tanh u_{a \to i} }{2} \ ,
\qquad
\eta_{i \to a}(\s_i) = \frac{1 - J_i^a \s_i \tanh h_{i \to a} }{2} \ .
\label{eq_param_sat}
\end{equation}
As we included the coupling contant $J_i^a$ in these definitions a positive
value of, for instance $h_{i \to a}$, does not indicate a bias of $\s_i$
towards the value $+1$ in the absence of clause $a$, but rather towards the
value $-J_i^a$ that does satisfy $a$.
The message sent by a clause to one of its variables is then found to be
\begin{equation}
u_{a \to i} = f(\{h_{j \to a}\}_{j \in \dami}) \ , \qquad
f(h_1, \dots, h_{k-1}) = -\frac{1}{2} \ln \left( 
1- \prod_{i=1}^{k-1} \frac{1-\tanh h_i}{2} \right) \ .
\label{eq_BP_sat_1}
\end{equation}
To give the explicit form of the other set of BP equations it is advisable
to introduce some further definitions. We shall call $\dpi$ (resp. $\dmi$)
the set of clauses in $\di$ which are satisfied by $\s_i=+1$ (resp. $\s_i=-1$),
that is $\dsi = \{a \in \di | J_i^a = -\s \}$. Moreover we let
$\dpima$ (resp. $\dmima$) denote the set of clauses in $\dima$ agreeing
(resp. disagreeing) with $a$ on the value $i$ should take. In formulae,
$\dpima = \{ b \in \dima | J_i^b = J_i^a \}$, 
$\dmima = \{ b \in \di | J_i^b = -J_i^a \}$. With these notations the
message sent by a variable to a clause reads
\begin{equation}
h_{i \to a} = 
\sum_{b \in \dpima} u_{b \to i} - \sum_{b \in \dmima} u_{b \to i} \ ,
\label{eq_BP_sat_2}
\end{equation}
while the marginal probability of a variable $i$ reads from 
(\ref{eq_marginal}):
\begin{equation}
\mu_i(\s_i) = \frac{1+\s_i \tanh(H_i)}{2} \ , \qquad 
H_i = \sum_{a \in \dpi} u_{a \to i} - \sum_{a \in \dmi} u_{a \to i} \ .
\end{equation}
Finally when a subset of variables $D$ is fixed to a reference configuration
$\ut_D$, the BP equations (\ref{eq_BP_sat_1},\ref{eq_BP_sat_2}) are
complemented with the boundary condition 
$h_{i \to a}^{\ut_D}= - J_i^a \t_i \infty$ when $i \in D$. Note that in all
the numerical implementations of these equations we keep the hyperbolic
tangent of the messages $u$ and $h$ which are free from this apparent 
singularity in the definition of $h$ around a decimated variable.

\subsection{The usual cavity method}

The replica symmetric version of the cavity method for non-decimated random
satisfiability formulas, following the general formalism recalled in
Sec.~\ref{sec_cav_usual}, corresponds to a probabilistic interpretation of 
the BP equations (\ref{eq_BP_sat_1},\ref{eq_BP_sat_2}) applied to
random factor graphs. As the cardinality of $\dima$ converges to a
Poisson random variable of parameter $\alpha k$ and the sign $J_i^a$
in the constraints definition are $\pm 1$ with probability one half,
it follows that $|\dpima|$ and $|\dmima|$ converge to two independent
Poisson random variables with parameter $\alpha k /2$, denoted $l_+$
and $l_-$ below. One has thus to look for the solution of the
distributional equations corresponding to
(\ref{eq_BP_sat_1},\ref{eq_BP_sat_2}) for the random variables $h$ and $u$:
\begin{equation}
h \eqd \sum_{i=1}^{l_+} u_i - \sum_{i=1}^{l_-} v_i \ , \qquad
u \eqd f(h_1,\dots,h_{k-1}) \ .
\label{eq_sat_RS}
\end{equation}
In these equations $h_1,\dots,h_{k-1}$ are independent copies of $h$,
$u_1,\dots,u_{l_+}$ and $v_1,\dots,v_{l_-}$ are independent copies of $u$.

A numerical determination of the fixed point distributions
solutions of these equations can be achieved by the population dynamics
method, revivified in this context by~\cite{MePa}.
This consists in representing the random variable $u$ (resp. $h$) by a sample
of $\cN$ elements $u_1,\dots,u_\cN$ (resp. $h_1,\dots,h_\cN$). The sample
representing $h$ is initialized arbitrarily, for instance $h_j=0$ for all
$j \in [1,\cN]$, then the two samples are updated alternatively as follows.
A new sample representing $u$ is obtained from the representation of $h$ by,
independently for each $j \in [1,\cN]$:
\begin{itemize}
\item[$\bullet$] drawing $k-1$ indices $i_1,\dots,i_{k-1}$ independently, 
uniformly in $[1,\cN]$
\item[$\bullet$] setting $u_j=f(h_{i_1},\dots,h_{i_{k-1}})$
\end{itemize}
Subsequently the sample of $h$ is updated, for each $j$, by:
\begin{itemize}
\item[$\bullet$] drawing $l_+$ and $l_-$, two Poisson random variables of
mean $\alpha k/2$
\item[$\bullet$] drawing $l_+ + l_-$ indices 
$i^+_1,\dots,i^+_{l_+},i^-_1,\dots,i^-_{l_-}$ independently, 
uniformly in $[1,\cN]$
\item[$\bullet$] setting 
$h_j=\sum_{n=1}^{l_+} u_{i^+_n} - \sum_{n=1}^{l_-} u_{i^-_n}$
\end{itemize}

The replica symmetric description of the solution space of random
satisfiability formulas (first obtained with replica computations 
in~\cite{ksat_RS})
is only valid for low enough values of $\alpha$.
The 1RSB analysis at $m=1$, described in generic terms in 
Sec.~\ref{sec_cav_usual}, has been performed on the satisfiability ensemble
in~\cite{pnas,sat_long}. For $k \ge 4$ one finds a clustering transition at
$\alpha_{\rm d}$ and a condensation one at $\alpha_{\rm c}$, before the
satisfiability transition $\alpha_{\rm s}$ determined
in~\cite{MeZe,MePaZe,MeMeZe} (for instance $\alpha_{\rm d}\approx 9.38$,
$\alpha_{\rm c}\approx 9.55$ and $\alpha_{\rm s}\approx 9.93$  for $k=4$). In
the intermediate regime $[\alpha_{\rm d},\alpha_{\rm c}]$ the complexity of
the relevant clusters is positive and vanishes at $\alpha_{\rm c}$, a point
beyond which most of the solutions are contained in a sub-exponential number
of clusters. The value $k=3$ happens to be a particular case, for which the
intermediate regime with a positive complexity is absent, that we shall not
consider in the following.

\subsection{The computation of $\omega(\theta)$}
\label{sec_sat_omega}

Let us now apply the formalism of Sec.~\ref{sec_cav_dec} to ensemble of
decimated random satisfiability formulas. Using the parametrization
(\ref{eq_param_sat}) of the messages the random variables 
$(\eta,\teta^{\, \t})_\ell$, $(\eta,\eta^\t)_\ell$ and $(\nu,\nu^\t)_\ell$ becomes
random pairs of reals, respectively 
$(h,\th^\t)_\ell$, $(h,h^\t)_\ell$ and $(u,u^\t)_\ell$ for $\t=\pm 1$.
In the same spirit as we include the coupling constants in the definitions
(\ref{eq_param_sat}) of the fields $u$ and $h$, we also ``gauge'' the
definition of these random variables such that $u^+$ (resp. $u^-$) corresponds
to the message $u_{a \to i}^{\ut_D}$ where $\ut_D$ is drawn conditional on
$\t_i$ satisfying (resp. not satisfying) clause $a$. 
The recursion equations 
(\ref{eq_pair_teta},\ref{eq_pair_eta},\ref{eq_pair_nu}) then reads
\begin{equation}
(h,\th^\t)_\ell \eqd \begin{cases}
(h,h^\t)_\ell & \text{with probability} \ 1- \theta \\
(h,\t \infty) & \text{otherwise}
\end{cases} \ , \quad
(h,h^\t)_\ell \eqd \left( \sum_{i=1}^{l_+} u_i - \sum_{i=1}^{l_-} v_i,
\sum_{i=1}^{l_+} u^\t_i - \sum_{i=1}^{l_-} v^{-\t}_i   \right) \ ,
\label{eq_sat_pair_1}
\end{equation}
where $l_\pm$ are two independent Poisson random variables of parameter
$\alpha k/2$ and the $(u_i,u_i^\t)$ and $(v,v_i^\t)$ are independent copies
of $(u,u^\t)_\ell$. Finally Eq.~(\ref{eq_pair_nu}) translates into
\begin{equation}
(u,u^\t)_{\ell+1} \eqd \left( f(h_1,\dots,h_{k-1}) , 
f(\th^{\t_1}_1,\dots,\th^{\t_{k-1}}_{k-1}) \right) \ ,
\label{eq_sat_pair_2}
\end{equation}
where the configuration of the variables $\t_1,\dots,\t_{k-1}$ is drawn
with one of the two following probability laws according to the 
value of $\t$,
\begin{equation}
\text{Prob}[\t_1,\dots,\t_{k-1} | \t=+, h_1,\dots,h_{k-1} ] = 
\prod_{i=1}^{k-1} \frac{1+\t_i \tanh h_i }{2} \ ,
\label{eq_sat_broadp}
\end{equation}
or
\begin{equation}
\text{Prob}[\t_1,\dots,\t_{k-1} | \t=-,h_1,\dots,h_{k-1} ] = 
\frac{1-\I(\t_1=\dots=\t_{k-1}=-1 )}
{1-\prod_{i=1}^{k-1}\frac{1-\tanh h_i}{2}} 
\prod_{i=1}^{k-1} \frac{1+\t_i \tanh h_i }{2} \ .
\label{eq_sat_broadm}
\end{equation}
The fields $h_i$ are the same for the computation of $u$ in 
(\ref{eq_sat_pair_2}) and in the probability law of the $\t_i$'s expressed
in (\ref{eq_sat_broadp},\ref{eq_sat_broadm}).
The two initial conditions correspond to $(h,h^\t)_{\ell=0}\eqd (h,h)$ for the
initialization called $I_0$, and $(h,h^\t)_{\ell=0}\eqd (h,\t \infty)$ for
$I_1$.
Finally the average entropy of the decimated random formulas reads from  
(\ref{eq_omega_cav})
\begin{eqnarray}
\omega = &-&\alpha k (1-\theta) \, \E \left[ \sum_\t 
\frac{1 + \t \tanh(u+h)}{2} \ln\left( \frac{1 + \tanh u^\t \tanh h^\t}{2} \right) 
\right] \nonumber \\
&+& \alpha \, \E \left[ \sum_{\t_1,\dots,\t_k} 
\frac{1-\I(\t_1=\dots=\t_k=-1 )}
{1-\prod_{i=1}^k\frac{1-\tanh h_i}{2}} 
\prod_{i=1}^k \frac{1+\t_i \tanh h_i }{2}
\ln \left( 1-\prod_{i=1}^k\frac{1-\tanh \th^{\t_i}_i}{2}
\right) \right] \nonumber \\
&+& (1-\theta) \, \E \left[ \sum_\t 
\frac{1+\t \tanh\left(\sum_{i=1}^{l_+} u_i - \sum_{i=1}^{l_-} v_i\right)}{2}
\ln\left( \sum_\s \prod_{i=1}^{l_+} \frac{1+\s \tanh u^\t_i}{2}
\prod_{i=1}^{l_-} \frac{1- \s \tanh v^{-\t}_i}{2} \right)\right] \ .
\label{eq_omega_sat}
\end{eqnarray}

A numerical determination of the distribution of the random variables
$(h,h^\t)_\ell$ and $(u,u^\t)_\ell$ can be performed by a population dynamics
algorithm. We introduce two population of $\cN$ triplets of reals, 
$\{(h_i,h_i^+,h_i^-)\}_{i=1}^\cN$ and $\{(u_i,u_i^+,u_i^-)\}_{i=1}^\cN$, such
that, for instance, the empirical distribution of $(h_i,h_i^+)$ after $\ell$
steps of the algorithm is a good approximation of the random variable 
$(h,h^+)_\ell$. 
In the initialization step of the algorithm the $h_i$'s are drawn
according to the fixed point solution of Eq.~(\ref{eq_sat_RS}), which is
obtained from a preliminary RS population dynamics procedure. For the initial
condition $I_0$ (resp. $I_1$) one sets $h_i^+=h^-_i=h_i$ 
(resp. $h_i^\pm = \pm \infty$) for all $i \in [1,\cN]$.
Then the following two kind of updates are iterated $\ell$ times.
A new sample of $\{(u_i,u_i^+,u_i^-)\}_{i=1}^\cN$ is obtained by,
independently for each $j \in [1,\cN]$:
\begin{itemize}
\item[$\bullet$] drawing $k-1$ indices $i_1,\dots,i_{k-1}$ independently, 
uniformly in $[1,\cN]$
\item[$\bullet$] setting $u_j=f(h_{i_1},\dots,h_{i_{k-1}})$
\item[$\bullet$] independently for $n=1,\dots,k-1$
\begin{itemize}
\item with probability $\theta$ set $\th^+_n= + \infty$ and $\th^-_n= - \infty$
\item otherwise set $\th^+_n=h_{i_n}^+$, and $\th^-_n=h_{i_n}^-$
\end{itemize}
\item[$\bullet$] generating a configuration $\t_1,\dots,\t_{k-1}$ from the
law $\text{Prob}[\t_1,\dots,\t_{k-1} | \t=+, h_{i_1},\dots,h_{i_{k-1}} ]$ 
defined in Eq.(\ref{eq_sat_broadp})
\item[$\bullet$] setting $u^+_j=f(\th^{\t_1}_1,\dots,\th^{\t_{k-1}}_{k-1})$
\item[$\bullet$] generating a configuration $\t_1,\dots,\t_{k-1}$ from the
law $\text{Prob}[\t_1,\dots,\t_{k-1} | \t=-, h_{i_1},\dots,h_{i_{k-1}} ]$ 
defined in Eq.(\ref{eq_sat_broadm})
\item[$\bullet$] setting $u^-_j=f(\th^{\t_1}_1,\dots,\th^{\t_{k-1}}_{k-1})$
\end{itemize}
Subsequently the sample of $\{(h_i,h_i^+,h_i^-)\}_{i=1}^\cN$ is updated, 
for each $j$, by:
\begin{itemize}
\item[$\bullet$] drawing $l_+$ and $l_-$, two Poisson random variables of
mean $\alpha k/2$
\item[$\bullet$] drawing $l_+ + l_-$ indices 
$i^+_1,\dots,i^+_{l_+},i^-_1,\dots,i^-_{l_-}$ independently, 
uniformly in $[1,\cN]$
\item[$\bullet$] setting 
$h_j=\sum_{n=1}^{l_+} u_{i^+_n} - \sum_{n=1}^{l_-} u_{i^-_n}$,
$h_j^+=\sum_{n=1}^{l_+} u^+_{i^+_n} - \sum_{n=1}^{l_-} u^-_{i^-_n}$ and
$h_j^-=\sum_{n=1}^{l_+} u^-_{i^+_n} - \sum_{n=1}^{l_-} u^+_{i^-_n}$
\end{itemize}
After a large number of these iterations has been performed the determination
of the residual entropy (\ref{eq_omega_sat}) is easily obtained: the 
expectation values can be interpreted as empirical averages over the 
population.

\begin{figure}
\includegraphics[width=8cm]{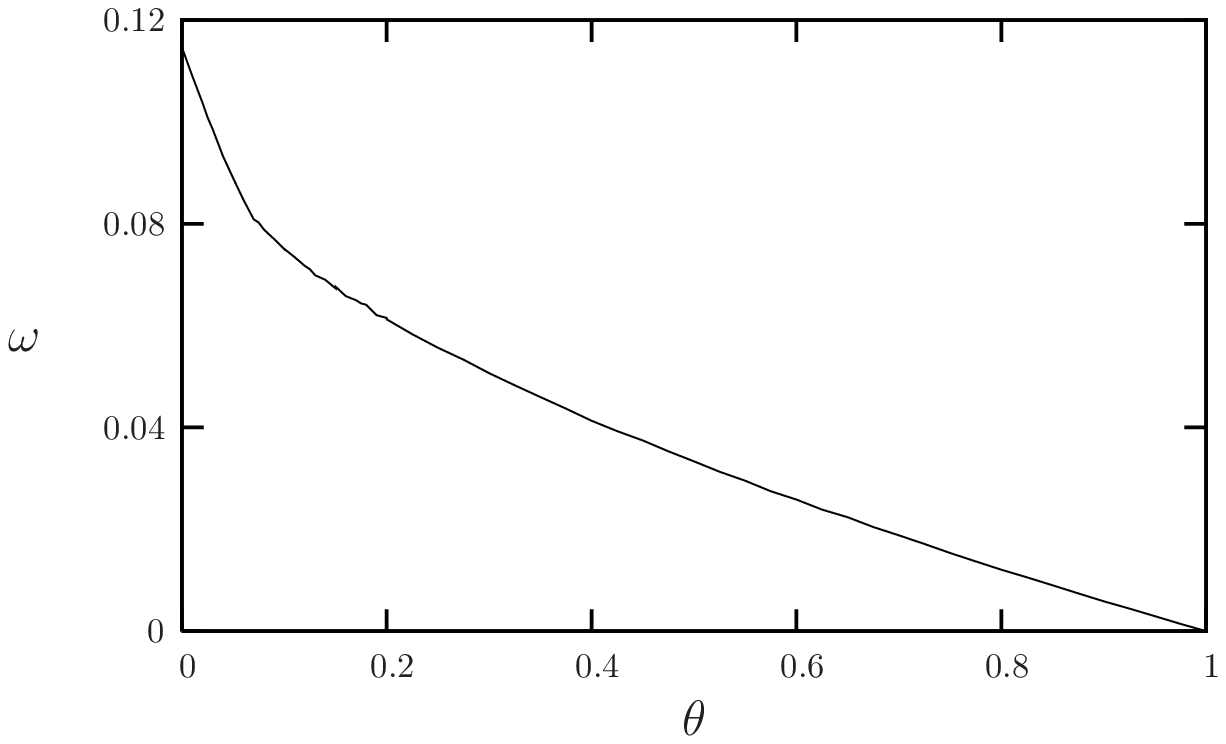}
\hspace{1cm}
\includegraphics[width=8cm]{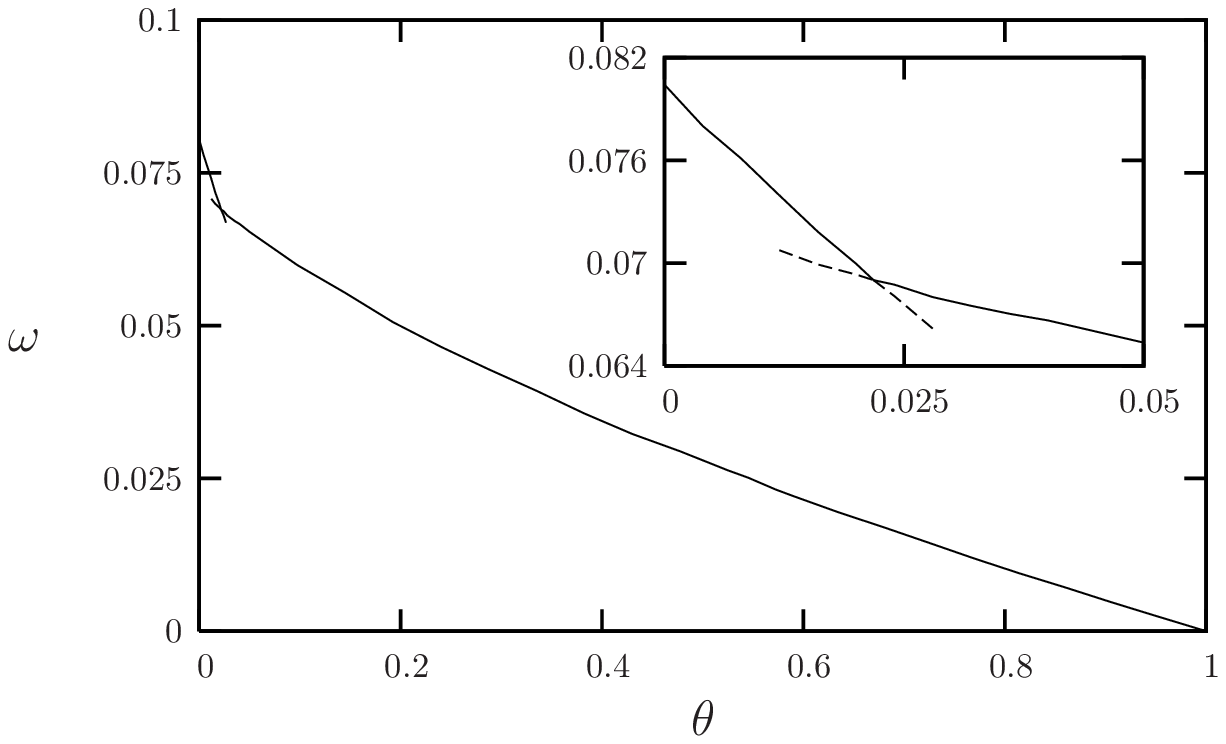}
\caption{Residual entropy $\omega(\theta)$ of 4-sat random formulas. 
Left panel: $\alpha=8.8$, $\omega(\theta)$ is smoothly decreasing.
Right panel: $\alpha=9.3$, the inset is a zoom around the singular point
of $\omega(\theta)$.}
\label{fig_sat_omega}
\end{figure}

\begin{figure}
\includegraphics[width=8cm]{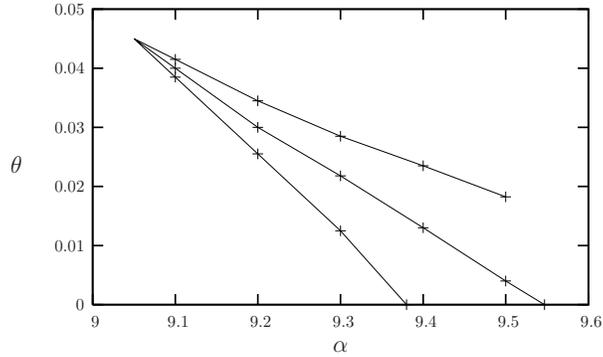}
\caption{Phase diagram of 4-sat random formulas in the $(\alpha,\theta)$ plane,
from bottom to top $\theta_-(\alpha)$, $\theta_{\rm c}(\alpha)$ and 
$\theta_+(\alpha)$. Symbols result from the population dynamics algorithm, 
lines are guides to the eye.}
\label{fig_sat_lines_thermo}
\end{figure}

We have implemented this numerical procedure and performed the computation for
various values of $\alpha$ and $\theta$. The results for $k=4$ are as follows.
For small enough values of $\alpha$ the large $\ell$ limit of the recursion
relations~(\ref{eq_sat_pair_1},\ref{eq_sat_pair_2}) is found to be independent
of the initial condition $I_0$ or $I_1$ used, and the residual entropy density
$\omega(\theta)$ is a smoothly decreasing function. This quantity is plotted
for $\alpha=8.8$ on the left panel of Fig.~\ref{fig_sat_omega}. For larger
values of $\alpha$ there appears a regime
$\theta \in [\theta_-(\alpha),\theta_+(\alpha)]$ in which the two initial
conditions leads to different fixed point solutions
of~(\ref{eq_sat_pair_1},\ref{eq_sat_pair_2}), signaling the presence of
non-trivial long range point-to-set correlations in the decimated formula. The
two branches of $\omega(\theta)$ are plotted in the right panel of
Fig.~\ref{fig_sat_omega} for $\alpha=9.3$, and are found to cross each other
at $\theta_{\rm c}(\alpha) \in [\theta_-(\alpha),\theta_+(\alpha)] $.
For $\theta \in [\theta_-(\alpha),\theta_{\rm c}(\alpha)]$ the branch with the
highest value of $\omega$ corresponds to the $I_0$ initialization, the
situation being reversed for 
$\theta \in [\theta_{\rm c}(\alpha),\theta_+(\alpha)]$.
As explained on the simpler xorsat example, we interpret these results as
following from the existence of a positive complexity of relevant clusters in
the regime $[\theta_-,\theta_{\rm c}]$. In this case the highest branch of
$\omega(\theta)$ is the total entropy of the decimated formula, while the
difference between the two branches is its complexity. On the contrary for
$[\theta_{\rm c},\theta_+]$ the upper branch is the only relevant one, the
total entropy is dominated by the subexponential number of clusters around a
typical reference solution $\ut$. The three critical lines are displayed in
the $(\alpha,\theta)$ of Fig.~\ref{fig_sat_lines_thermo}, which also shows
that, as follows from their definitions, $\theta_-(\alpha)$ 
(resp. $\theta_{\rm c}(\alpha)$)
reaches the horizontal axis $\theta=0$ at the usual dynamic transition
$\alpha_{\rm d}$ (resp. condensation threshold $\alpha_{\rm c}$). We estimated
the location of the critical point where $\theta_\pm$ and $\theta_{\rm c}$ merge
to be $\alpha_*=9.05$, $\theta_*=0.045$,
by interpolation of the results obtained for values of $\alpha$ slightly 
larger.

\subsection{The computation of $\phi(\theta)$}
\label{sec_sat_phi}

We proceed now with the computation of the fraction of logically implied
variables, following the lines sketched in Sec.~\ref{sec_cav_phi} (the same
results were presented in~\cite{allerton} with a slightly different
formulation). 

Let us first discuss the Warning Propagation equations for satisfiability
formulas. According to the projection equations~(\ref{eq_projection}) one has
to identify the situations in which a single value of a variable $\s_i$ is
allowed by a BP message. For the messages sent by a clause to a variable this
can only happen when the variable is forced to satisfy the clause,
i.e. $\nu_{a\to i}(\s_i) = \delta_{\s_i,-J_i^a}$, in which case we define the
WP message to be $\uW_{a \to i}=1$, otherwise $\uW_{a \to i}=0$. A message
$\eta_{i \to a}$ sent from a variable to a clause can allow both values of
variable $\s_i$, or force it to the value satisfying $a$ ($\eta_{i \to
  a}(\s_i)=\delta_{\s_i,-J_i^a}$), or to the value disastifying it
($\eta_{i \to a}(\s_i)=\delta_{\s_i,J_i^a}$). It is only the latter case that
shall be propagated by the WP equations, we thus affect the value $\hW_{i \to
  a}=0$ to the first two situations and $\hW_{i \to a}=1$ to the latter. The
WP equations read with these definitions:
\begin{equation}
\uW_{a \to i} = \prod_{j \in \dami} \hW_{j \to a} \ , \qquad
\hW_{i \to a} = 1 - \prod_{b \in \dmima} (1 - \uW_{b \to i}) \ .
\end{equation}
For a variable $i\in D$ the boundary condition reads 
$\hW_{i \to a} = \I(\t_i = J_i^a)$. 

In order to compute the average fraction of logically implied variables, within
the assumptions of the RS cavity method on the local description of the
uniform probability measure $\mu(\cdot)$, we introduce the sequences of random
variables $(h,\thW^\t)_\ell$, $(h,\hW^\t)_\ell$ and $(u,\uW^\t)_\ell$ as
defined in Sec.~\ref{sec_cav_phi}. It turns out that not all the values of
$\t$ have to be considered. Consider for instance the random variable
$(u,\uW^\t)_\ell$. Its distribution is by definition the one of 
$(u_{a \to i},\uW_{a \to i}^{\ut_D})$, in the random tree model of depth
$\ell$, rooted at variable $i$ which appears solely in the clause
$a$. Depending on $\t$ the reference configuration $\ut$ is drawn conditional
on $\t_i$ either satisfying (if $\t=+1$) or not satisfying ($\t=-1$) the
constraint $a$. In the latter case $\uW_{a \to i}^{\ut_D}$ is necessarily
equal to 0: at least one of the variables in $\dami$ must satisfy $a$ in
$\ut$, and this variable cannot be forced to its opposite value by $\ut_D$. We
can hence restrict our attention to $(u,\uW^+)_\ell$, which is found to obey
\begin{equation}
(u,\uW^+)_{\ell +1} \eqd (f(h_1,\dots,h_{k-1}), 
\zeta\, \thW_1^- \dots \thW_{k-1}^-)
\ , \qquad 
\zeta \eqd \begin{cases} 1 & \text{with probability} \  
\prod_{i=1}^{k-1}\frac{1-\tanh h_i}{2} \\
0 & \text{otherwise}
\end{cases} \ .
\label{eq_sat_uW_dist}
\end{equation}
The probability that the random variable $\zeta$ equals $1$ is indeed the 
probability that, conditional on $\t_i$ satisfying the root clause $a$, the 
configuration of the $k-1$ other variables in $\dami$ are drawn to the values 
unsatisfying $a$. 

For similar reasons the right hand side of this equation does not depend on
$(h,\thW^+)$ and we can complete this equation with the recursion on 
$(h,\thW^-)$ and $(h,\hW^-)$, which read
\begin{equation}
(h,\thW^-)_\ell \eqd \begin{cases}
(h,\hW^-)_\ell & \text{with probability} \ 1- \theta \\
(h,1) & \text{otherwise}
\end{cases} \ , \quad
(h,\hW^-)_\ell \eqd \left( \sum_{i=1}^{l_+} u_i - \sum_{i=1}^{l_-} v_i,
1 - \prod_{i=1}^{l_-}(1-\vW_i^+)   \right) \ ,
\label{eq_sat_hW_dist}
\end{equation}
where as usual $l_\pm$ are two Poisson random variables of parameter $\alpha
k/2$ and the $(u_i,\uW_i^+)$ and $(v_i,\vW_i^+)$ are independent copies
of $(u,\uW^+)_\ell$. Finally the average fraction of either decimated or
directly implied variables can be obtained as
\begin{equation}
\phi(\theta) = \E [(1-\tanh h) \thW^- ] \ .
\label{eq_sat_phi}
\end{equation}
These recursion equations can be solved numerically using the same kind of
population dynamics as explained above, updating in turns populations of pairs
$\{(h_i,\hW_i^-)\}_{i=1}^\N$ and $\{(u_i,\uW_i^-)\}_{i=1}^\N$. The two kind of
initial conditions already discussed correspond here to $(h,\hW^-)_{\ell=0}
\eqd (h,0)$ for $I_0$, and $(h,\hW^-)_{\ell=0} \eqd (h,1)$ for $I_1$.

\begin{figure}
\includegraphics[width=8cm]{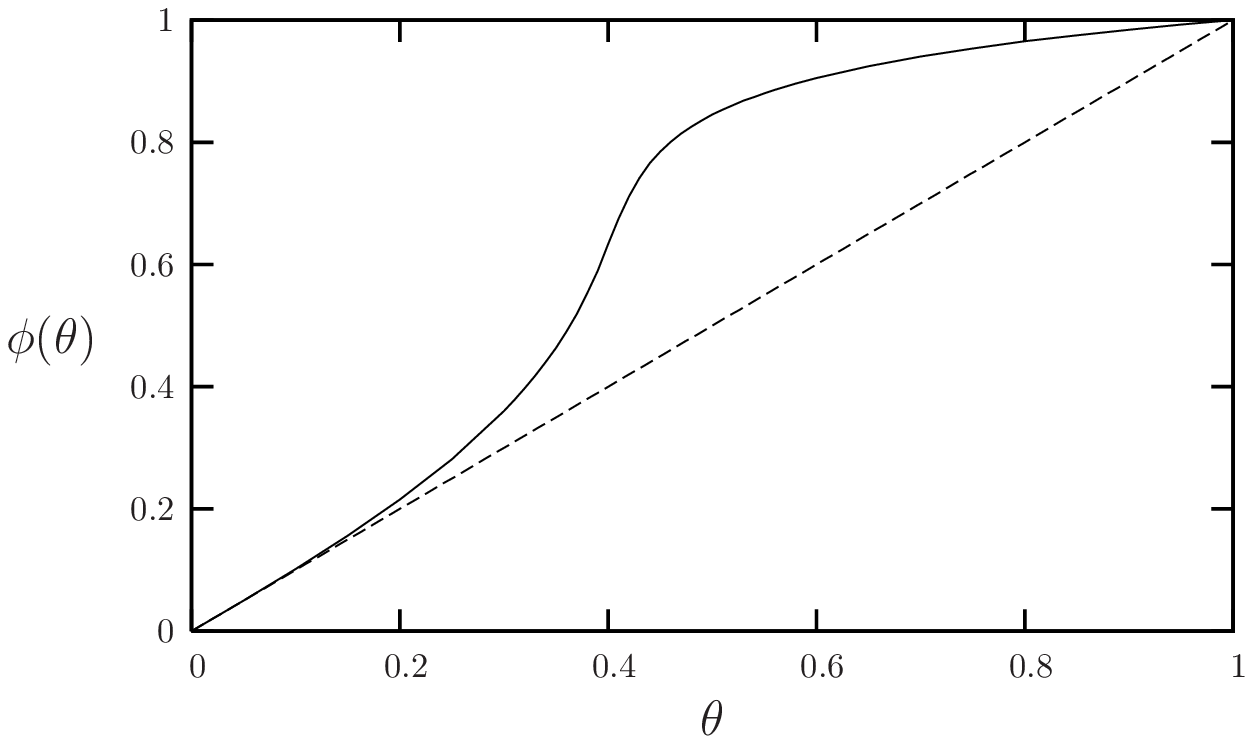}
\hspace{1cm}
\includegraphics[width=8cm]{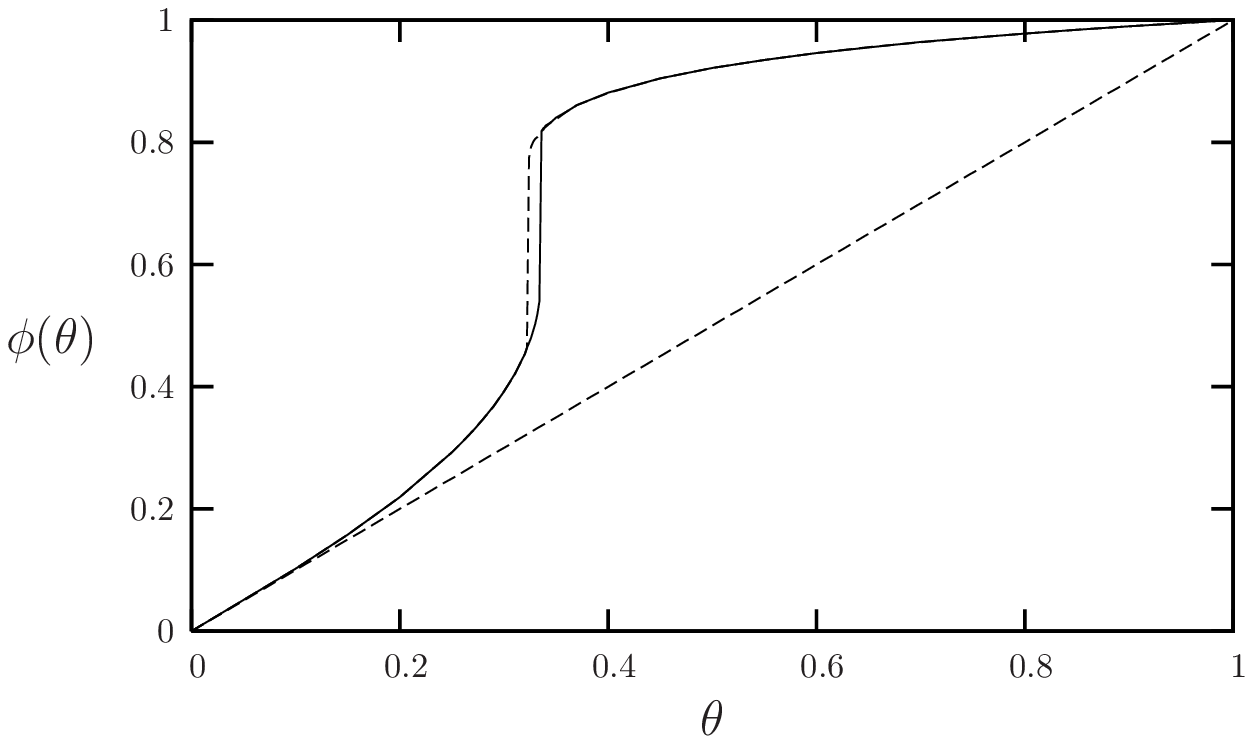}
\caption{Fraction $\phi(\theta)$ of assigned and logically implied variables
in 4-sat random formulas , left: $\alpha=7.0$, right: $\alpha=8.4$.}
\label{fig_sat_phi}
\end{figure}

\begin{figure}
\includegraphics[width=8cm]{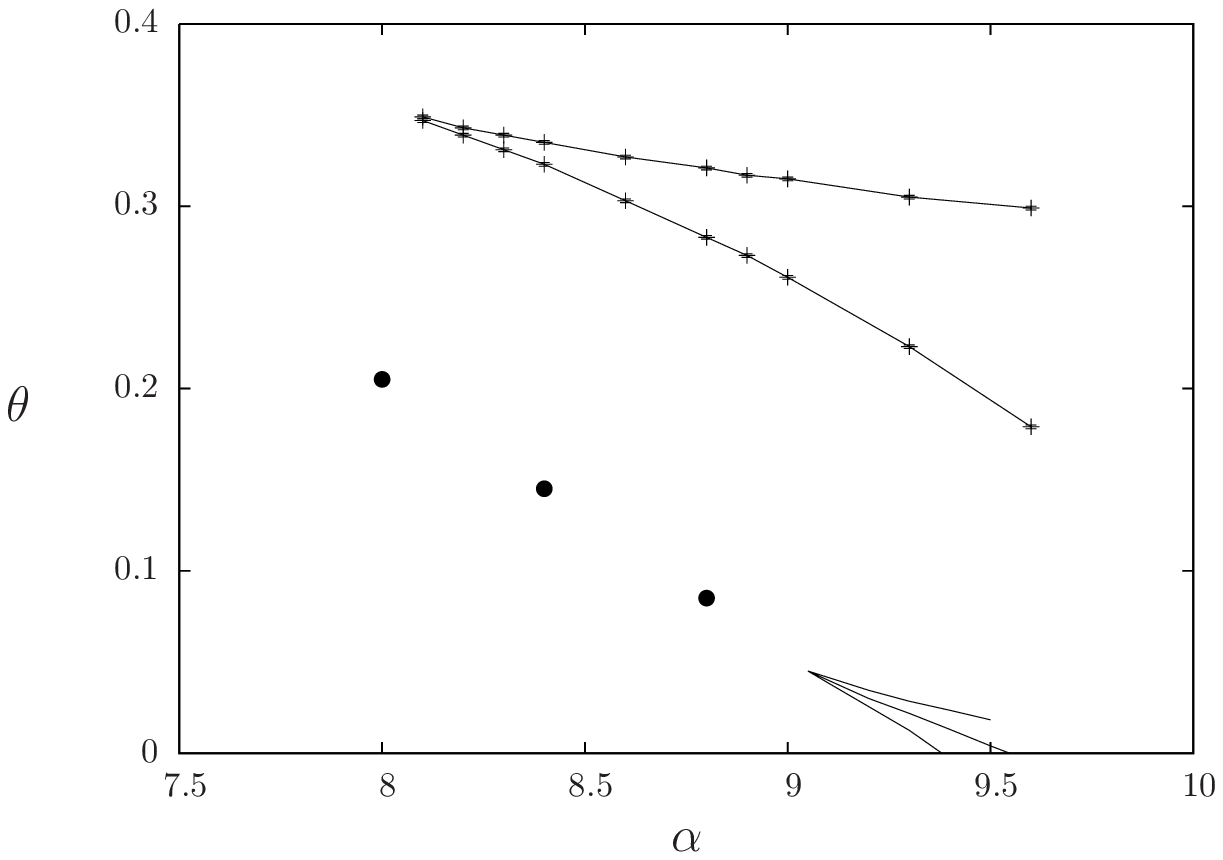}
\caption{Phase diagram of 4-sat random formulas in the $(\alpha,\theta)$ plane.
The three lines in the bottom of the figure are those of 
Fig.~\ref{fig_sat_lines_thermo}, the upper two are 
$\theta'_-(\alpha)<\theta'_+(\alpha)$ defined in Sec.~\ref{sec_sat_phi} from
the discontinuities of $\phi(\theta)$ and $\psi(\theta)$. The filled circles
show the location $\theta_{\rm max}(\alpha)$
of the slowest convergence of the BP iterations in the BP
guided decimation algorithm, see Sec.~\ref{sec_sat_results_cvg}.}
\label{fig_sat_lines_both}
\end{figure}

This numerical resolution leads to the following results for $k=4$. At small
enough values of $\alpha$ the two initial conditions lead to the same large
$\ell$ limit, the function $\phi(\theta)$ is smoothly increasing (see left
panel of Fig.~\ref{fig_sat_phi}). For larger values of $\alpha$ there exists a
range of parameter $[\theta'_-(\alpha),\theta'_+(\alpha)]$ where the quantity 
(\ref{eq_sat_phi}), computed from the initial condition $I_1$, is strictly 
greater than the one reached from $I_0$. In this coexistence regime we shall
call $\psi(\theta)$, in analogy with the notations used for the xorsat model,
the upper branch obtained from $I_1$, see for instance the right panel of 
Fig.~\ref{fig_sat_phi}. The function $\phi$ (resp. $\psi$) is discontinuous at
$\theta'_+$ (resp. $\theta'_-$). These two thresholds are the two upper curves
in the phase diagram of Fig.~\ref{fig_sat_lines_both}, which also contains
for comparison a repetition of the phase diagram of 
Fig.~\ref{fig_sat_lines_thermo}. The two
regimes for the behaviour of $\phi(\theta)$ are separated by the value
$\alpha'_*\approx 8.05$.

At this point the reader might be puzzled by the apparent contradiction
between these results and those of the previous subsection. Consider indeed
some parameters $\alpha > \alpha'_*$ and 
$\theta \in [\theta'_-(\alpha),\theta'_+(\alpha)]$. We claimed in the previous
subsection that the large $\ell$ limit of the random variable $(h,h^\t)_\ell$ 
was independent of the initial condition in $\ell=0$, whereas we just found
that $(h,\hW^-)_\ell$ does depend on it. As the latter variable is a projection
of the former, this statement is at first sight paradoxical. This apparent
contradiction can however be resolved by a closer inspection of the
relationship between the two random variables. One has indeed
\begin{equation}
(h,\hW^-)_\ell \eqd \lim_{\ve \to 0} (h,\I(\tanh h^- \le -1 + \ve))_\ell \ , 
\end{equation}
while the two apparently contradictory statements are
\begin{equation}
\lim_{\ell \to \infty} (h,h^-)_\ell^{I_0} \eqd 
\lim_{\ell \to \infty} (h,h^-)_\ell^{I_1} \ , \qquad
\lim_{\ell \to \infty} (h,\hW^-)_\ell^{I_0} \neqd 
\lim_{\ell \to \infty} (h,\hW^-)_\ell^{I_1} \ . \qquad
\label{eq_puzzle}
\end{equation}
The resolution of the paradox relies on the non-commutativity of the limits
$\ell \to \infty$ and $\ve \to 0$. More explicitly, under the 
initialization $I_0$ there is a positive probability for a field $\tanh h^-$ to 
have -1 as its large $\ell$ limit, yet remaining strictly superior to -1
as long as $\ell$ is finite. If the limit $\ve \to 0$ is taken before
$\ell \to \infty$ these fields do not
participate to $\hW^-$, which is thus found to be smaller in the
initialization $I_0$ with respect to $I_1$. Yet if the limit $\ell \to \infty$ 
is performed first this positive fraction of the fields $\tanh h^-$ 
(with initialization $I_0$) reach their limit $-1$, hence making possible
the first statement of Eq.~(\ref{eq_puzzle}).
We checked explictly this
phenomenon by constructing a coupling of the two initializations and solved it
with the population dynamics algorithm.

\subsection{Numerical experiments on BP guided decimation}
\label{sec_sat_results}

We have run the belief propagation guided decimation algorithm for
many random 4-sat formulas.  The sizes of the formulas studied are
$N=10^3, 3 \times 10^3, 10^4, 3 \times 10^4$, with $\alpha$ varying
between $6.0$ and $9.2$.  The number of formulas analyzed varies with
$\alpha$, but it is always larger than 2000 for $N=10^3$, larger than
1200 for $N = 3 \times 10^3$, larger than 400 for $N=10^4$ and between
360 and 25 (increasing $\alpha$) for $N = 3 \times 10^4$.

\subsubsection{Details on the practical implementation}

Some technical details about the numerical implementation of the BP
guided decimation algorithm were given in~\cite{allerton} (see also
Appendix A of~\cite{sat_long} for details on the representation of BP
messages and probabilities).
The main numerical bottleneck in applying the BP guided decimation
algorithm is the convergence of the iterative method for solving the
BP equations, described in Section~\ref{sec_BPguided}.  This iterative
scheme is known to be a fast way of finding a fixed point of the BP
equations, although sometimes it may not converge.  Lack of
convergence may be due to different reasons: in case long range
correlations develop, multiple BP fixed points appear and the
convergence of BP to one of these fixed point cannot be guaranteed;
on the other hand, when a single BP fixed point exists, convergence
problems can be typically cured by the use of a damping
term~\cite{Pretti}. In all our numerical simulations we have used a
damping term of intensity 0.1, that is, when we update a message {\tt
  x}, we do not assign to it directly the new value {\tt xNew}, but
rather the weighted sum {\tt 0.9 * xNew + 0.1 * x}. We have verified
that under these conditions the convergence (if any) is always
exponentially fast in the number of iterative steps (although
sometimes with an exponent very small). Because of the exponentially
fast convergence, our arbitrary choice of considering BP equations
solved when the maximal change in any BP message is below $10^{-4}$
turns out to be very reasonable: indeed an accuracy of $10^{-8}$ can
be reached by simply doubling the running time.  Anyhow, in order to
avoid entering a never-ending loop we have also fixed a maximum number
of iterations equal to 1000; when this limit is reached non-converged
BP messages are used to compute marginals and to proceed with the
decimation.

The last comment about technical issues regards the
initialization of BP messages before the iterative solving procedure is
applied.  At the beginning, when the formula is still not decimated,
BP messages are initialized in a random way assigning to each
$\tanh(h_{i \to a})$ a random value uniformly distributed between
$-0.2$ and $0.2$.  After each variable decimation, one can choose to
keep the BP messages obtained from the last iterative procedure or to
re-initialize them along the same random way described above.  In
principle, if a single BP fixed point exists and if this is reached by
the iterative method, then the starting point should be irrelevant.
Moreover, one would expect that the BP fixed points of two formulas
differing in just a variable are very close, and that starting from
the one already reached should help convergence (with respect to start
from random messages). This intuition turns out to be wrong. We have
strong numerical evidence that a random re-initialization of BP
messages after each decimation strongly enhances the performances of
the algorithm. A possible explanation is the following. Our numerical
procedure does not produce a perfect estimation of the marginal probabilities 
(in particular when the stopping criterion used is the maximal number of
iterations); if messages are not re-initialized small errors may easily 
accumulate in the same direction, while a random re-initialization of BP
messages results in a partial neutralization of these errors.

\subsubsection{Algorithm performances and convergence probabilities}
\label{sec_sat_results_cvg}

\begin{figure}
\centering
\includegraphics[width=0.6\textwidth]{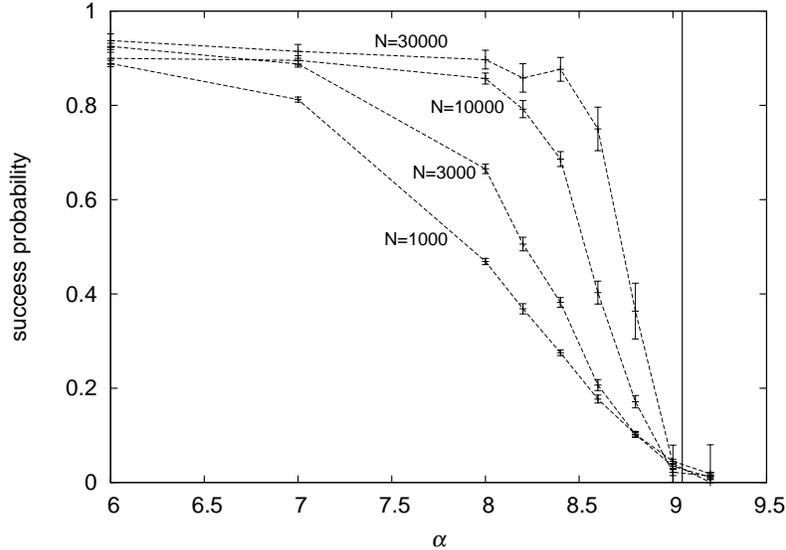}
\caption{Success probability of the BP guided decimation algorithm as
  a function of $\alpha$ for random 4-sat formulas of various
  sizes. The vertical line marks the value $\alpha_*=9.05$ beyond
  which the analytical computations predicted a condensation
  transition in the residual entropy $\omega(\theta)$.}
\label{fig:succProb}
\end{figure}

\begin{figure}
\centering
\includegraphics[width=8cm]{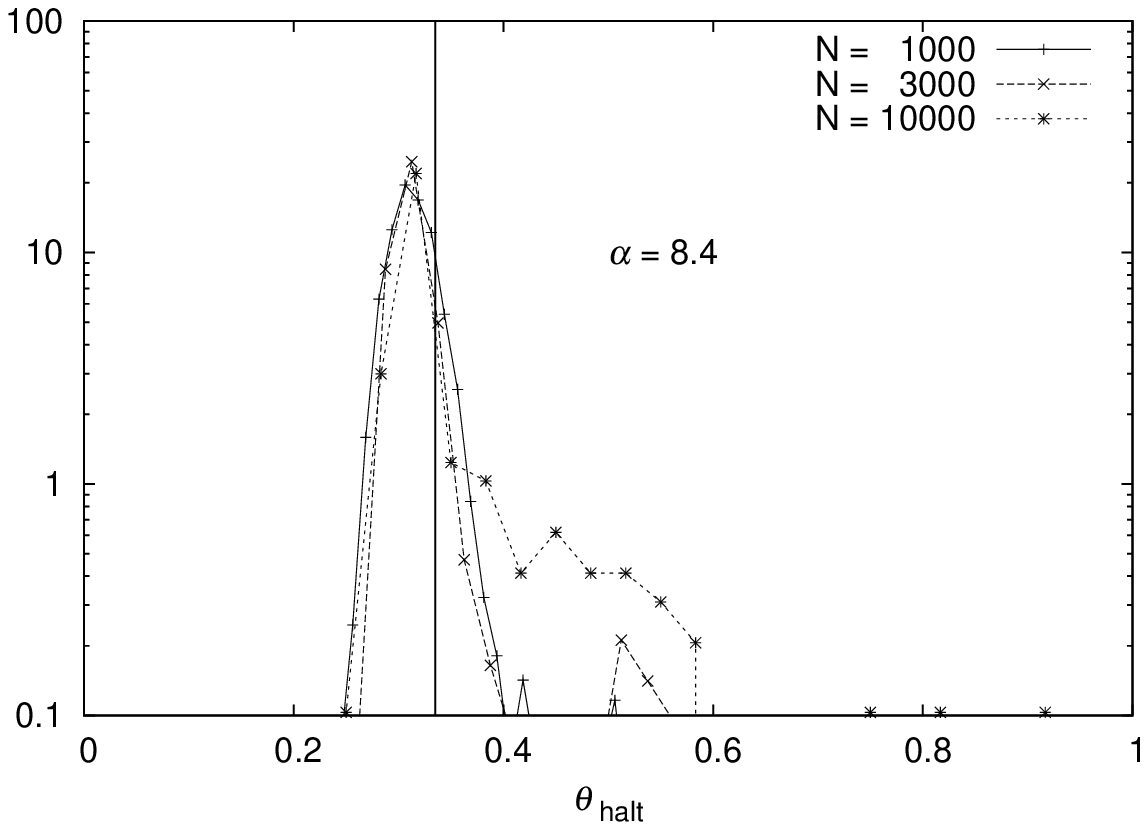}
\hspace{1cm}
\includegraphics[width=8cm]{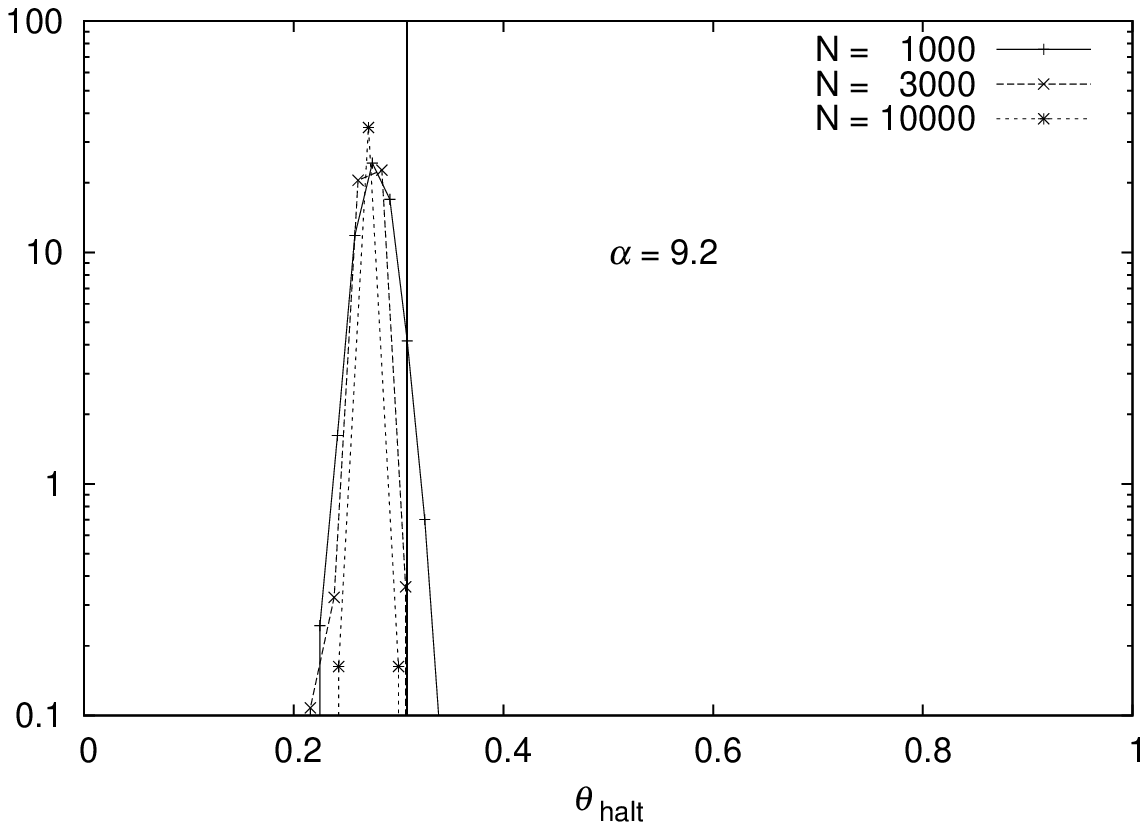}
\caption{Distribution of the halting time of the BP guided decimation
algorithm on 4-sat random formulas with $\alpha=8.4$ (left panel) and
$\alpha=9.2$ (right panel). Vertical lines show the value of
$\theta'_+(\alpha)$ where $\phi(\theta)$ is discontinuous.}
\label{fig_sat_halting}
\end{figure}

As a first result, we show in Fig.~\ref{fig:succProb} the success
probability for the BP guided decimation algorithm, i.e. the fraction
of formulas which have been solved by this algorithm.
The numerical data clearly point to an algorithmic threshold 
$\alpha_{\rm a}$ very close to the theoretical prediction of the
point $\alpha_*=9.05$ (marked by a vertical line in
Fig.~\ref{fig:succProb}) above which phase transitions occur in the
thermodynamic properties of the decimated ensemble of random formulas.
For $\alpha < \alpha_{\rm a}$ a large $N$
formula is solved with positive probability by the BP guided decimation
algorithm.  The appearance of a jump in the function
$\phi(\theta)$ at $\alpha \simeq 8.1$ [see below for a more detailed
  analysis of $\phi(\theta)$], with a consequent avalanche of directly
implied variables during the decimation of formulas with $\alpha >
8.1$, does not have any visible effect on the success probability. This 
phenomenon has however a trace in the random variable $\theta_{\rm halt}$,
that is the fraction of variables assigned before the discovery of a 
contradiction during the unsuccesful runs. The distribution of this
random variable is shown in Fig.~\ref{fig_sat_halting} for two values of
$\alpha$ (below and above $\alpha_{\rm a}$). One can see a maximum in this
distribution for values of $\theta$ slightly smaller than $\theta'_+(\alpha)$,
the point of discontinuity of $\phi(\theta)$.

\begin{figure}
\centering
\includegraphics[width=0.8\textwidth]{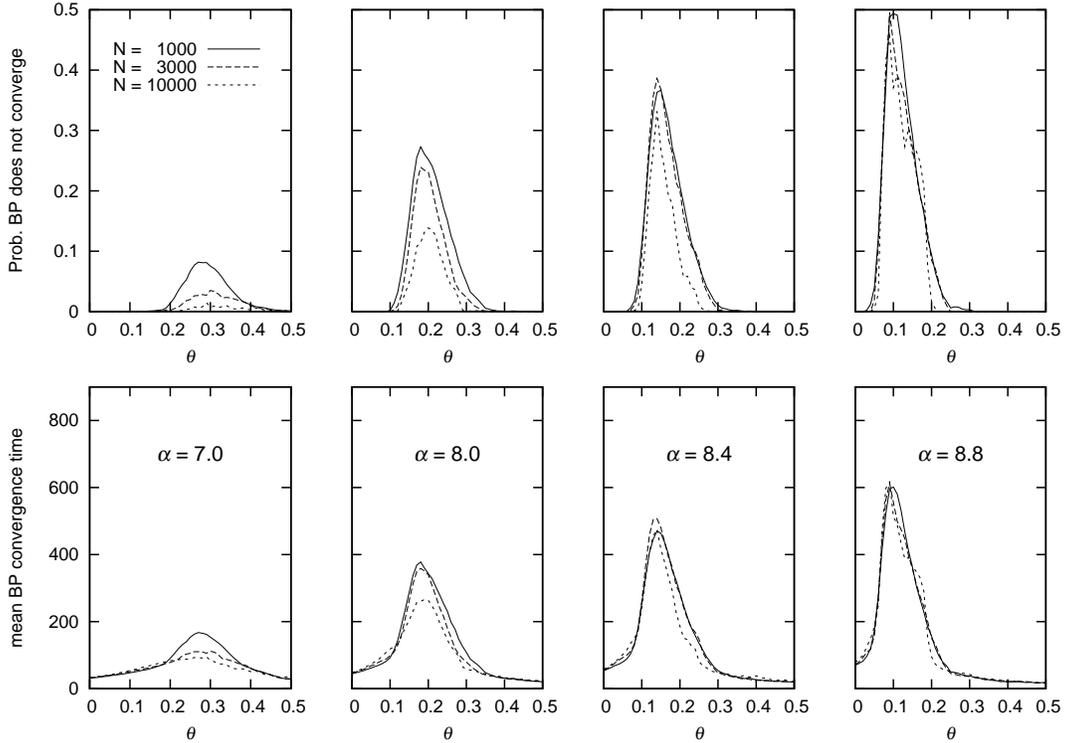}
\caption{Probability of non convergence in 1000 iterations (upper
  panels) and average number of iterations required to reach
  convergence (lower panels) for the BP part of the BP guided
  decimation algorithm, as a function of the fraction $\theta$ of
  decimated variables. From left to right $\alpha=7$, $\alpha=8$,
  $\alpha=8.4$, $\alpha=8.8$.}
\label{fig:convTimes}
\end{figure}

In the following we are going to present data only in the region
$\alpha < \alpha_{\rm a}$. In order to reduce finite size effects we
will concentrate only on formulas which have been actually solved by
our algorithm.
The study of the convergence probability and of the average
convergence time for the iterative method used to solve the BP
equations provides very useful information, as it allows
to identify the most difficult formulas, which should appear close to
the threshold. In Fig.~\ref{fig:convTimes} we show both the
probability that the BP fixed point is not reached after 1000
iterations (upper panels) and the average number of iterations
required to converge (lower panels). Non-converged instances count
with 1000 in the average. Four values of $\alpha$ are shown (from left
to right), $\alpha=7.0$, $8.0$, $8.4$ and $8.8$, and $\theta > 0.5$ is
not shown since in that region nothing of interest takes place.

We see that, for small values of $\alpha$, by increasing the size of
formulas the probability that BP does not converge in 1000 steps
reduces considerably, thus suggesting that in the large $N$ limit the
typical running time of BP is below 1000 for any $\theta$ value. On
the contrary, for larger values of $\alpha$, the probability that BP
does not converge is not varying very much with $N$ and seems to
remain positive even in the large $N$ limit, thus suggesting that the
typical number of iterations required to make BP converge is larger
than 1000 for some values of $\theta$.

The overall picture we get from Fig.~\ref{fig:convTimes} is very
clear.  For any $\alpha$ value, the decimation procedure initially
produces formulas which are more and more difficult to solve and the
running time of BP thus increases with $\theta$. The running time (or
equivalently the probability of not converging in a fixed number of
iterations) has a maximum at a value $\theta_{\rm max}(\alpha)$ and
then decreases again. By increasing $\alpha$, $\theta_{\rm
  max}(\alpha)$ decreases and the running time at $\theta_{\rm max}$
increases. It is natural to expect that the maximum running time
should diverge at the threshold $\alpha_{\rm a}$; moreover, if one
assumes that this phenomenon is related to the critical point
$\alpha_*$ marking the end of the (first-order) condensation
transition line $\theta_{\rm c}(\alpha)$ for $\alpha>\alpha_*$, one
should expect that $\theta_{\rm max}(\alpha)$ is a precursor of the
transition line $\theta_{\rm c}(\alpha)$ in the phase $\alpha<\alpha_*$. 
The data of
$\theta_{\rm max}(\alpha)$ plotted with filled circles in
Fig.~\ref{fig_sat_lines_both} are in agreement with this intuition,
showing in particular that $\theta_{\rm max}(\alpha)$ reaches values
very close to $\theta_*$ for the largest values of $\alpha$ the
algorithm is able to handle.

\subsubsection{Entropy of decimated formulas}

We measured the entropy of decimated formulas along the execution
of the BP guided decimation algorithm, using the Bethe approximation
stated in (\ref{eq_lnZ_ut_D}).
When comparing these results with the analytic
prediction presented in Section~\ref{sec_sat_omega}, a lot of care is
required in dealing with cases where BP did not converge to a fixed
point. Indeed in these cases the marginal probabilities do not satisfy
the consistency constraints and the resulting value for the entropy
may be quite far from the correct one.  In order to avoid this problem
we have adopted a drastic, but safe approach: we take the average over
only those formulas for which the algorithm always converged before
reaching the 1000 iterations limit.  A possible criticism to this
approach is that we are taking the average over the simplest formulas,
thus obtaining a biased estimate for the residual entropy. If this
criticism is well founded we should observe a dependence of the
average residual entropy upon the value of maximal number of
iterations. Actually we do not observe any variation by doubling the
maximal number of iterations.

\begin{figure}
\centering
\includegraphics[width=0.6\textwidth]{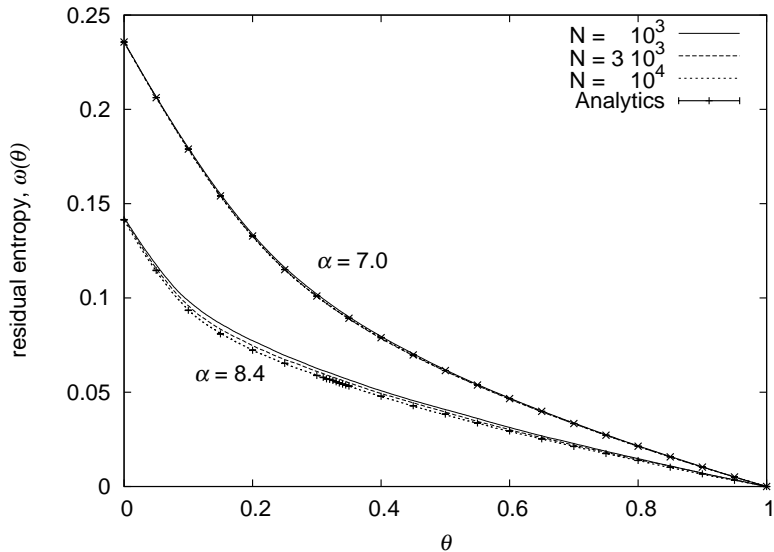}
\caption{Residual entropy as a function of the fraction $\theta$ of
  decimated variables for $\alpha=7$ (upper curves) and $\alpha=8.4$
  (lower curves).  The symbols correspond to the analytical
  predictions presented in Sec.~\ref{sec_sat_omega}, the lines to the
  numerical simulations of the BP guided decimation algorithm for
  various sizes.}
\label{fig:entropy}
\end{figure}

In Fig.~\ref{fig:entropy} we plot the residual entropy as a function
of $\theta$ for three different sizes (lines) together with the analytical
predictions of Sec.~\ref{sec_sat_omega} (points with errors).  For
$\alpha = 7.0$, even the $N=10^3$ data are almost superimposed to the
analytic result. On the contrary, for $\alpha = 8.4$ finite size
effects are much more evident and only $N=10^4$ data start to be
compatible with the analytical computations in the thermodynamic limit.

It is also worth noticing that the function $\omega(\theta)$ shows its
point of maximum curvature close to $\theta_{\rm max}$. More in
general the curvature of $\omega(\theta)$ seems to be somehow related
to the typical running time of BP. 

\subsubsection{Forced variables and multiple WP fixed points}

\begin{figure}
\centering
\includegraphics[width=0.7\textwidth]{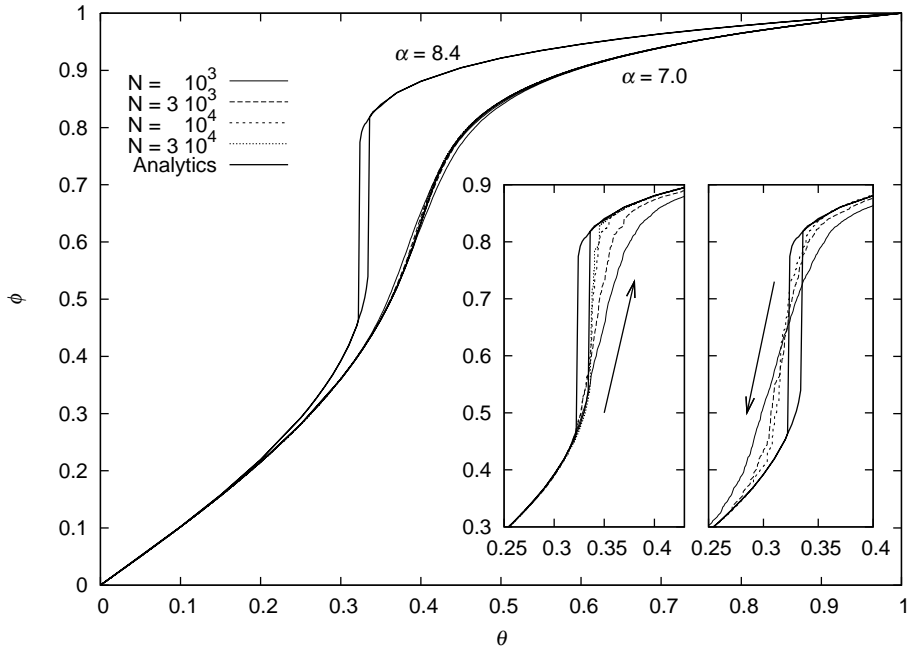}
\caption{The fraction $\phi(\theta)$ of assigned and logically implied
  variables for $4$-sat formulas of density $\alpha=7$ and $\alpha=8.4$. The
  insets detail the results at $\alpha=8.4$ for the decimation and backward
  algorithm, see the text for details.}
\label{fig:phiOfTheta}
\end{figure}

We also keep track of the fraction of logically implied variables
in the partially decimated formulas, measuring it with the WP 
algorithm (equivalent to UCP as explained in Sec.~\ref{sec_BPguided}).
In the main panel of Fig.~\ref{fig:phiOfTheta} we plot the function
$\phi(\theta)$ computed analytically (solid curves). For $\alpha=7.0$
the function $\phi(\theta)$ is smoothly increasing and the numerical
data follow this curve so nicely that only the $N=10^3$ data is hardly
visible, the rest being perfectly superimposed to the analytic curve.
For $\alpha=8.4$ the function $\phi(\theta)$ is multivalued in the
interval $[\theta'_-(\alpha),\theta'_+(\alpha)]$ and the two curves
plotted in the main panel correspond to the two branches
$\phi(\theta)$ and $\psi(\theta)$ defined in Sec.~\ref{sec_sat_phi}.
The numerical data for $\alpha=8.4$ are shown in the insets for
clarity. The left inset corresponds to the decimation algorithm (which
indeed increases $\theta$ during the run). The right inset reports the
data gathered while running the {\em backward} algorithm which works
as follows. 

\begin{figure}
\centering
\includegraphics[width=0.6\textwidth]{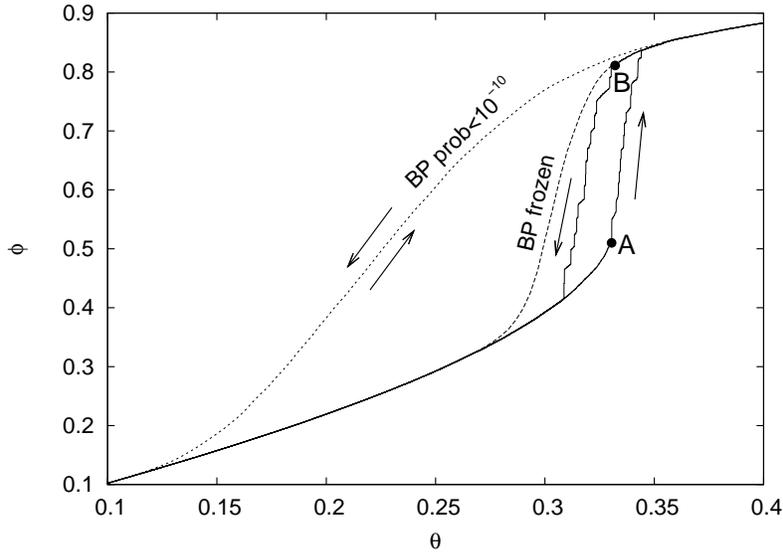}
\caption{Fraction of frozen variables as a function of $\theta$, for
  $\alpha=8.4$. The full lines correspond to the WP results in the
  decimation and backward algorithm, the dashed lines to the analysis
  of the BP messages.}
\label{fig:WPandBP}
\end{figure}

After a succesful run of the BP guided decimation algorithm we use
the solution $\ut$ constructed as the reference one, and variables are 
unfixed one by one in the reverse
fixing order: in this way at any $\theta$ value the residual formula
is exactly the same the decimation algorithm had to work with. The
only differences between the two algorithms are the initial values for
the WP and BP messages: in the backward algorithm all messages are set
initially according to the solution found in $\theta=1$, namely
$\tanh h_{i \to a}= - J_i^a \t_i$ and $\hW_{i \to a} = \I(\t_i = J_i^a)$
for all edges. Then the updates of WP and BP are run as usual for the
edges outside the decimated ones.
The numerical data shown in the insets suggest
that, although finite size effects are huge, in the large $N$ limit
the decimation (resp. the backward) algorithm follow the lower branch
$\phi(\theta)$ (resp. upper branch $\psi(\theta)$) curve shown in the
main panel.  As predicted by the analytical computation, for
$\alpha>\alpha'_*\simeq 8.05$ the fraction of variables which are
frozen (either assigned or directly implied by WP), $\phi(\theta)$,
has a hysteresis loop when the number of assigned variables $\theta$
is increased and decreased across the interval
$[\theta'_-(\alpha),\theta'_+(\alpha)]$, the backward algorithm used
when decreasing $\theta$ corresponding to the $I_1$ boundary condition
of the infinite tree computation. The hysteresis loop obtained by
looking at the WP messages is reported with a full line in
Fig.~\ref{fig:WPandBP} (where data for $N=3\times 10^4$ and
$\alpha=8.4$ are used).

The apparent paradox discussed at the end of Sec.~\ref{sec_sat_phi}
shows up here again, at the level of the single sample analysis
instead of the computation on the infinite tree: the forward and
backward procedure allows us to construct two distinct fixed points of
the WP equations on the same partially decimated formula (see points
{\sf A} and {\sf B} in Fig.~\ref{fig:WPandBP}), while the success
probability of the algorithm is still positive for this value of
$\alpha$ and the residual entropy of Fig.~\ref{fig:entropy} has no
singularity. In addition to the above discussion on the
non-commutativity between the projection from WP to BP and the limit
of infinite depth tree/number of iterative updates, it is worth noting
that with the initial condition used in the backward algorithm the
Warning Propagation procedure corresponds actually to the whitening
construction (see for instance~\cite{Maneva}), starting from the
solution found at the end of the forward algorithm. The interpretation
of the number of frozen variables in {\sf A} and {\sf B} is thus
different: {\sf A} corresponds to the variables which are logically
implied (in the UCP sense) in the partially decimated formula. On the
other hand {\sf B} counts the number of constrained variables in the
core~\cite{Maneva} of the reference solution of the partially
decimated formula (the one reached by the decimation). The existence
of distinct WP fixed points would be a sign of a positive SP complexity
if one assumed these fixed points to be exponentially numerous. Even
when this assumption is correct it does
however not lead to a contradiction with the inexistence of thermodynamically
relevant clusters or long-range correlations as defined
in~(\ref{eq_correl2}). The former corresponds indeed to a 1RSB
computation with Parisi parameter $m=0$ and the latter to $m=1$. We
checked indeed that computing the residual entropy with the backward
algorithm yields the same result as with the decimation one, whereas
in presence of an extensive thermodynamical complexity we would have
obtained only the contribution from the internal entropy of the
cluster around the reference solution.

More indications come from the study of BP fixed point messages.  In
Fig.~\ref{fig:WPandBP} we plot the fraction of variables that receive
BP messages forcing it to a unique value (BP frozen) and the fraction
of variables which are extremely biased under the BP messages,
i.e. the less probable value has a probability smaller than $10^{-10}$
(BP prob$<10^{-10}$).  The fraction of BP frozen variables again shows
hysteresis: in the decimation algorithm BP frozen variables perfectly
coincide with WP frozen variables, while in the backward algorithm BP
frozen variables are a little more than WP frozen variables because of
the numerical impossibility of keeping marginals arbitrarily small.
What is more interesting is that the fraction of extremely biased
variables (BP prob$<10^{-10}$) does not show any sign of hysteresis:
the same smooth curve is followed by the decimation algorithm as well
as by the backward algorithm. This observation suggests that BP
marginals obtained by the two algorithms are exactly the same, except for
(almost) completely frozen variables.

\subsection{Large $k$ behaviour}

We have seen that the behaviour of the ensemble of decimated random
formulas is richer in the satisfiability case than for
xor-satisfiability, with in particular the appearance of two sets of
critical lines, one describing the thermodynamic properties,
$\omega(\theta)$, and the other the singularities of the logical
implications, $\phi(\theta)$. The common wisdom is however that when
the length $k$ of the clauses gets large the satisfiability model gets
simpler, notably allowing some tight rigorous
results~\cite{largek_1,largek_2,largek_3,largek_4}, 
and in fact becomes very similar to
the xor-satisfiability one. We shall hence briefly discuss now the
large $k$ limit of our results for the decimated ensembles.

Let us begin with the xor-satisfiability case; the results of
Sec.~\ref{sec_xor} having an explicit form, they can be easily turned
in asymptotic expansions for large $k$. For instance the threshold
$\alpha_*$ given in Eq.~(\ref{eq_xor_alpha_*}) is found to behave as
$\frac{e}{k}(1+O(k^{-1}))$, while the clustering threshold
$\alpha_{\rm d}$ is $\frac{\ln k}{k}(1+O(\ln \ln k/\ln k))$ and the
condensation threshold $\alpha_{\rm c}$ goes to 1 in the large $k$
limit. One can also study the behaviour of the transition line
$\theta_{\rm c}(\alpha)$ in the last of these three asymptotic scales
(i.e. for $\alpha$ constant with respects to $k$).  After a short
computation, which consists in expanding the fixed point solutions
$\phi(\theta)$ and $\psi(\theta)$ of Eq.(\ref{eq_xor_phi}), and the
associated entropies (\ref{eq_xor_homega}), one finds that the two
leading orders are $\theta_{\rm c}(\alpha) \sim 1-\alpha-\alpha
e^{-\alpha k}$.

\begin{figure}
\begin{center}
\includegraphics[width=8cm]{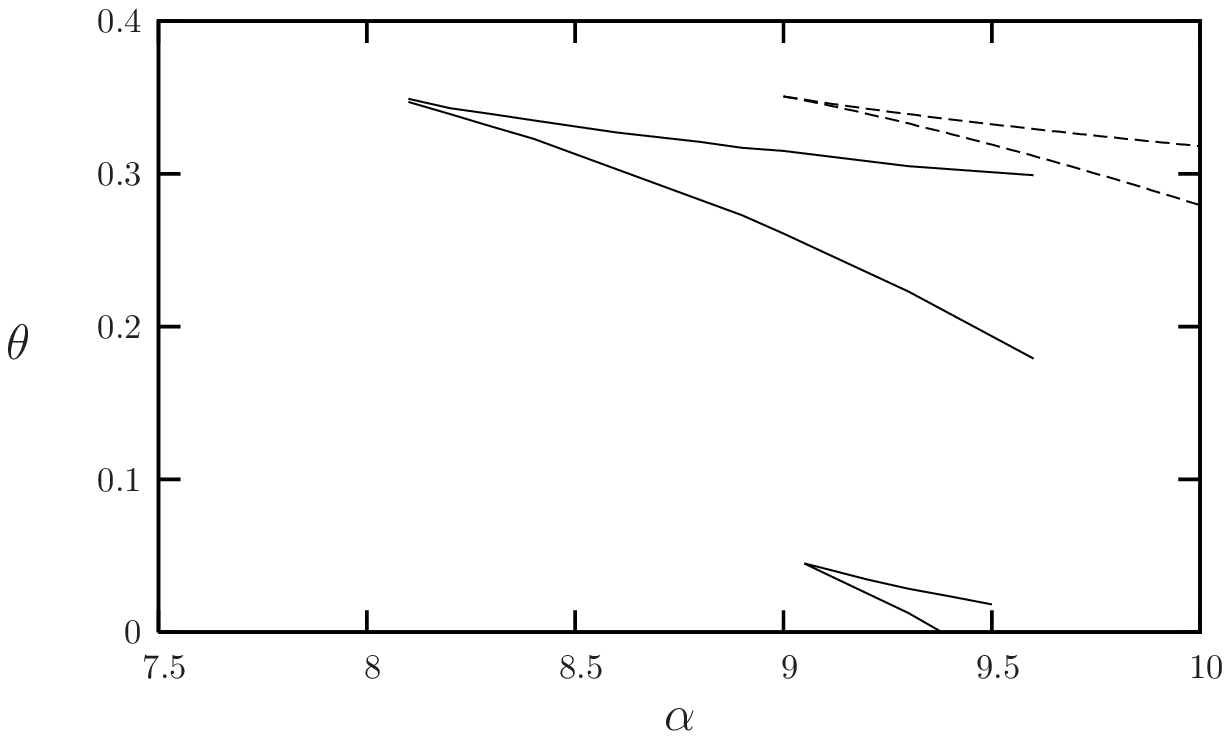}
\hspace{1cm}
\includegraphics[width=8cm]{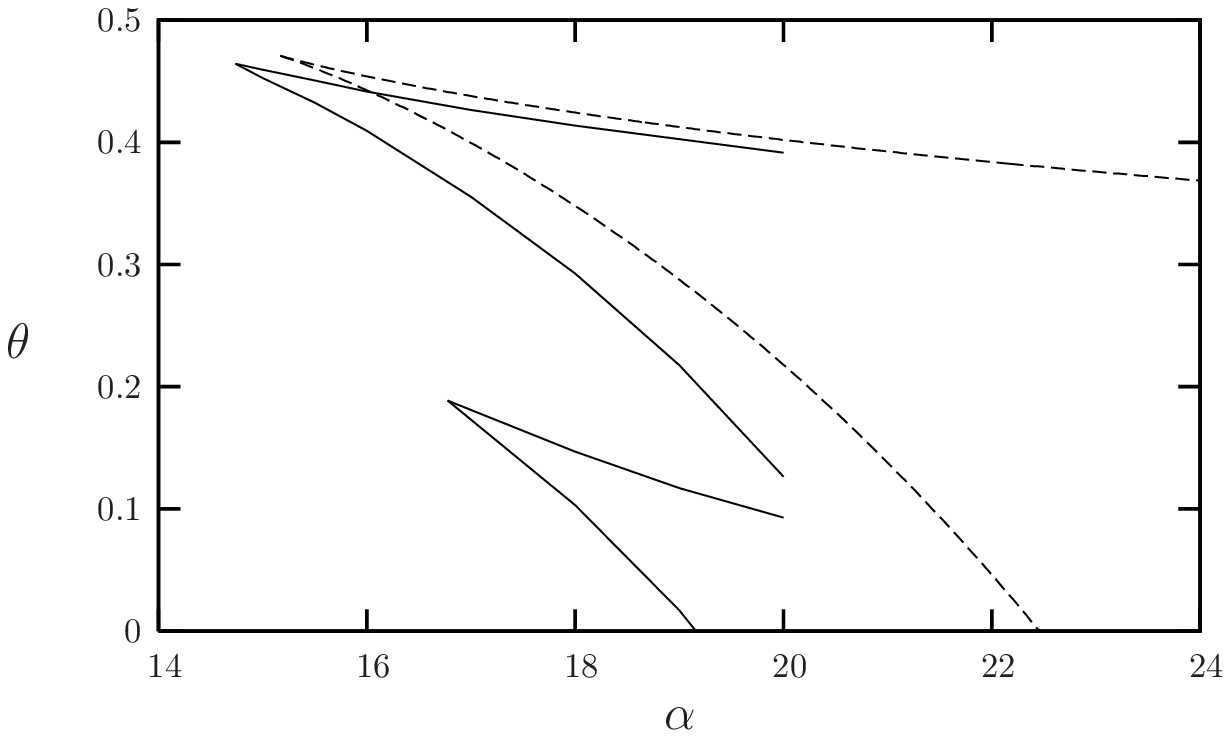}
\end{center}
\caption{Critical lines $\theta_\pm(\alpha)$ and $\theta'_\pm(\alpha)$
  for $k$-satisfiability, left: $k=4$, right: $k=5$.  The dashed lines
  are the $k$-xorsatisfiability critical lines, with a rescaling of
  $2^k$ on their clause density.}
\label{fig_compax}
\end{figure}

Large $k$ asymptotic expansions of the non-decimated ($\theta=0$)
ensemble of $k$-satisfiability were performed for the clustering and
condensation thresholds in~\cite{sat_long} and in~\cite{MeMeZe} for
the satisfiability one.  The leading order of the clustering threshold is
$2^k \frac{\ln k}{k}$, while both condensation and satisfiability
occurs for values of $\alpha$ around $2^k \ln 2$
(see~\cite{sat_long,MeMeZe} for the subleading corrections, which are
different for $\alpha_{\rm c}$ and $\alpha_{\rm s}$). Consider now the
fraction of decimated or implied variables $\phi(\theta)$, computed
for the satisfiability ensemble according to
Eqs.~(\ref{eq_sat_uW_dist},\ref{eq_sat_hW_dist}). In the large $k$
limit a crucial simplification occurs thanks to the concentration, at
the leading order, of the $h$ and $u$ random variables solutions of
the RS equation (\ref{eq_sat_RS}) around 0. From this fact follows
easily (compare with Eqs.~(\ref{eq_xor_ansatz},\ref{eq_xor_x_y_phi}))
that the functions $\phi(\theta)$ and $\psi(\theta)$, for
satisfiability random formulas of clause density $\alpha$, approach
the corresponding functions of xor-satisfiability formulas of clause
density $\alpha/2^k$. In consequence this correspondence holds for the
lines $\theta'_\pm(\alpha)$, and the critical point $\alpha'_*$ of
satisfiability is expected to scale as $\frac{e
  2^k}{k}(1+O(k^{-1}))$. The asymptotic study of the thermodynamic
lines $\theta_\pm(\alpha)$ is slightly more involved because of the
continuous nature of the second member of the pair in the random
variable $(h,h^\t)$, whereas it can only take two values in
$(h,\hW^\t)$. One can however notice that a consistent Ansatz in the
large $k$ limit is to assume $(h,\tanh h^\t) \approx (h, \t \hW^\t)$,
that is all non strictly forcing messages are approximated as
completely unbiased. If this hypothesis is correct the distinction
between $\theta_+(\alpha)$ (resp. $\theta_-(\alpha)$, $\alpha_*$) and
$\theta'_+(\alpha)$ (resp. $\theta'_-(\alpha)$, $\alpha'_*$) should
vanish in the large $k$ limit. We have not attempted to obtain a
formal proof of this statement but repeated the determination of the
satisfiability phase diagram for $k=5$. The results presented in
Fig.~\ref{fig_compax} (and in Tab.~\ref{tab_thr} for the values of the 
thresholds) confirms the intuition stated above. The two
sets of critical lines are much closer for $k=5$ than $k=4$, and also
in better agreement with the xor-satisfiability values (dashed lines,
with a rescaling factor of $2^k$ on the $\alpha$ axis). Finally an
expansion of the residual entropy at the leading order leads us to
conjecture that asymptotically $\theta_{\rm c}(\alpha) \sim 1 -
\frac{\alpha}{\alpha_{\rm c}(k)}$, as obtained explicitly in the
xor-satisfiability case. The datas of $\theta_{\rm c}(\alpha)$ obtained
by the population dynamics algorithm for $k=5$ (not shown) are already
in good agreement with this asymptotic form.

We have also performed BP guided decimation simulations for $k=5$, and
found an algorithmic threshold $\alpha_{\rm a}$ between 16
and $16.5$. This range is clearly above the appearance of the jump in
$\phi(\theta)$, thus confirming the results presented in
Sec.~\ref{sec_sat_results_cvg} for $k=4$, but it also a little bit
below the thermodynamic triple point (mainly because for $k=5$ the
constraint on the maximum number of iterations produces a more drastic
effect).

\begin{table}
\begin{tabular}{| c || c | c || c | c |}
\hline
$k$ & $\alpha_*$ & $\theta_*$ & $\alpha'_*$ & $\theta'_*$ \\
\hline
4 & 9.05 & 0.045 & 8.05 & 0.35 \\
\hline   
5 & 16.8 & 0.188 & 14.7 & 0.46 \\
\hline   
\end{tabular}
\caption{Values of the thresholds $\alpha_*$ and $\alpha'_*$, and the 
associated fraction of decimated variables $\theta_*$ and $\theta'_*$, 
for random $k$-satisfiability formulas, $k=4,5$.}
\label{tab_thr}
\end{table}

\section{Conclusions}
\label{sec_conclu}

In this article we have introduced analytical tools that allows
computations similar to the Franz-Parisi quenched potential for
diluted mean-field systems. We have precisely defined ensembles
of partially decimated random CSP and refined their analytical
description initiated in~\cite{allerton}. These methods have been
applied to xor-satisfiability random formulas, putting known 
results~\cite{xor1,xor2,xor3,Maxwell} in a slightly different 
perspective, and to the satisfiability case which presents a much
richer phase diagram. These computations are expected to describe
the behaviour of an hypothetic ideal decimation algorithm based
on an oracle able to compute exact marginal probabilities in large
graphical models.

We have then confronted these results with the outcomes of extensive
simulations of the BP guided decimation algorithm, which is a
practical, approximate implementation of the ideal procedure. In the
case of xor-satisfiability formulas the interpretation of the
comparison is very clear and can in fact be confirmed by rigorous
calculations. The satisfiability problem is much more difficult; the
interpretation of the results of the BP decimation should be based on
a precise description of the ``small'' errors made by BP, which
somehow accumulate along the decimation for large enough values of
$\alpha$ and cannot avoid conflicting choices in the decimated
variables. Lacking such a refined control of BP we have to turn to a
more intuitive explanation, based on the analysis of the ideal
algorithm. The algorithmic threshold $\alpha_{\rm a}$ for BP
decimation on random 4-satisfiability formulas is found to be very
close to the value $\alpha_*$ above which clustering and condensation
transitions do occur in the $(\alpha,\theta)$ plane. One is thus led
to conjecture more generally that the presence of a condensed regime,
in which BP is expected to fail because of replica symmetry breaking
effects, will coincide with the BP decimation threshold for generic
CSP.

It is fair to say that we first found the numerical results reported in 
this work quite surprising. We initially
expected~\cite{allerton} the BP guided decimation algorithm to fail
when $\phi(\theta)$ develops a jump, that is when the assignment of a
single variable produces an avalanche of $O(N)$ forced variables: in
this situation, we expected that a contradiction would be generated with
high probability.  Our numerical results clearly show this is not the
case: the algorithm we have studied is able to fix {\em at the same
  time} a finite fraction of variables without entering a
contradiction. Moreover, as soon as a thermodynamical condensation
transition is taking place, for $\alpha \ge \alpha_*$, the success
probability falls down to zero sharply.  The most natural explanation
to these observations is that the marginals used by the algorithm to
fix variables are extremely close to true marginals for any $\alpha <
\alpha_*$ (and this is clearly a very good news for the use of BP even
very close to the clustering threshold) and not so good above
$\alpha_*$.  Still some small differences between BP marginals and
true marginals are expected even below $\alpha_*$, mainly given by
$1/N$ corrections to the Bethe
approximation~\cite{MontRizzo,SlaninaParisi} and to the precision used
to solve BP equations ($\epsilon<10^{-4}$ in our case).  Then, why
these small differences between true marginals and BP estimations do
not affect the success probability of the algorithm before $\alpha_*$,
while they become relevant above $\alpha_*$? Maybe because the nature
of these errors changes crossing $\alpha_*$. Roughly speaking,
below $\alpha_*$ they have a statistical origin and produce random 
perturbations of
intensity $1/N$ and $\epsilon$ (in a sense that should be precised): 
if these errors are largely
uncorrelated, when summing $O(N)$ of these we still get errors of
order $1/\sqrt{N}$ and $\epsilon\sqrt{N}$, which are very small.  On
the contrary, above $\alpha_*$, deviations from true marginals are
{\em systematic}, because of long range correlations: in this case
errors are strongly correlated and summing $O(N)$ of these produces a
contradiction with high probability.

Let us also sketch a few possible directions for future research.
Apart from the computation of the usual Franz-Parisi potential, the
analytical formalism can be adapted to other CSP models besides the
satisfiability and xor-satisfiability cases treated here. For instance
the case of coloring should be relatively easy because of the
triviality of the RS description (as in xorsat), and relates to the
recent study of~\cite{quiet,quiet2}. It would also be interesting
to perform simulations of the BP decimation algorithm on coloring
formulas (this was done in~\cite{col} but with a bias in the choice of
the decimated variable), and check whether our conjecture on the closeness
of $\alpha_{\rm a}$ and $\alpha_*$ holds in this case.

A more rigorous analysis of the BP guided decimation algorithm would
be welcome, and should be easier to perform in the large $k$
limit. Until very recently the highest clause densities where
algorithms were rigorously shown to succeed in finding solutions were
$O(2^k /k)$~\cite{FrSu}.  Our rough analysis suggest that the
threshold of success for BP guided decimation should be on that scale
too, with $\alpha_{\rm a}(k) \sim e 2^k / k$. 
It has been shown very recently in~\cite{largek_3}
that a polynomial time algorithm
can find solutions of formulas with densities up to $2^k \ln k /k$,
which corresponds to the scale of the dynamic threshold $\alpha_{\rm d}(k)$.

On the algorithmic side many variations on the simplest procedure
studied in this paper are more efficient (and notably the Survey
Propagation algorithm~\cite{MeZe,SID}), yet seems much more difficult
to tackle analytically. A slight modification of the BP decimation
algorithm where the order of the assignments is not uniformly random
but treats in priority variables with the most biased marginals
already changes substantively the highest value of $\alpha$ where
formulas are solved with positive probability~\cite{pnas}.  Other
interesting directions to explore would be more efficients procedures
for the resolution of the BP equations (for instance double-loop
algorithms~\cite{doubleloop}), the use of reinforcement
strategies~\cite{reinforcement} instead of explicit decimation, or
the coordinated decimation of groups of variables~\cite{suscp}.

\acknowledgments

We warmly thank Andrea Montanari with whom part of this work was done
and who made important suggestions. We are also grateful to Francesco Zamponi
and Lenka Zdeborova for a critical reading of the manuscript.
This work has been partially funded by the PHC Galileo program for
exchanges between France and Italy.

\appendix

\section{Details on the computations of Sec.~\ref{sec_xor_direct}}
\label{app_xor}

We present in this appendix some details on the computations discussed
in Sec.~\ref{sec_xor_direct}. The properties of decimated xorsat formulas 
will be derived through the analysis of two algorithms which act on the formula
and which can be described by the differential equation
method~\cite{Kurtz,Wormald,Ac}. Two successive steps will be performed~: the
logical implications of the decimation of a fraction $\theta$ of the variables
are first drawn with Unit Propagation. Then the structure of the set of
solutions of the reduced formula is analyzed by the leaf removal algorithm.
This approach has been developed for $k$-xorsat formulas in~\cite{xor1,xor2} 
and generalized to arbitrary degree distributions in~\cite{xor3}, we reproduce
here their results for the sake of self-containedness.

\subsection{Unit propagation}

As explained at the beginning of Sec.~\ref{sec_xor_direct} we can assume
here the formula to be an unfrustrated ($J_a=1$) set of $M=\alpha N$
equations of the form~(\ref{eq_psi_xorsat}), each involving $k$ indices chosen
uniformly at random among the $N$ variables. A fraction $\theta$ of the
variables are then set to $+1$, and can then be removed from the clauses where
they appeared. Let us call $R_\kappa=N \rho_\kappa$ the number of clauses with 
$\kappa$ non decimated variables, and $L_l =N \lambda_l$ the number of
variables which appears in $l$ clauses (note that a decimated variable does
not appear in any clause after the above simplification step). One obtains:
\begin{eqnarray}
\lambda_l &=& (1-\theta) e^{-\alpha k} \frac{(\alpha k)^l}{l!} + 
\theta \, \delta_{l,0} \ , \label{eq_xor_init_1}\\
\rho_\kappa &=& \alpha \binom{k}{\kappa} (1-\theta)^\kappa \theta^{k-\kappa} 
\ \ \text{for} \ \kappa \le k \ . \label{eq_xor_init_2}
\end{eqnarray}
Consider now the action of the Unit Propagation algorithm. As long as clauses
of length $\kappa=1$ are present in the formula, one of them is chosen
randomly, the single variable it contains is fixed to $+1$, and is then
removed from the other clauses it appeared in. The formula obtained after
$T$ steps of this procedure is uniformly random, conditional on the values of
$\{R_\kappa(T), L_l(T) \}$, so that the analysis of the process amounts to
follow these random variables. At each time step $T \to T+1$ 
they vary by a bounded random increment whose distribution depends only on
the current values $\{R_\kappa(T), L_l(T) \}$ and not on the previous history
of the process. In consequence the reduced
quantities $\rho_\kappa(t) = R_\kappa(T=Nt)/N$ and $\lambda_l(t)=L_l(T=Nt)/N$
concentrate around their average values~\cite{Kurtz,Wormald,Ac}, solutions of
the following set of differential equations:
\begin{eqnarray}
\frac{\dd\phantom{t}}{\dd t} \lambda_l(t) &=& \delta_{l,0}
-\frac{l \lambda_l(t)}{\sum_{l'} l' \lambda_{l'}(t)} \ , \label{eq_xor_up_1} \\
\frac{\dd\phantom{t}}{\dd t} \rho_\kappa(t) &=& - \delta_{k,1}
+ \left(\frac{\sum_l l(l-1) \lambda_l(t)}{\sum_l l \lambda_l(t)} \right)
\left[\frac{(\kappa+1) \rho_{\kappa+1}(t)}
{\sum_{\kappa'}\kappa' \rho_{\kappa'}(t) } 
- \frac{\kappa \rho_\kappa(t)}
{\sum_{\kappa'}\kappa' \rho_{\kappa'}(t) }\right] \ . \label{eq_xor_up_2}
\end{eqnarray}
These expressions arise because the variable selected in the unit clause is
present in $l$ clauses with probability proportional to $l \lambda_l(t)$;
apart from the unit clause, the $l-1$ other occurences of the variable take
place in clauses of length $\kappa$ with probability proportional to
$\kappa \rho_\kappa(t)$.

In order to solve these equations we introduce the generating function of the
initial distribution of degrees of the variables,
$\lambda(x) = \sum_l \lambda_l x^l$. Equation~(\ref{eq_xor_up_1}) is
solved for any $l\ge 1$ by $\lambda_l(t) = \lambda_l a(t)^l$,
where $a(t)$ is the solution of
\begin{equation}
\frac{\dd \phantom{t}}{\dd t} a(t) = - \frac{1}{\lambda'(a(t))} \ , \ \ 
\text{with the initial condition} \ a(t=0)=1 \ .
\label{eq_xor_up_a}
\end{equation}
One can then insert this expression of $\lambda_l(t)$ in (\ref{eq_xor_up_2})
and solve to obtain
\begin{equation}
\rho_\kappa(t) = \sum_{\kappa' \ge \kappa} \binom{\kappa'}{\kappa} 
\rho_{\kappa'} \left( \frac{\lambda'(a(t))}{\lambda'(1)} \right)^\kappa
\left(1 - \frac{\lambda'(a(t))}{\lambda'(1)} \right)^{\kappa'-\kappa}
- \delta_{\kappa,1} \lambda'(a(t)) (1-a(t)) \ .
\label{eq_xor_up_rho}
\end{equation}
The differential equations~(\ref{eq_xor_up_1},\ref{eq_xor_up_2}) only make
sense if $\rho_1(t) > 0$~: the procedure stops when no more unit clause can be
selected, i.e. at the reduced stopping time $t_*=\min \{t : \rho_1(t)=0 \}$.

For the initial degree distribution given in 
(\ref{eq_xor_init_1}) one obtains for the derivative
of the generating function:
$\lambda'(x) = \alpha k (1-\theta)e^{-\alpha k (1-x)}$, hence by integration
of (\ref{eq_xor_up_a})
\begin{equation}
a(t) = 1 - \frac{1}{\alpha k}\ln\left(\frac{1-\theta}{1-\theta-t} \right) \ .
\end{equation}
Plugging this result in the expression (\ref{eq_xor_up_rho}) of $\rho_1(t)$,
one finds that $t_*$ is the smallest solution of
\begin{equation}
\alpha k (\theta + t)^{k-1} = \ln\left(\frac{1-\theta}{1-\theta-t} \right)
\ .
\end{equation}
Defining finally $\phi=\theta+t_*$, one realizes that $\phi$ is the smallest 
solution of Eq.~(\ref{eq_xor_phi}): this quantity gives the fraction of
variables that are either fixed by the decimation or by the propagation of
logical implications. When all these implications are taken into account, the
degree distributions of the reduced formula reads
\begin{eqnarray}
\lambda_l(t_*) &=& (1-\phi) e^{-\alpha k (1-\phi^{k-1})} 
\frac{(\alpha k (1-\phi^{k-1}))^l}{l!}  \ \ \text{for} \ l \ge 1 \ ,
\label{eq_xor_init2_1}\\
\rho_\kappa(t_*) &=& \alpha \binom{k}{\kappa} (1-\phi)^\kappa \phi^{k-\kappa} 
\ \ \text{for} \ \kappa \ge 2 \ .\label{eq_xor_init2_2}
\end{eqnarray}
In consequence the number of non-trivial clauses is
\begin{equation}
M' = N \sum_{\kappa=2}^k \rho_\kappa(t_*) = 
\alpha N \left(1-\phi^k - k (1-\phi) \phi^{k-1} \right) \ .
\label{eq_xor_up_Mprime}
\end{equation}

\subsection{Leaf removal}

We shall now analyze the set of solutions of random unfrustrated formulas
with degree distributions given by 
(\ref{eq_xor_init2_1},\ref{eq_xor_init2_2}). Following~\cite{xor1,xor2}, we
consider the action of the leaf removal algorithm on such a hypergraph. 
Each leaf removal step consists in picking at random one variable of degree
1 (a leaf), and remove the single clause it appeared in. This simplification 
is repeated until no leaf is left in the graph, which provokes the stopping of
the algorithm. There are two possible situations at that point: either all
clauses have been removed, or there remains a non empty 2-core, that is the
maximal subgraph of the original formula in which all variables have degree
at least 2. In both cases the total entropy is given (in units of $\log 2$)
by the initial number of variables minus the initial number of clauses.
In the former case the set of solutions is unclustered, while in the latter 
the solutions are split into an exponential number of clusters. Each cluster
corresponds to one solution of the 2-core formula, hence the complexity, 
i.e. the exponential rate of the number of clusters, is given by the 
difference between the number of variables and of clauses in the 2-core.
Each cluster contains an exponential number of solutions, this internal
entropy being associated to the degeneracy arising from the freedom in
the choice of the value of the leaf variables when reinserted in the formula
in the reverse order with respect to their removal.

The evolution of the degree connectivities during the execution of the leaf 
removal algorithm can be computed in a very similar manner with respect to
the Unit Propagation case sketched above. With a slight abuse of notation
we denote again $N\lambda_l(t)$ and $N\rho_\kappa(t)$ the average number of
variables and constraints of degrees $l$ and $\kappa$ after $Nt$ steps of
the leaf removal algorithm. These quantities obey the following set of
differential equations:
\begin{eqnarray}
\frac{\dd \phantom{t}}{\dd t} \rho_\kappa(t) &=& -\frac{\kappa \rho_\kappa(t)}
{\sum_{\kappa'} \kappa' \rho_{\kappa'}(t)}  \ , \label{eq_xor_leaf_1} \\
\frac{\dd\phantom{t}}{\dd t} \lambda_l(t) &=& - \delta_{l,1} + \delta_{l,0}
+ \left(\frac{\sum_\kappa \kappa(\kappa-1) \rho_\kappa(t)}
{\sum_\kappa \kappa \rho_\kappa(t)} \right)
\left[\frac{(l+1) \lambda_{l+1}(t)}
{\sum_{l'}l' \lambda_{l'}(t) } 
- \frac{l \lambda_l(t)}
{\sum_{l'} l' \lambda_{l'}(t) }\right] \ . \label{eq_xor_leaf_2}
\end{eqnarray}
These equations are essentially the same as 
(\ref{eq_xor_up_1},\ref{eq_xor_up_2}) with the role of $\lambda$ and $\rho$
being exchanged (in the leaf removal algorithm one picks a variable of
degree 1, in Unit Clause Propagation it is a clause of degree 1). 
They can thus be
solved with the same technique. Let us define the generating function of
the clause lengths at the beginning of the leaf removal,
$\rho(x)=\sum_\kappa \rho_\kappa x^\kappa$. At all times 
$\rho_\kappa(t) =\rho_\kappa b(t)^\kappa $, where $b(t)$ is solution of
\begin{equation}
\frac{\dd \phantom{t}}{\dd t} b(t) = - \frac{1}{\rho'(b(t))} \ , \ \ \ \ 
\text{with} \ b(t=0)=1 \ .
\label{eq_xor_leaf_b}
\end{equation}
The distribution of the variable degrees is then found to be
\begin{equation}
\lambda_l(t) = \sum_{l' \ge l} \binom{l'}{l} 
\lambda_{l'} \left( \frac{\rho'(b(t))}{\rho'(1)} \right)^l
\left(1 - \frac{\rho'(b(t))}{\rho'(1)} \right)^{l'-l}
- \delta_{l,1} \rho'(b(t)) (1-b(t)) \ .
\label{eq_xor_leaf_lambda}
\end{equation}
The stopping time of the leaf removal algorithm is given by 
$t_*=\min \{t : \lambda_1(t)=0 \}$. A non-trivial 2-core exists at this
stopping time if and only if $b(t_*) > 0$. As the function $b(t)$ is
decreasing in time (see Eq.~(\ref{eq_xor_leaf_b})), the value $b_*=b(t_*)$
can also be defined as the largest solution in $[0,1]$ of
\begin{equation}
\rho'(b_*) (1-b_*) = \frac{\rho'(b_*)}{\rho'(1)} 
\sum_{l=1}^\infty l \lambda_l 
\left(1 - \frac{\rho'(b_*)}{\rho'(1)} \right)^{l-1} \ .
\label{eq_xor_leaf_bstar}
\end{equation}

Let us now apply these results to the degree distributions 
(\ref{eq_xor_init2_1},\ref{eq_xor_init2_2}) of the formula obtained at the end
of the Unit Propagation. These imply the following form of the derivative
of the clause length generating function:
\begin{equation}
\rho'(x) = \alpha k (1-\phi) \left((\phi+x(1-\phi))^{k-1} - \phi^{k-1} 
\right) \ .
\end{equation}
The solution of the equation (\ref{eq_xor_leaf_bstar}) is either $b_*=0$, or
the largest strictly positive solution of
\begin{equation}
1-b_* = \exp\left[-\alpha k \left(\phi + b_*(1-\phi) \right)^{k-1} 
- \phi^{k-1} \right] \ .
\end{equation}
Defining $b_* = (\psi - \phi)/(1-\phi)$, one realizes that the
equation on $b_*$ is equivalent to $\psi$ being the largest solution of 
Eq.~(\ref{eq_xor_phi}). We have thus justified one of the statement made
in Sec.~\ref{sec_xor_direct}~: when there is only one
solution to Eq.~(\ref{eq_xor_phi}), $\psi=\phi$ or in other terms $b_*=0$.
This means that the leaf removal algorithm does not stop before having 
emptied the complete formula, there is no 2-core and the solution space is not
clustered. On the contrary the existence of the multiple solutions $\psi >
\phi$ corresponds to $b_* > 0$ and hence to the presence of a non-trivial
2-core in the hypergraph of constraints. This latter case corresponds to
$\alpha > \alpha_*$ and $\theta \in [\theta_-(\alpha),\theta_+(\alpha)]$.

Let us conclude with the justification of the expressions of the entropy and
complexity given in Sec.~\ref{sec_xor}. We have seen that the number of
non-implied variables at the end of the Unit Propagation procedure is
$N(1-\phi)$, while the number of non-trivial clauses is given in 
Eq.~(\ref{eq_xor_up_Mprime}). In the region of the $(\alpha,\theta)$ plane
where there is a single solution of Eq.~(\ref{eq_xor_phi}) the entropy
(\ref{eq_xor_homega}) is given (in units of $\ln 2$) by the difference between
the number of variables and constraints, in agreement with the results
of~\cite{xor1,xor2}. When a non-empty 2-core subsists at the end of the leaf
removal algorithm, one can compute from the solution of 
(\ref{eq_xor_leaf_1},\ref{eq_xor_leaf_2}) at the stopping time $t_*$ the
number of variables and clauses in the 2-core:
\begin{eqnarray}
N_{\rm core} &=& N 
\left[\psi-\phi - \alpha k (1-\psi) (\psi^{k-1} - \phi^{k-1}) \right] \ ,\\
M_{\rm core} &=& N \alpha 
\left[\psi ^k -\phi^k - k \phi^{k-1} (\psi - \phi) \right] \ .
\end{eqnarray}
It is then easy to verify that 
$\homega(\phi) - \homega(\psi) = \ln(2) (N_{\rm core} - M_{\rm core}  )/N$.
When this quantity is positive, that is in the interval 
$[\theta_-(\alpha),\theta_{\rm c}(\alpha)]$, it is equal to the entropy
density associated with the exponential number of solutions of the 2-core. On
the contrary when it is negative the 2-core with more clauses than variables
has only a subexponential number of solutions (recall that we conditioned from
the beginning on the formula being satisfiable and on the reference
configuration unveiled being a solution). One can show in this case that the
entropy arising from the variables outside the 2-core is given by 
$\homega(\psi)$, hence the total entropy of the decimated formula is always
given by $\max[\homega(\phi),\homega(\psi)]$, the largest branch as plotted
in the right panel of Fig.~\ref{fig_xor_omega}.







\begin{thebibliography}{99}

\bibitem{Beyond} M.~M\'ezard, G.~Parisi and M.A.~Virasoro, 
\emph{Spin glass theory and beyond}, World Scientific (1987).

\bibitem{Talagrand} M.~Talagrand,
\emph{Spin glasses: a challenge for mathematicians}, Springer (2003).

\bibitem{IPC} M.~M\'ezard and A.~Montanari, 
\emph{Information, Physics, and Computation}, Oxford University Press (2009).

\bibitem{Friedgut} E.~Friedgut, J. Amer. Math. Soc. {\bf 12}, 1017 (1999).

\bibitem{lb} J.~Franco,
Theoretical Computer Science {\bf 265}, 147 (2001).

\bibitem{ub} O.~Dubois,
Theoretical Computer Science {\bf 265}, 187 (2001).

\bibitem{MeZe} M.~M\'ezard and R.~Zecchina,
Phys. Rev. E {\bf 66}, 056126 (2002).

\bibitem{MePaZe} M.~M\'ezard, G.~Parisi, and R.~Zecchina,
Science {\bf 297}, 812 (2002).

\bibitem{MeMeZe} S.~Mertens, M.~M\'ezard and R.~Zecchina,
Random Struct. Alg. {\bf 28}, 340 (2006).

\bibitem{BiMoWe} G.~Biroli, R.~Monasson and M.~Weigt,
Eur. Phys. J. B {\bf 14}, 551 (2000).

\bibitem{pnas}
F.~Krzakala, A.~Montanari, F.~Ricci-Tersenghi, 
G.~Semerjian and L.~Zdeborova, 
Proc.~Natl.~Acad.~Sci. {\bf 104}, 10318 (2007).

\bibitem{wsat1} G.~Semerjian and R.~Monasson, 
Phys. Rev. E {\bf 67}, 066103 (2003).

\bibitem{wsat2} W.~Barthel, A.~Hartmann and M.~Weigt,
Phys. Rev. E {\bf 67}, 066104 (2003).

\bibitem{wsat3} S.~Seitz, M.~Alava and P.~Orponen,
J. Stat. Mech. P06006 (2005).

\bibitem{DPLL} M.~Davis and H.~Putman, 
J. Ass. Comput. Mach. {\bf 7}, 201 (1960).
M.~Davis, G.~Logemann and D.~Loveland, Commun. ACM {\bf 5}, 394 (1962).

\bibitem{SID} A.~Braunstein, M.~M\'ezard and R.~Zecchina,
Random Struct. Alg. {\bf 27}, 201 (2005).

\bibitem{MosselPlanted} U.~Feige, E.~Mossel and  D.~Vilenchik, 
Proceedings of RANDOM, Barcelona,  (2006).

\bibitem{unif_sat} A.~Coja-Oghlan, M.~Krivelevich and D.~Vilenchik,
Proceedings of the 13th International Conference on Analysis of Algorithms
(2007). 

\bibitem{unif_sat2} F. Altarelli, R. Monasson and F. Zamponi,
J. Phys. A {\bf 40}, 867 (2007).

\bibitem{Pretti} M.~Pretti, J. Stat. Mech. P11008 (2005).

\bibitem{Aurell} E.~Aurell, U.~Gordon and S.~Kirkpatrick,
Eighteenth Annual Conference on Neural Information Processing Systems 
(NIPS 2004).

\bibitem{allerton} A.~Montanari, F.~Ricci-Tersenghi and G.~Semerjian,
Proceedings of the 45th Allerton Conference, pp. 352--359 (2007).

\bibitem{ZdMe} L.~Zdeborova and M.~M\'ezard,
J. Stat. Mech. P12004 (2008).

\bibitem{factor} F.~Kschischang, B.J.~Frey and H.-A.~Loeliger,
IEEE Transactions on Information Theory {\bf 47}, 498 (2001).

\bibitem{Yedidia} J.~S.~Yedidia, W.~T.~Freeman and Y.~Weiss, in 
\emph{Advances in Neural Information Processing Systems} {\bf 13}, 689 (2001).

\bibitem{potential} S.~Franz and G.~Parisi,
J. Phys. I France {\bf 5}, 1401 (1995).

\bibitem{NP} M.R.~Garey and D.S.~Johnson, 
\emph{Computers and Intractability: 
A Guide to the Theory of NP-Completeness}, W.~H.~Freeman (1983).

\bibitem{xor1} M. M\'ezard, F. Ricci-Tersenghi and R. Zecchina,
J.~Stat.~Phys. {\bf 111}, 505 (2003).

\bibitem{xor2} S. Cocco, O. Dubois, J. Mandler and R. Monasson,
Phys.~Rev.~Lett. {\bf 90}, 047205 (2003).

\bibitem{sat_long} A.~Montanari, F.~Ricci-Tersenghi and G.~Semerjian,
J. Stat. Mech. P04004 (2008).

\bibitem{MeMo} M.~M\'ezard and A.~Montanari, 
J. Stat. Phys. {\bf 124}, 1317 (2006).

\bibitem{MePa} M.~M\'ezard and G.~Parisi, 
Eur. Phys. J. B {\bf 20}, 217 (2001).

\bibitem{xor3} F.~Altarelli, R.~Monasson and F.~Zamponi,
Journal of Physics: Conference Series {\bf 95}, 012013 (2008).

\bibitem{Maxwell} C.~Measson, A.~Montanari and R.~Urbanke,
IEEE Transactions on Information Theory {\bf 54}, 5277 (2008).

\bibitem{global} A.~Braunstein, M.~Leone, F.~Ricci-Tersenghi and R.~Zecchina,
J. Phys. A {\bf 35}, 7559 (2002).

\bibitem{FrSu} A.~Frieze and S.~Suen, 
J. Algorithms {\bf 20}, 312 (1996).

\bibitem{DeMo} C.~Deroulers and R.~Monasson,  
Eur. Phys. J. B {\bf 49}, 339 (2006).

\bibitem{ksat_RS} R.~Monasson and R.~Zecchina, 
Phys. Rev. E {\bf 56}, 1357 (1997).

\bibitem{Maneva} E.~Maneva, E.~Mossel and M.J.~Wainwright,
Journal of the ACM {\bf 54}, 1 (2007).

\bibitem{largek_1} D.~Achlioptas and Y.~Peres, 
Journal of the AMS {\bf 17}, 947 (2004).

\bibitem{largek_2} D.~Achlioptas and A.~Coja-Oghlan, 
{\tt arXiv:}0803.2122 (2008).

\bibitem{largek_3} A.~Coja-Oghlan, {\tt arXiv:}0902.3583 (2009).

\bibitem{largek_4} A.~Montanari, R.~Restrepo and P.~Tetali,
{\tt arXiv:}0904.2751 (2009).

\bibitem{MontRizzo} A.~Montanari and T.~Rizzo,
J. Stat. Mech., P10011 (2005).

\bibitem{SlaninaParisi} G.~Parisi and F.~Slanina,
J. Stat. Mech., L02003 (2006).

\bibitem{quiet} F.~Krzakala and L.~Zdeborova, 
{\tt arXiv:}0901.2130 (2009).

\bibitem{quiet2} L.~Zdeborova and F.~Krzakala, 
{\tt arXiv:}0902.4185 (2009).

\bibitem{col} L.~Zdeborova and F.~Krzakala, 
Phys. Rev. E {\bf 76}, 031131 (2007).

\bibitem{doubleloop} A.~Yuille, Neural Computation {\bf 14}, 1691 (2002).

\bibitem{reinforcement} J.~Chavas, C.~Furtlehner, M.~M\'ezard and R. Zecchina, 
J. Stat. Mech. P11016 (2005).

\bibitem{suscp} S.~Higuchi and M.~M\'ezard, {\tt arXiv:}0903.1621 (2009)

\bibitem{Kurtz} T.G.~Kurtz, J. Appl. Probab. {\bf 7}, 49 (1970).

\bibitem{Wormald} N.C.~Wormald, in Lectures on Approximation and Randomized 
Algorithms (M. Karonski and H.J. Proemel, eds), pp. 73-155. PWN, Warsaw, 
(1999).

\bibitem{Ac} D.~Achlioptas,
Theoretical Computer Science {\bf 265}, 159 (2001).

\end{thebibliography}
\end{document}